\newcommand{\project}{Measuring what matters: \\A scalable framework for application-level quantum benchmarking}
\newcommand{\reporttype}{White paper}

\newcommand{\disclosuretype}{
Confidential 
}

\documentclass[11pt]{article}
\usepackage[utf8]{inputenc}

\usepackage[letterpaper, top=.25in,bottom=1in,right=1in,left=1in,headsep=1in,includehead]{geometry}

\usepackage{fancyhdr}
\usepackage{graphicx}
\usepackage{xcolor}
\usepackage{comment}
\usepackage{float}
\usepackage{placeins}
\usepackage{bbm}

\usepackage{booktabs}
\RequirePackage[colorlinks=true, allcolors=blue]{hyperref}
\usepackage{array}
\usepackage{multirow}
\usepackage{algorithm}
\usepackage{algpseudocode}
\usepackage{bm}
\usepackage{amsmath,amssymb}
\usepackage[toc,page]{appendix}
\usepackage[pages=all,placement=top]{background}
\usepackage{todonotes}
\usepackage[braket, qm]{qcircuit}
\usepackage{braket}
\usepackage{tabularx}
\usepackage{makecell}
\usepackage{threeparttable}
\usepackage{longtable}
\usepackage{siunitx}
\usepackage[absolute]{textpos}
\usepackage{datetime}
\usepackage{enumitem}

\usepackage{minted}

\usepackage[version=4]{mhchem}

\usepackage{subcaption}

\usepackage{cleveref}

\newdateformat{monthyeardate}{%
  \monthname[\THEMONTH] \THEYEAR}

\newcommand{\copyrightdisclaimer}{  Copyright \textcopyright \the\year{} IonQ, Inc. All Right Reserved.\\
  IonQ, Inc. \disclosuretype Information.}

\title{\project
\\\vskip 12pt\large{\reporttype}}
\date{April 14th, 2026}

\usepackage{listings}
\usepackage{color}
\usepackage{pagecolor}

\usepackage{pythonhighlight}

\definecolor{dkgreen}{rgb}{0,0.6,0}
\definecolor{gray}{rgb}{0.5,0.5,0.5}
\definecolor{mauve}{rgb}{0.58,0,0.82}

\definecolor{ionqorange}{HTML}{FF5000}
\definecolor{ionqbackground}{RGB}{19,25,31}

\definecolor{jsonStrings}{RGB}{42,0.0,255}
\definecolor{jsonKeywords}{RGB}{127,0,85}
\colorlet{numb}{magenta!60!black}

\lstdefinelanguage{json}{
    basicstyle=\small\ttfamily,
    commentstyle=\color{jsonStrings}, 
    stringstyle=\color{jsonKeywords}, 
    numbers=none,
    numberstyle=\scriptsize,
    stepnumber=1,
    numbersep=8pt,
    showstringspaces=false,
    breaklines=true,
    frame=lines,
    backgroundcolor=\color{white}, 
    string=[s]{"}{"},
    comment=[l]{:\ "},
    morecomment=[l]{:"},
    literate=
        *{0}{{{\color{numb}0}}}{1}
         {1}{{{\color{numb}1}}}{1}
         {2}{{{\color{numb}2}}}{1}
         {3}{{{\color{numb}3}}}{1}
         {4}{{{\color{numb}4}}}{1}
         {5}{{{\color{numb}5}}}{1}
         {6}{{{\color{numb}6}}}{1}
         {7}{{{\color{numb}7}}}{1}
         {8}{{{\color{numb}8}}}{1}
         {9}{{{\color{numb}9}}}{1}
}

\fancyheadoffset[R]{1.2cm}
\fancyfootoffset{1.4cm}

\begin{document}

\clearpage 
\color{black} 

\backgroundsetup{scale=1,contents={
    \includegraphics[width=\paperwidth]{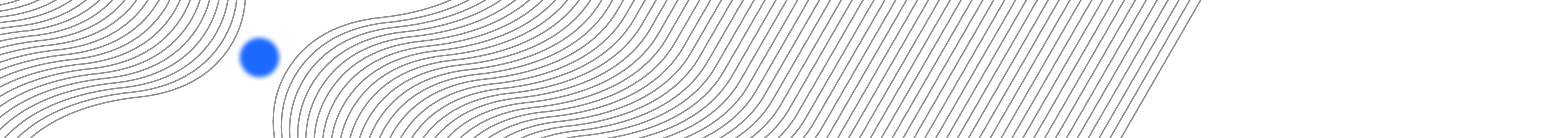}}
}


\pagestyle{fancy}
\fancyhf{}
\cfoot{\thepage}

\begin{center}

{\LARGE \bf Measuring what matters: A scalable framework for application-level quantum benchmarking\par}

\vspace{4ex}

Willie Aboumrad,
Claudio Girotto,
Joshua Goings,
Luning Zhao,
Miguel Angel Lopez-Ruiz,
Daiwei Zhu,
Ananth Kaushik,
Sayonee Ray,
Samwel Sekwao,
Jason Iaconis,
Andrew Arrasmith,
Andrii Maksymov,
Yvette de Sereville,
Felix Tripier,
Far McKon,
Coleman Collins,\\
Evgeny Epifanovsky,
Masako Yamada,
Martin Roetteler

\vspace{2ex}
IonQ Inc, College Park, MD, 20740, USA\\
\vspace{2ex}
April 14th, 2026
\vspace{2ex}

\end{center}

\begin{abstract}
As quantum computing systems continue to mature, there is an increasing need for benchmarking methodologies that capture performance in terms of meaningful, application-level metrics. In this work, we present a scalable framework for application-level quantum benchmarking that is designed to support internal system evaluation and cross-platform comparison across technology providers. Our framework is guided by a set of core principles, including measurability, simplicity, scalability, and extensibility. We present 13 benchmark families that reflect realistic workloads across multiple domains. This enables the systematic evaluation of the quality of solutions, the total execution time, total used energy, as well as Time-to-Solution. The benchmarks are designed to be reproducible, interpretable across stakeholder groups, and adaptable to evolving system capabilities. The framework aims to bridge the gap between low-level performance metrics and real-world value, providing a unified approach to assessing quantum systems. The resulting benchmarks support development and validation and contribute to the foundation of industry-wide benchmarking standards.
\end{abstract}

\newpage
\tableofcontents

\section{Introduction}

The development of application-level benchmarks is essential for systematically evaluating and communicating the performance of quantum computing systems. Benchmarks serve multiple roles across internal development, external communication, and industry-wide comparison, each with distinct objectives and requirements. The benefits of benchmarks include the ability to monitor system readiness and guide platform development, perform engineering tests at the system level (``integration tests''), communicate system value to customers and stakeholders, and enable cross-platform comparisons across technology providers. Considerable efforts have already been made to identify and develop application-level benchmarks, see e.g.~\cite{
siekierski2025scalablebenchmarks,
lubinski2024applicationbenchmarks,
patel2025modularbenchmarking,
chatterjee2025crossmodel,
koch2025decathlon,
lubinski2023optimizationbenchmarks,
lubinski2023applicationbenchmarks,
sawaya2024hamlib}. The following principles guided the choices made in the development of the application-level benchmarking framework presented in this document:  

\begin{itemize}
    \item \textbf{Measurability}: Benchmarks must be inexpensive enough to run frequently for continuous monitoring. Physical quantities such as the execution time, execution energy, and Time-to-Solution should be reported. Metrics should be actionable and be usable to directly inform system characterization, integration testing, and improvement efforts.
    \item \textbf{Simplicity}: Outputs should be easy to understand, mathematically precise. It should be easy to integrate them into dashboards, internal reports, build pipelines. There should be a simple score produced by each benchmark which allows to assess performance. 
    \item \textbf{Scalability}: Benchmarks should consist of families of instances for various input problem sizes and be able to be scaled to larger problems, covering the NISQ regime but also be extensible into the FTQC regime. 
    \item \textbf{Extensibility}: The benchmark collection should be extensible and the community should be able to compete for each reported benchmark result and be able to report their own findings. 
\end{itemize}
A key challenge in selecting benchmarks is ensuring that they capture system-level performance in ways that are meaningful to end users, i.e., they must reflect real-world computational workloads, results must be reproducible across similarly capable hardware platforms, and performance must be reported consistently in terms that connect directly to the cost and effort of obtaining a solution. To address these needs, we selected a total collection of 13 benchmark families which span a wide variety of both, underlying application technology as well as use cases. Table \ref{tab:benchmark-overview} gives an overview of the selection. 

Application benchmarks should evaluate the full system stack: hardware, compiler, runtime, performance management, and software layers, all working in concert as this reflects real-world workflows. The term performance management is used in this document broadly, encompassing error mitigation, error correction, circuit optimization, and other algorithmic optimizations, each of which represents a distinct capability that can enhance how well a system executes a workload in practice. Just like with MLPerf, all techniques are acceptable, as long as they are actually available as value to users and are disclosed in the benchmark.

Component-level metrics such as gate fidelity, qubit count, and logical qubit count are critical and meaningful in their own right, but in isolation they do not capture the complete picture in terms of impact on applications or commercial value. When designing quantum computers, there is a complex trade-off space across component-level metrics driven by architectural choices. For example, higher gate fidelity may come at the expense of slower gate speeds, increasing logical qubit count through error correction consumes physical qubits, and compiler optimizations interact with hardware constraints in ways that vary by algorithm. An application-level benchmark suite of sufficient size and coverage measures how all of these components work in concert across the full problem stack, inclusive of the hybrid quantum-classical components of the workload. This is the complete articulation of performance that component-level metrics alone fail to provide.

The framework we have chosen is inspired by MLPerf~\cite{mattson2020mlperf}, the standard for AI benchmarking, due to its emphasis on fairness, transparency, and real-world relevance. Following MLPerf's structure, this framework distinguishes between two benchmark divisions: 
\begin{itemize}
    \item {\bf Closed benchmarks} fix the implementation and enable direct cross-platform, cross-vendor comparison. The variable is the system, not the algorithm.
    \item {\bf Open benchmarks} hold the success criterion constant but permit algorithmic innovation beyond the reference implementation, allowing teams to demonstrate their latest results without sacrificing intellectual property.
\end{itemize}
Both divisions are represented in the benchmark collection.\footnote{
Note the slightly counterintuitive use of the terms ``open'' and ``closed,'' which is the reverse of their meaning in the open-source software context. Closedness of a benchmark refers to the fact that the actual implementation is locked down (closed), whereas openness of a benchmark refers to the fact that only the problem definition is given and it is left unspecified (open) as to how exactly the problem is solved.} Like MLPerf, this framework is designed to evolve alongside the technology it measures. Benchmark problems can be proposed and launched as new application domains emerge, and existing problems can be modified or retired as hardware capabilities and relevant business problems shift. This keeps the framework grounded in what quantum systems are actually being asked to do at any given point in the technology's development. 
We developed a Python-based framework that comes with a code runner, a collection of instances, reference implementations of all benchmarks, as well as reference measurements of all benchmarks. The code, covering both quantum and classical components, is made available in a public repository at \texttt{\href{https://github.com/ionq-publications/apps-benchmark}{github.com/ionq-publications/apps-benchmark}} using a standard development environment (Qiskit), so that anyone can (a) reproduce the results on IonQ hardware, (b) run these benchmarks on other systems for independent comparison, and (c) contribute their measured performance metrics back to the benchmark. Next, we give a short overview of the main metrics reported. 

\paragraph{Solution quality.}
The primary performance metric for this framework is {\em solution quality}, defined as the figure of merit that determines what constitutes a valid answer in a given benchmark. Solution quality takes different forms depending on the problem. For quantum chemistry benchmarks, it could be energy error relative to an exact reference. For combinatorial optimization, it is often an approximation ratio. For machine learning applications, it could be classification accuracy. The quality threshold is established before results are reported. Meeting it is the condition for a benchmark to be considered as having reached a solution. As solution quality generally is a multi-dimensional metric, we distill the overall collection of solution quality performance indicators into a single numerical score. 

\paragraph{Execution time.}
Execution time measures the compute time a program spends running against the hardware, i.e., the active time the QPU spends executing the quantum programs. It intentionally does not include waiting times, pre-flight times, compilation times, and similar overheads. Execution time is best for evaluating algorithm efficiency, while time to solution dictates user experience and task deadlines. Further detail is provided in Section~\ref{sec:exec}

\paragraph{Execution energy.}
Execution energy measures the total energy consumed to reach the solution quality threshold across both quantum and classical components. IonQ's Forte and Forte Enterprise systems are instrumented to measure power consumption using meters and overall electric energy consumption of these systems is reported.

\paragraph{Time-to-solution.}
In contrast to execution time, the time-to-Solution (TTS), often also called ``wall-clock time'' or ``response time,'' measures the total elapsed time from task start to completion, including IO, waiting, and setup. We included measurements of TTS where the particular benchmark structure permits a meaningful definition of a task to be measure completion for. Typically, these tasks are more complex than a single run and may involve, for example, searching for a solution of a pre-defined quality. Some benchmarks in this suite involve hybrid quantum-classical algorithms. Both quantum execution time and classical co-processing are included in TTS measurements where applicable. Further detail and examples are provided in Section~\ref{sec:tts}. Similar to TTS, total Energy-to-Solution (ETS) and Cost-to-Solution (CTS) can be defined.

\paragraph{Extensibility.} 
The framework is designed to incorporate new metrics without structural change. An example is Cost-to-Solution (CTS), which is not currently reported.  
CTS treats time and energy as direct inputs to the economic cost of running a quantum workload. It connects hardware performance to procurement and operational decisions in terms that matter to quantum computing end users. We expect that future iterations of the benchmarking framework will incorporate CTS and other metrics. 

\subsection{Overview of the Benchmark Families}

The benchmarks in this suite span optimization (QAOA, LR-QAOA, varQITE), quantum chemistry (VQE with the UpCCD ansatz, QC-AFQMC), machine learning (QCNN, quantum copula), data loading (tensor network-based image loading), simulation (Quantum Lattice Boltzmann), and foundational subroutines (QFT, Hidden Shift, Fixed-Point Amplitude Amplification, High Energy Physics). Each benchmark is described in terms of its rationale for inclusion, the algorithm itself, problem instances, and results. Hardware results are reported for IonQ systems using standardized problem instances and shot counts. 

The benchmarking code is publicly available, and any party wishing to perform cross-platform comparisons may do so using the tools provided in the repository (using their own system access arrangement). Because performance degrades with circuit depth on all NISQ platforms, we present the results as a function of problem size and circuit depth so that the noise regime of each system is visible rather than obscured by selective reporting. For Open Benchmarks (where the benchmark code can be modified to demonstrate algorithmic innovation), we provide the definition of the problem and a description of the method we followed, but not necessarily IonQ's proprietary algorithm - it is up to each benchmarking participant to try and outperform through innovation. Error mitigation has been used throughout the benchmarks, following broadly the methods outlined in \cite{symm2023}. These methods are either already integrated into IonQ's compiler and runtime system, or will be made available in future releases.

The goal is a benchmarking framework that scales with the technology, evolves with new real-world workflows and algorithm developments, and helps the ecosystem with information they need to make real development, procurement and deployment decisions. We aim to advance a shared understanding of where quantum systems and algorithms perform today and what remains to be solved.

\begin{table}[H]
\centering
\footnotesize
\sffamily 
\setcellgapes{5pt}   
\makegapedcells
\begin{tabular}{cclll}
\Xhline{4\arrayrulewidth}
\textbf{Algorithm} & \textbf{Name} & \makecell{\textbf{Relevant}\\ \textbf{Industries}} & \makecell{\textbf{Type}} & \textbf{Division}\\
\hline
\makecell{VQE with \\ UpCCD ansatz} & \makecell{Variational Quantum Eigensolver \\ (hydrogen chains)} & \makecell{Pharmaceuticals \\ Chemical Manufacturing \\ Materials Science } & \makecell{Quality} & Closed \\
QAOA & \makecell{Quantum Approximate Optimization \\ Algorithm (MaxCut)} & \makecell{Financial Services \\ Logistics \\ Energy } & \makecell{Quality} & Closed\\
\medskip
LR-QAOA & \makecell{Linear-Ramp QAOA \\ (MaxCut)} & \makecell{Financial Services \\ Logistics \\ Manufacturing } & \makecell{Quality \\ Time to Solution} & Closed\\
FAA & \makecell{Fixed Point Amplitude Amplification} & \makecell{Financial Services \\ Pharmaceuticals \\ Cybersecurity } & \makecell{Quality} & Closed \\
QCNN & \makecell{Quantum Convolutional Neural Network \\ (image classification)} & \makecell{Healthcare \\ Automotive \\ Defense } & \makecell{Quality} & Closed \\
\makecell{Quantum \\ Copula} & \makecell{Quantum Circuit Born Machine Copula \\ (portfolio risk/VaR)} & \makecell{Financial Services \\ Insurance } & \makecell{Quality} & Closed \\
\makecell{Image \\ Loading} & \makecell{Tensor Network-based MPS image loading} & \makecell{Automotive \\ Healthcare \\ Financial Services } & \makecell{Quality} & Closed \\
QFT & \makecell{Quantum Fourier Transform \\ (Cosine, and Hidden Phase)} & \makecell{Cybersecurity \\ Financial Services \\ Pharmaceuticals } & \makecell{Quality \\ Time to Solution} & Closed \\
HSBP & \makecell{Hidden Shift Benchmark Problem} & \makecell{Quantum Hardware \\ Defense \\ Government } & \makecell{Quality \\ Time to Solution} & Closed \\
VarQITE & \makecell{Variational Quantum Imaginary-\\ Time Evolution (MaxCut)} & \makecell{Manufacturing \\ Energy \\ Financial Services } & \makecell{Quality} & Open\\
QC-AFQMC & \makecell{Quantum-Classical Auxiliary Field \\ Quantum Monte Carlo} & \makecell{Pharmaceuticals \\ Chemical Manufacturing \\ Materials Science } & \makecell{Quality} & Open \\
QLBM & \makecell{Quantum Lattice Boltzmann Method \\ (advection-diffusion)} & \makecell{Aerospace \\ Automotive \\ Energy } & \makecell{Quality} & Open \\
\makecell{High Energy \\ Physics} & \makecell{Neutrinoless double beta decay simulation \\ (lattice QCD)} & \makecell{National Labs \\ Defense \\ Nuclear Energy } & \makecell{Quality} & Open\\
\Xhline{4\arrayrulewidth}
\end{tabular}
\caption{\label{tab:benchmark-overview} \footnotesize Overview of the families of application benchmarks analyzed in this paper. For each benchmark, we report a Quality score and the execution time. For certain benchmarks that look for a 'needle-in-a-haystack' solution, we provide Time-to-Solution (TTS) measurements. We distinguish between two Divisions: Closed benchmarks in which an open source implementation of the entire benchmark along with a complete specification of all circuits is provided, and Open benchmarks in which the target problem is specified but in which the implementation is left open.}
\end{table}

\subsection{Execution Time Definition}
\label{sec:exec}

Execution time, represented in green in this figure, is recorded on the QPU and represents the total time the job spent on the QPU, including codegen of waveforms with templated calibration parameters, execution on-ions via the real-time subsystem, and measurement thresholding and aggregation. 

\begin{figure}[H]
    \centering
    \includegraphics[width=1.0\textwidth]{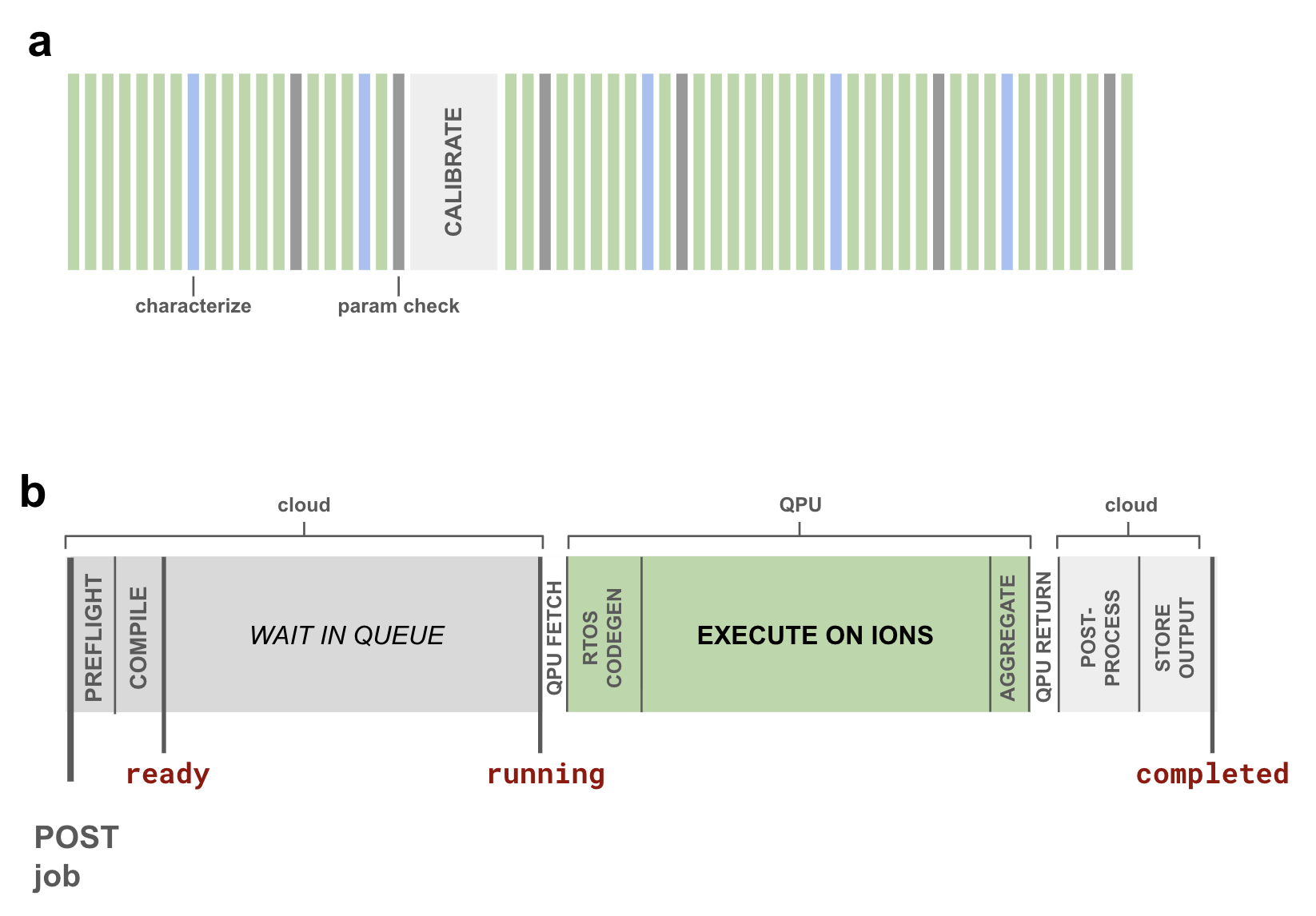}
    \caption{
Illustration of execution time as reported in the framework. Shown in a) is the execution duty cycle from the perspective of the QPU — individual end-user jobs are interleaved with periodic characterization jobs (blue) and parameter check jobs (dark gray). If a parameter check fails, the calibration system kicks off a short unscheduled calibration (order minutes) to bring the parameter back into an acceptable range. Forte-class systems see approximately 40–50 minutes of aggregate execution time per hour, depending on systematic drift, job structure, and other variable factors. Shown in b) are the end-to-end steps the job goes through from submission to results retrieval. All portions shown in gray are parallelized via horizontally-scaling cloud infrastructure. In both figures, bar lengths are representative but not to scale.}
\end{figure}

\subsection{Time-to-Solution Definition}
\label{sec:tts}
Although raw shot rate is a useful measure of hardware throughput, it does not by itself determine application-level performance. For practical quantum computing, the relevant question is how quickly a system produces a sample of sufficiently high quality, not simply how quickly it produces any sample at all. In this sense, shot quality is the more consequential metric of utility: a device that returns lower-quality samples at a high rate may be less useful in practice than one that produces high-quality samples more reliably. We therefore use Time-to-Solution (TTS), and in particular confidence time-to-first-solution, as the central metric for assessing quantum utility in real applications.

For each target quality threshold \(p\), we estimate the probability \(\hat{q}(p)\) of obtaining a sample with quality at least \(p\) from the empirical histogram of samples returned by the device. An expectation-based TTS proxy is then obtained by taking \(1/\hat{q}(p)\) as the expected number of shots required to observe the first successful sample, giving
\[
\mathrm{TTS}(p) = \frac{t_{\mathrm{shot}}}{\hat{q}(p)},
\]
where \(t_{\mathrm{shot}}\) is the effective time per shot. We also compute a confidence time-to-first-solution, defined as the runtime required to obtain at least one sample of quality at least \(p\) with confidence level \(c\). Assuming independent shots, this yields
\[
\mathrm{TTS}_{c}(p) = t_{\mathrm{shot}} \frac{\log(1-c)}{\log\!\bigl(1-\hat{q}(p)\bigr)}.
\]
This confidence-based TTS is the primary metric used below, as it captures the quantity most relevant to end users: the time required to obtain a solution of practical value.

\subsection{Organization of the Results}
\label{sec:org}
The rest of the paper is structured as follows: In \Cref{sec:closed} we present the methodology and the results of 9 Closed benchmarks. Python code for these benchmarks, including a benchmark running for streamlined execution and automatic computation of scores is available as described in Appendix~\ref{app:format}. In \Cref{sec:open} we present the methodology and the results of 4 Open benchmarks. Appendix~\ref{app:format} also provides further detail on how to contribute your own measurements and the requirements on reporting in the Closed and Open benchmark divisions. Appendix~\ref{app:tables} has a collection of measurements that were obtained on IonQ's quantum hardware. 

\section{Results: Closed Benchmarks}
\label{sec:closed}

\subsection{VQE for Chemistry with Unitary Pair-Coupled Cluster Doubles (UpCCD)}

\subsubsection{Rationale} 
This benchmark measures a quantum system's ability to compute molecular ground-state energies (specifically, the energy of hydrogen chains at their lowest quantum state)\cite{Peruzzo2014-xz}. The core task is to execute a deep Variational Quantum Eigensolver (VQE) quantum circuit with pre-optimized parameters and return the expectation value $\bra{\Psi}{H}\ket{\Psi}$ within chemical accuracy ($\pm$1.6 mHa).

The benchmark addresses the electronic structure problem, a foundational challenge in quantum chemistry focused on calculating a molecule's ground-state energy. We use one-dimensional chains of hydrogen atoms as our model system. While seemingly simple, these hydrogen chains are a conventional and powerful tool for studying electronic correlation (the quantum entanglement between electrons that makes many molecules classically intractable)\cite{stair2020}. By tuning the chain length and atomic spacing, we can create a spectrum of problems that range from weakly- to strongly-correlated, mimicking the diverse challenges found in industrial applications like drug discovery and materials science.

To rigorously benchmark the core quantum hardware, the VQE protocol is designed as an optimization-free test. By providing pre-optimized parameters, we isolate and measure the system's ability to execute a deep circuit faithfully. This directly assesses the accumulation of hardware errors---the true bottleneck for achieving chemical accuracy and near-term quantum advantage.

\subsubsection{Algorithm Description} 
The UpCCD ansatz \cite{Lee2019-bu,Elfving2021-pi,Zhao2023-ck} constructs the quantum state through a specific pattern of two-electron excitations.

This operator moves electron pairs from occupied orbitals (i) to virtual orbitals (a), maintaining spin pairing throughout. The restriction to paired excitations (what physicists call the ``seniority-zero subspace'') dramatically reduces circuit complexity while capturing the dominant physics of closed-shell molecules under bond breaking scenarios.

The circuit depth scales favorably with system size. For $N$ qubits\footnote{ At half-filling, as is the case for the hydrogen chains in minimal basis STO-3G. This is an upper bound, though the asymptotics hold for the average case as well.}, single-qubit gate counts go as $\textonehalf(N^2+7N)$ while two-qubit gates go as $\textonehalf{N^2}$. This polynomial scaling contrasts with the exponential growth of the full configuration space, making UpCCD tractable for near-term devices.

The molecular Hamiltonian, after mapping to qubits via the paired-electron formalism, is partitioned for efficient measurement. All diagonal terms (containing only Z and Identity operators) can be measured simultaneously in the computational basis. The off-diagonal terms (the XX and YY components) are grouped separately and require basis-change rotations before measurement. This well-known grouping strategy limits the total measurement overhead to a small, constant number of circuit executions (three in this case), regardless of the molecule's size. This is in contrast to the more general UCC Ans\"atze where the number of terms to measure grows prohibitively.

The benchmark measures success through the absolute error in energy:
$\mathrm{Error} = |E_\mathrm{measured} - E_\mathrm{DOCI}|$
Here, $E_\mathrm{DOCI}$ represents the exact ground-state energy within the paired-electron model. Because UpCCD operates entirely within this restricted space, doubly-occupied configuration interaction (DOCI) provides the theoretical limit of accuracy. Any deviation from $E_\mathrm{DOCI}$ directly measures hardware imperfection rather than ansatz approximation.

\subsubsection{Problem Instances} 
The specific instances consist of linear chains of 2 to 18 hydrogen atoms, with the interatomic distance varying from $0.75\mathrm{\AA}$ to $2.0\mathrm{\AA}$. Each instance is delivered as a self-contained JSON file that encodes the complete computational task, including the molecular geometry, the qubit Hamiltonian, and importantly, the pre-optimized ansatz parameters, which eliminates classical optimization variance and ensures reproducible benchmarking.

\begin{lstlisting}[language=json,firstnumber=1, caption={A representative JSON data structure for a problem instance from the quantum chemistry benchmark suite. This example, h002\_chain\_1\_25, defines the simulation of a 2-qubit hydrogen molecule (H2) at a $1.25\mathrm{\AA}$~bond distance using the STO-3G basis. The data includes the qubit Hamiltonian, optimizer settings, and final results from a noiseless Variational Quantum Eigensolver (VQE) simulation.}]
{
"benchmark_category": "chemistry",
"problem_type": "hydrogen_chain_vqe", 
"instance_name": "h002_chain_1_25", 
"solution_algorithms": ["vqe_puccd"], 
"num_qubits": 2,
"data": {
  "geometry": [["H", 0.0, 0.0, -0.625], ["H", 0.0, 0.0, 0.625]], 
  "description": "H2 1D chain, 1.25 Angstroms", 
  "basis": "sto-3g", "mapper": "PairedElectron",
  "hf_energy": -0.989113814090892, "nuclear_repulsion_energy": 0.42334176873600005, 
  "num_spatial_orbitals": 2, "num_alpha": 1, "num_beta": 1,
  "paired_hamiltonian_dict": {
    "II": -0.13566483543472385, 
    "IZ": 0.22512501915719244, 
    "ZI": -0.14722118724272235, 
    "ZZ": 0.4811027722562538, 
    "XX": 0.10655120065707539, 
    "YY": 0.10655120065707539},
  "reference_energy_doci": -1.045783144549802, "reference_energy_fci": -1.0457831445498016,
  "optimizer_config": {...},
  "vqe_final_energy": -1.0457831445494497, "optimal_parameters": [0.25990952197965067],
  "optimization_info": {...},
 }
}
\end{lstlisting}

The difficulty of these instances scales in two ways. First, increasing the system size (the number of atoms) requires deeper circuits with quadratically more gates, amplifying the effects of decoherence and gate errors. Second, stretching the bond lengths (particularly in the $1.5-2.0\mathrm{\AA}$~range) increases the electronic correlation, creating more complex quantum states that are inherently more sensitive to hardware noise. This dual scaling of size and correlation allows the benchmark to comprehensively probe different hardware failure modes, from predictable error accumulation with gate count to the state's increased sensitivity to phase errors under strong correlation. The benchmark thus answers a precise question: given perfect classical pre-processing (optimal parameters), can a quantum system execute the required circuit faithfully enough to achieve chemical accuracy?

\subsubsection{Results} 

The VQE benchmark sets a rigorous bar, with the ``solved'' criterion defined as achieving accuracy within 1 mHa of the exact solution. As this standard currently remains unmet across the industry, the benchmark provides a critical measure of hardware fidelity. To isolate and quantify hardware noise, we execute a fixed, optimized circuit and measure the absolute deviation from the known exact classical answer, ensuring minimal statistical error. This method provides a direct, unvarnished view of a system's true error accumulation.

\begin{figure}[H]
    \centering
    \includegraphics[width=0.75\linewidth]{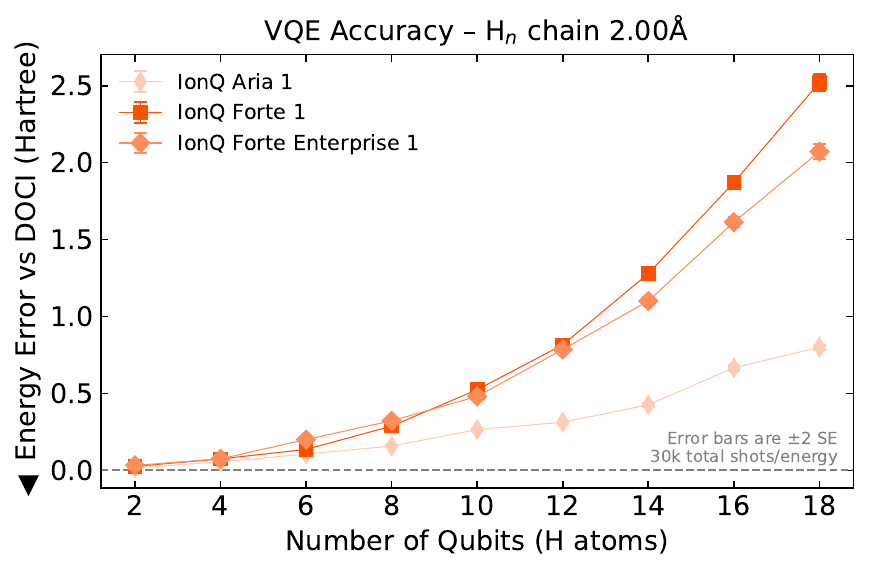}
\caption{VQE benchmark performance for hydrogen chains at $2.00\mathrm{\AA}$~bond distance across IonQ QPU backends. Absolute energy error relative to exact DOCI energy versus system size (2--18 qubits). Three backends are compared: \texttt{Aria-1}, \texttt{Forte-1}, and \texttt{Forte-Enterprise-1}. Error bars represent $\pm 2$ standard errors; each evaluation uses 30{,}000 total shots across three measurement bases (ZZ, YY, XX) by nature of the pUCCD ansatz.}
    \label{fig:vqe_results}
\end{figure}

\subsection{Quantum Approximate Optimization Algorithm (QAOA)}\label{sec:qaoa}

\subsubsection{Rationale}
QAOA is a universal hybrid (classical-quantum) algorithm capable of addressing a wide range of combinatorial optimization problems relevant to many industries \cite{farhi2014}. For this benchmark, we focus specifically on the MaxCut problem, a canonical NP-hard optimization problem that can be directly mapped onto a quantum Hamiltonian, providing a clear and rigorous testbed for evaluating quantum hardware performance.

QAOA's performance is notably sensitive to underlying hardware characteristics, such as qubit connectivity, gate fidelities, and coherence times. A key parameter influencing this sensitivity is the number of circuit layers $p$, which directly controls circuit depth, impacting both solution quality and susceptibility to hardware-induced errors. Therefore, QAOA serves as an effective benchmark for characterizing and comparing quantum computing systems \cite{lall2025}.

\subsubsection{Algorithm Description}
QAOA is a variational approach that finds approximate solutions to optimization problems by encoding them into a cost Hamiltonian $H_c$ and then preparing a parameterized quantum state $\ket{\psi(\boldsymbol{\gamma}, \boldsymbol{\beta})}$, to minimize the cost function $C(\boldsymbol{\gamma}, \boldsymbol{\beta})$ defined by the expectation value

\begin{align*}
    C(\boldsymbol{\gamma}, \boldsymbol{\beta}) = \bra{\psi(\boldsymbol{\gamma}, \boldsymbol{\beta})}{H_c}\ket{\psi(\boldsymbol{\gamma}, \boldsymbol{\beta})}.
\end{align*}

The trial state, or ansatz, $\ket{\psi(\boldsymbol{\gamma}, \boldsymbol{\beta})}$ is prepared by applying $p$ alternating layers of unitary operators to an initial uniform superposition state, $\ket{+}^{\otimes{n}}$. The circuit is parametrized by $2p$ angles: $\boldsymbol{\gamma} = (\gamma_1, \ldots, \gamma_p)$ and $\boldsymbol{\beta} = (\beta, \ldots, \beta_p)$. The operators of the $k$th layer are the cost unitary, $U(H_c , \gamma_k) = \exp[-i\gamma_kH_c]$, and the mixer unitary, $U(H_B , \beta_k) = \exp[-i\beta_kH_B]$ , where $H_B$ is chosen such that its ground state is easily prepared, e.g. $H_B= \sum_{i=1}^{n}{X_i}$ whose ground state is the initial state $\ket{+}^{\otimes{n}}$. The full $p$-layer QAOA ansatz is thus given by

\begin{align*}
    \ket{\psi(\boldsymbol{\gamma}, \boldsymbol{\beta})} = U(H_B , \beta_p)U(H_c , \gamma_p)\ldots U(H_B , \beta_1)U(H_c , \gamma_1)\ket{+}^{\otimes{n}}.
\end{align*}

While QAOA typically employs classical optimization methods to iteratively update the parameters $(\boldsymbol{\gamma}, \boldsymbol{\beta})$, this benchmark employs an optimization-free variant known as the fixed-angle conjecture \cite{wurtz2021,shaydulin2023}. In this approach, pre-determined sets of angles are used, proven to yield universally strong performance guarantees for specific classes of graphs. For example, for MaxCut on 3-regular graphs, these angles are established up to $p=11$ layers, and for 4-regular graphs up to $p=5$. Eliminating the classical optimization step reduces overhead and variability, enabling more direct comparisons of quantum hardware performance.

The primary score metric for this benchmark is the approximation ratio $\text{AR} \in [0,1]$, defined as
\begin{align}\label{eq:apx-ratio}
    \text{AR}= \frac{C(\boldsymbol{\gamma}, \boldsymbol{\beta})}{C_\text{opt}},
\end{align}
where $C_\text{opt}$ is the exact optimal solution obtained classically, which is feasible for these problem sizes. Ideally, without hardware noise, the AR should monotonically approach one as the number of layers $p$ increases. However, without error correction or mitigation, hardware noise typically causes performance degradation at larger circuit depths, observed as a drop in the AR after reaching a peak. Thus, this optimization-free QAOA benchmark provides direct insights into quantum hardware performance independent of classical optimization overhead.

\subsubsection{Problem Instances} The MaxCut problem involves partitioning the vertices of an undirected graph $G(V, E)$ with vertex set $V$ and edge set $E$, into two disjoint subsets to maximize the number of edges connecting these subsets. The objective function of the problem can be mapped to a cost Hamiltonian of the form

\begin{align*}
    H_c=\frac{1}{2}\sum_{(i,j)\in E}{(Z_iZ_j - 1)}.
\end{align*}

For this benchmark, we consider randomly generated 3- and 4-regular graphs with sizes ranging from 8 to 36 vertices. These are, in turn, mapped to Hamiltonians acting on 8 to 36 qubits, respectively. Each problem instance is thus defined by a $d$-regular graph of $n$ vertices, which is solved using a fixed $p$-layer QAOA ansatz of $n$ qubits.

\subsubsection{Results} 

\begin{figure}[H]
    \centering
    \includegraphics[width=\linewidth]{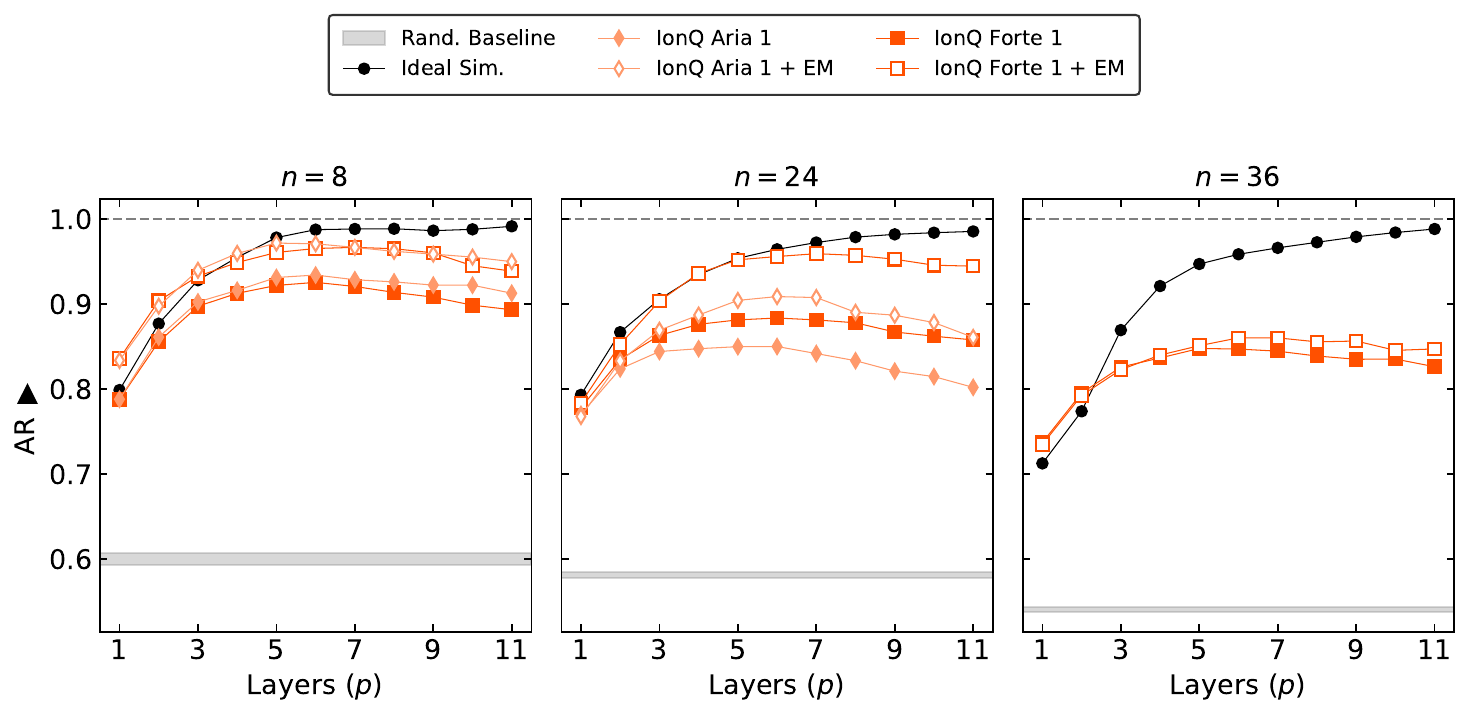}
    \caption{Performance of the fixed-angle QAOA benchmark for solving the MaxCut problem on 3-regular graphs of varying sizes ($n=8$, 24, and 36 vertices). The results in each subplot compare ideal noiseless simulations with empirical data from IonQ devices, presented in both raw and error-mitigated (EM) forms. Each hardware execution used 5,000 measurement shots. A dashed horizontal line indicates the theoretical maximum $\mathrm{AR}=1$, where the optimal solution is consistently sampled, and the shaded area shows the $\pm 3\sigma$ confidence interval for a random-sampling baseline. The ideal simulations were performed using a statevector simulator for the $n=8,24$ cases and a Matrix Product State (MPS) simulation with a bond dimension of $\chi=256$ for the $n=36$ case.}
    \label{fig:qaoa_results}
\end{figure}

The results in \Cref{fig:qaoa_results} show that the IonQ backends follow the expected ideal trend at small depth, where the approximation ratio initially improves as $p$ increases, and then gradually depart from the noiseless curves as circuit depth grows. Across all three problem sizes, the measured approximation ratios remain well above the random-sampling band over a substantial depth range, demonstrating that the hardware retains a clear optimization signal rather than collapsing to random output. Here, the random baseline corresponds to the AR obtained by drawing bit strings uniformly at random with the same number of shots, and the shaded band indicates its $\pm 3\sigma$ confidence interval.

The error-mitigated curves further narrow the gap to ideal performance in the shallow-to-intermediate depth regime, while the eventual rollover beyond roughly $p \approx 6$ reflects the accumulated impact of hardware noise at larger depths. This makes the benchmark useful both for identifying the onset of the noise-dominated regime and for quantifying how much optimization headroom above random sampling survives on quantum hardware.

\subsection{Linear-Ramp Quantum Approximate Optimization Algorithm (LR-QAOA)}

\subsubsection{Rationale}
The Linear-Ramp QAOA (LR-QAOA) offers a powerful, non-variational approach for characterizing quantum hardware. Instead of a layer-by-layer parameter search, it employs a pre-defined linear schedule for its angles. This schedule is governed by two fixed global parameters that scale the linear ramp of the individual layer angles across the layers of the circuit, completely eliminating the classical optimization loop. By removing the classical optimizer, LR-QAOA provides a method to benchmark different hardware platforms on an equal footing. Moreover, since every circuit instance is fully specified a priori, the protocol allows for a direct probe of how performance scales with circuit depth.

Although non-variational, LR-QAOA functions as a universal quantum optimization algorithm, capable of tackling different problems using the same or a similar parameter schedule \cite{montanezbarrera2024,montanezbarrera2025}. Furthermore, it has been shown that effective schedules for large problem instances can be estimated by extrapolating from the results of smaller, classically simulable problems \cite{dehn2025}. These features make LR-QAOA an attractive tool for both robust hardware characterization and resource-limited applications.

\subsubsection{Algorithm Description}
The LR-QAOA protocol uses the same ansatz as the standard QAOA, but it replaces the variational parameter optimization with a fixed, linear schedule. The parameters $\gamma_k$ and $\beta_k$ are determined by a linear ramp that depends on only two global hyperparameters, $\Delta_\gamma$ and $\Delta_\beta$, and the total number of layers $p$. The schedule for the parameters at layer $k$ (for $k = 0 ,\ldots, p−1$) is given by

\begin{align*}
    \beta_k = \left(1-\frac{k}{p}\right)\Delta_\beta, \hspace{0.5in} \gamma_k=\frac{k+1}{p}\Delta_\gamma.
\end{align*}

Since the parameters are pre-determined, the algorithm is non-variational and can be executed directly on the hardware without a classical optimization overhead. 

Similarly to the QAOA benchmark, the score metric we use is also the approximation ratio AR

\begin{align*}
    \text{AR}=\frac{C(\boldsymbol{\gamma}, \boldsymbol{\beta})}{C_\text{opt}},
\end{align*}

where $C_\text{opt}$ is the exact optimal solution obtained with a classical solver.

In addition to reporting the raw approximation ratio, we also use an effective approximation ratio, denoted $\text{AR}_\text{eff}$, which normalizes performance against a random baseline. Following the LR-QAOA benchmarking protocol of Ref.~\cite{montanezbarrera2025}, we first construct a random-sampling baseline by drawing batches of bit strings uniformly at random (with the same number of samples as the hardware experiment) and computing their approximation ratios. Let $\mu_\text{rand}$ and $\sigma_\text{rand}$ be the mean and standard deviation of these random-batch scores; we define a conservative random threshold  
\begin{align*}
\text{AR}_\text{rand} = \mu_\text{rand} + 3\sigma_\text{rand},  
\end{align*}
corresponding to the upper edge of a $3\sigma$ confidence interval for a purely random sampler. For each instance we then take the best hardware performance across depths,  
\begin{align*} 
\text{AR}_\text{max} = \max_p \text{AR}(p),  
\end{align*}
and define  
\begin{align*}  
\text{AR}_\text{eff} = \frac{\text{AR}_\text{max} - \text{AR}_\text{rand}}{1 - \text{AR}_\text{rand}}.  
\end{align*}

By construction, $\text{AR}_\text{eff} = 0$ means the best observed performance is statistically indistinguishable from random sampling, while $\text{AR}_\text{eff} = 1$ indicates that the device effectively closes the entire gap between the random baseline and the optimal solution for that instance. Values $0 < \text{AR}_\text{eff} < 1$ quantify the fraction of the available ``headroom'' over random that survives hardware noise. Conversely, $\text{AR}_\text{eff} < 0$ indicates that the device performs worse than the 3$\sigma$ random threshold (e.g., due to decoherence or strong bias), and is therefore considered to fail the benchmark. In this way, $\text{AR}_\text{eff}$ provides a simple, single scalar score that both normalizes performance across instances of differing difficulty and supports a clear pass/fail criterion.

\subsubsection{Problem Instances}

To maintain continuity with the QAOA benchmark, we again use the MaxCut problem. The problem instances consist of the same randomly generated 3- and 4-regular graphs with vertex counts of 24 and 36. We also add fully connected weighted graphs (FCW), since these instances can accurately gauge qubit connectivity in quantum hardware. The number of QAOA layers $p$ is varied up to a maximum of 53. This range is sufficient to observe the full performance curve, from the initial increase in the approximation ratio to the eventual decline into the noise-dominated regime on current NISQ hardware.

The global parameters for the linear schedule were determined empirically. Following an initial approximation of $\Delta_\gamma = \Delta_\beta$, a parameter search was conducted across different layers to find a single value that consistently optimized the AR. This search yielded an approximated optimal value of $\Delta_\gamma = \Delta_\beta = 1.25$, which was found to be robust across different problem sizes for both 3- and 4-regular graphs and was therefore used for all experiments.

\subsubsection{Results}

\begin{table}[H]
  \centering
  \small
  \caption{Effective approximation ratio $\text{AR}_\text{eff}$ for IonQ backends and problem instances.}
  \label{tab:r_eff}
\begin{tabular}{lll}
    \toprule
    Problem Instance & \texttt{aria-1} & \texttt{forte-1} \\
    \midrule
    3-reg, $n=24$ & \multicolumn{1}{c}{--} & 0.8390 \\
    3-reg, $n=36$ & \multicolumn{1}{c}{--} & 0.5280 \\
    4-reg, $n=24$ & \multicolumn{1}{c}{--} & 0.8668 \\
    4-reg, $n=36$ & \multicolumn{1}{c}{--} & 0.8009 \\
    FCW, $n=12$ & 0.3728 & 0.3821 \\
    FCW, $n=16$ & \multicolumn{1}{c}{--} & 0.2504 \\
    \bottomrule
\end{tabular}
\end{table}

\begin{figure}[H]
    \centering
    \includegraphics[width=\linewidth]{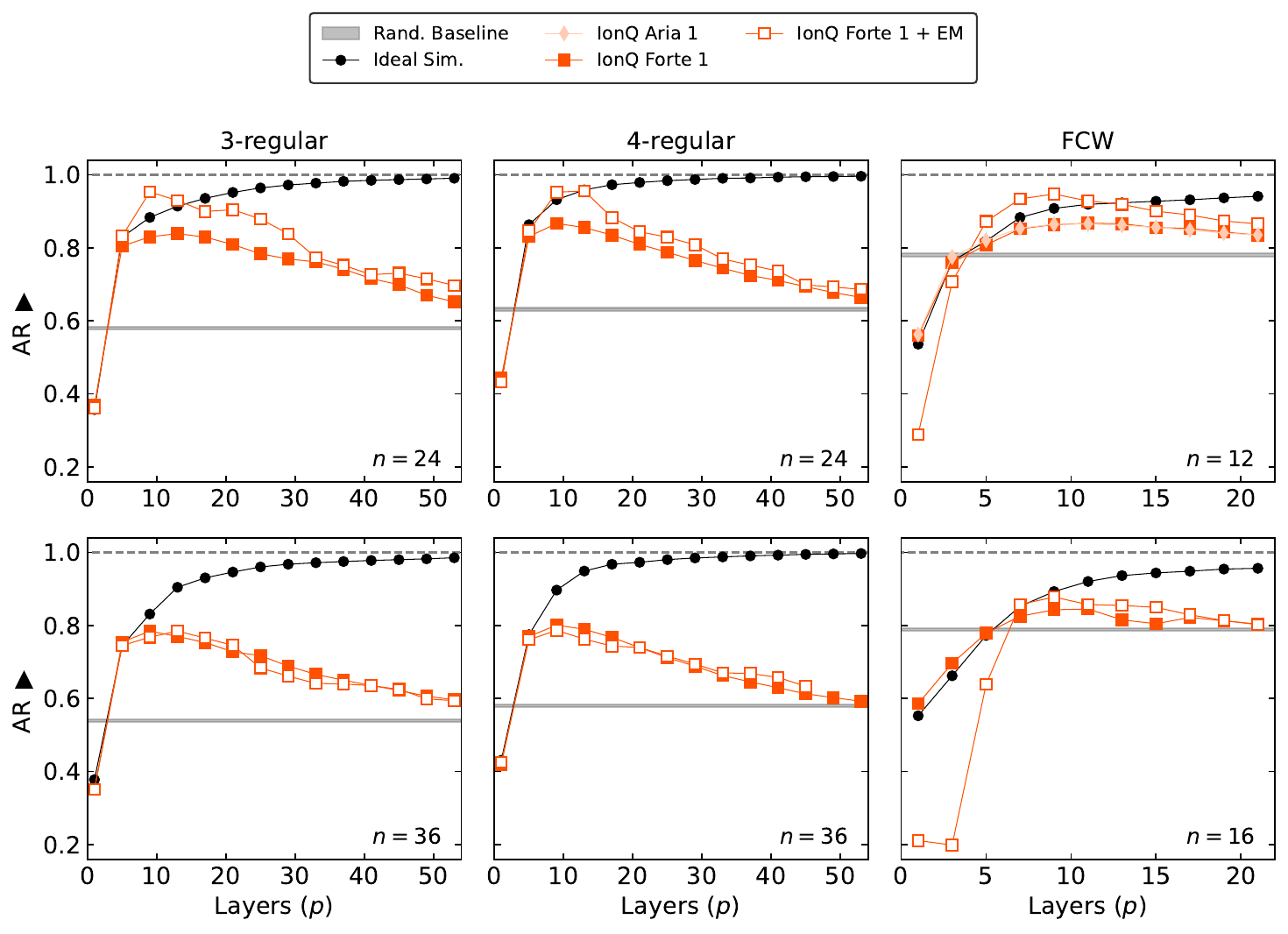}
    \caption{Performance of the LR-QAOA benchmark for solving the MaxCut problem on 3- and 4-regular graphs and on fully-connected weighted graphs (FCW) of varying sizes ($n=12$, 16, 24, and 36 vertices). The results in each subplot compare ideal noiseless simulations against empirical hardware data. Each hardware execution used 5,000 measurement shots. The horizontal dashed line at $\mathrm{AR}=1$ represents the theoretical maximum, and the shaded area delineates the $\pm 3\sigma$ confidence interval for a random-sampling baseline.}
    \label{fig:lrqaoa_results}
\end{figure}

The LR-QAOA results in \Cref{fig:lrqaoa_results} show the same qualitative pattern across the three problem families: the approximation ratio rises above the random-sampling band at small-to-intermediate depth, reaches a peak, and then gradually decays as circuit noise accumulates at larger $p$. Because the figure includes 3-regular, 4-regular, and FCW graph instances, it provides a broader view of how the hardware behaves across problems with different connectivity structure and depth requirements.

These trends are compactly summarized by the effective approximation ratio $\text{AR}_\text{eff}$ in \Cref{tab:r_eff}. IonQ backends achieve positive $\text{AR}_\text{eff}$ values across all reported instances, including the 3-regular, 4-regular, and FCW cases. Forte-1 reaches $\text{AR}_\text{eff}=0.8390$ and $0.8668$ on the 24-qubit 3- and 4-regular graphs, respectively, remains strongly above the random threshold for the 36-qubit instances, and retains a positive margin over random on the FCW graphs. Aria-1 also passes the reported FCW instance at $n=12$. Taken together, these results show that LR-QAOA preserves a substantial fraction of the available optimization headroom over random sampling on IonQ hardware across multiple graph classes.

\subsection{Fixed Point Amplitude Amplification (FAA)}

\subsubsection{Rationale}
At the heart of many quantum algorithms lies a powerful technique called amplitude amplification, which is the generalized underpinning of Grover's search algorithm \cite{grover2005fixed}. It allows for the significant increase of the probability of measuring a desired quantum state, known as the ``target state.'' However, standard amplitude amplification has a critical drawback of ``overshoot'' which occurs when the amplification is performed for more than the required number of times which can result in the probability of finding the target state actually decreasing. The fixed-point amplitude amplification (FAA) algorithm \cite{yoder2014fixed,martyn2021grand} is a refined version of this technique that solves this ``overshoot'' problem, guaranteeing that the probability of measuring the target state is confined within a predefined window close to 1.

The robustness of fixed-point amplitude amplification makes it a valuable tool in various areas of quantum computing, such as:

\begin{itemize}
\item Quantum Search: This is the most direct application. FAA allows for the creation of more reliable quantum search algorithms that find a marked item in an unstructured database without needing to know the number of target items in advance. This is a significant improvement over Grover's algorithm, which requires this prior knowledge to determine the optimal number of iterations \cite{ grover2005fixed,yoder2014fixed}.
\item Quantum Simulation: In simulations of quantum systems, it is often necessary to prepare specific initial states. FAA can be employed to efficiently prepare these states by amplifying the amplitude of the desired state from a more easily preparable superposition. This is crucial for simulations in areas like materials science and drug discovery \cite{zecchi2025improved}.
\item Quantum Optimization: In optimization problems, the goal is to find the best solution from a large set of possibilities. FAA can be used in conjunction with algorithms like the Quantum Approximate Optimization Algorithm (QAOA) to increase the probability of measuring the state that corresponds to the optimal solution \cite{zhang2024grover}.
\end{itemize}

\subsubsection{Algorithm Description}
Fixed-Point Amplitude Amplification (FAA) is designed to increase the probability of measuring a desired (``target'') quantum state. It's a crucial modification of the standard Grover-style amplitude amplification which guarantees that the probability of success converges and remains within a predefined window close to 1, eliminating the need to know the number of target states beforehand to avoid ``overshooting'' the solution. This is achieved by altering the standard reflection operators with controlled phase shifts using quantum signal processing (QSP), effectively adjusting the rotation angle at each step to ensure a steady approach to the target state.

The phase shifts are determined using the quantum signal processing package pyqsp \cite{martyn2021grand, cdghs_finding_qsp_angles_20, haah_decomposition_19, gslw_qsvt_19, dmwl_efficient_phases_21}. For this study, we used 5 layers of amplification, which resulted in 10 computed phase shift angles [$-1.44174911$, $2.96208034$, $3.64950635$, $2.62339909$, $5.22425252$, $5.22425252$, $2.62339909$, $3.64950635$, $2.96208034$, $-26.57449034$]. These angles are applied to each layer of amplification, alternating between the oracle and diffusion unitaries per layer. The multi-controlled CNOT gates are decomposed using the v-chain algorithm, which utilizes ancillae qubits but reduces the overall circuit depth compared to a naive decomposition of the multi-controlled gates \cite{bennakhi2024analyzing}.

\begin{figure}[H]
    \centering
    \includegraphics[width=\linewidth]{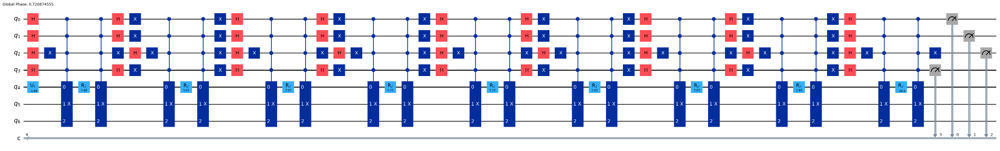}
    \caption{A quantum circuit implementing 5 layers of Fixed Point Amplitude Amplification (FAA) on 4 qubits. The circuit shows the extra ancillae qubits used for the v-chain decomposition of the multi-controlled CNOT gates in the oracle and diffusion unitaries.}
    \label{fig:faa_circuit}
\end{figure}

\subsubsection{Problem Instances}

The problem instances involve all possible binary bit strings that can be generated for a given number of qubits. The range of qubit sizes we studied here are 4, 5, and 6. So, this results in 16, 32, and 64 possible bit strings, respectively. Each bit string defines a problem instance and becomes the target for amplitude amplification. The FAA algorithm is applied to each problem instance with 5 layers of amplification, and the circuit is measured to sample the target bit string.

The score is computed as the ratio of the sampling probability of the target bit string and the maximum possible probability achievable for the target bit string with the same set number of layers of amplification. This range of qubits and problem instances is sufficient to observe the performance of the algorithm in the noise-dominated regime on current NISQ hardware.

\subsubsection{Results}

\begin{figure}[H]
    \centering
    \includegraphics[width=\linewidth]{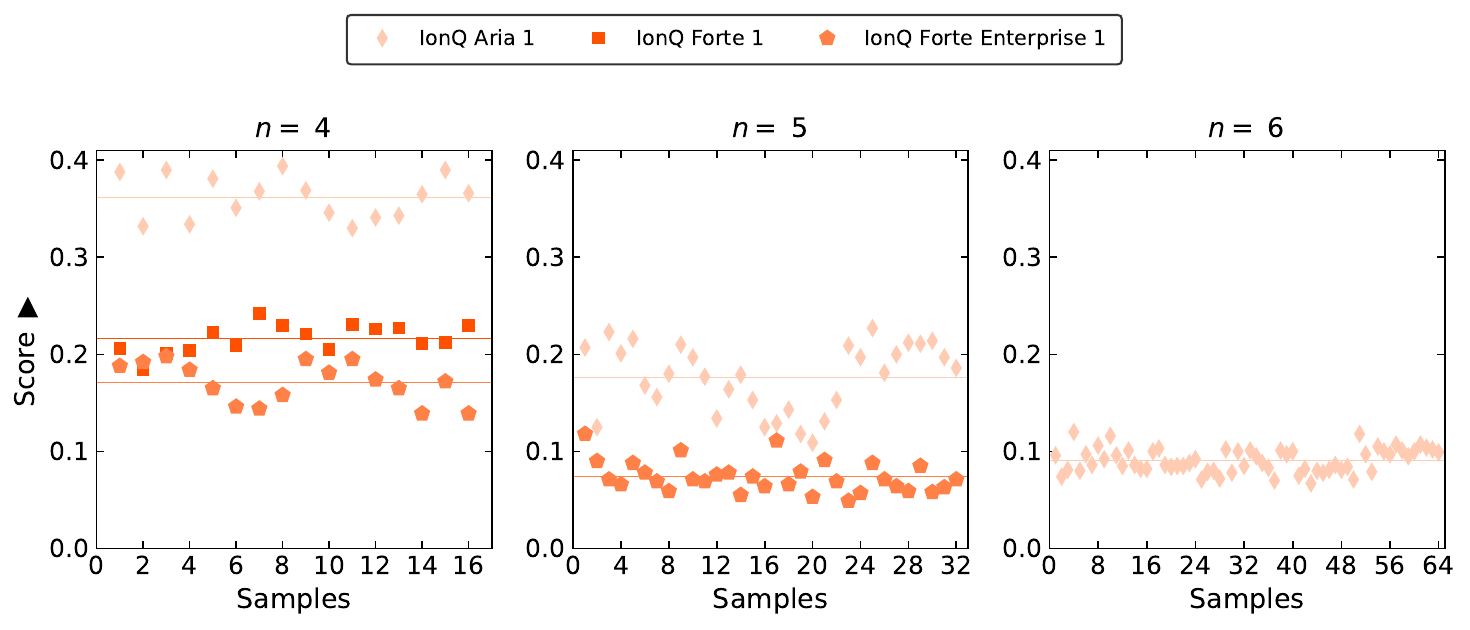}
    \caption{Performance of the Fixed Point Amplitude Amplification (FAA) benchmark for unstructured search on 4, 5, and 6 qubits. The results in each subplot compare empirical data from IonQ quantum hardware. 
    Each hardware execution utilized 1,000 measurement shots. The solid horizontal lines depict the average performance score across all target bit strings.}
    \label{fig:faa_results_odd}
\end{figure}

The FAA results are shown in \Cref{fig:faa_results_odd}. For all the problem sizes and across all bit strings measured, IonQ backends consistently produce high-quality results. At smaller problem sizes (4 qubits), all the backends perform better (produce higher scores) than at larger problem sizes due to the relatively short circuit depth. As the problem size increases to 5 qubits and 6 qubits, the circuit for the FAA algorithm gets deeper and deeper. Thus, the performance score of all backends degrades due to increased QPU noise. The separation in performance between the IonQ Aria and IonQ Forte backends decreases noticeably. This is because the gate fidelity of the IonQ Aria backend decays more rapidly into the noise-dominated region of increased circuit depth than the gate fidelity of IonQ Forte backends. Although the performance score is low (around 0.1) on the plot on the right ($n=6$), which has the deepest circuits, the variation in the score across different bitstrings tested is small showing robustness in the IonQ quantum hardware.

\subsection{Image Classification with Quantum Convolutional Neural Network (QCNN)}

\subsubsection{Rationale} This benchmark is based on the quantum Convolutional Neural Network (QCNN) circuit, which is trained for binary classification of MNIST handwritten digits. QCNNs are variational quantum circuits whose architecture is motivated by classical Convolutional Neural Networks (CNNs) \cite{oshea2015, lukinnature2019}. They have been widely researched and studied for image processing type tasks like recognition, classification, or object detection ~\cite{ieeexplore2021, springer_qmi_2022}. The quantum hardware is benchmarked by evaluating the inference performance of a pretrained QCNN, specifically measuring its test accuracy. QCNNs have shown certain promising features like good generalization capability over unseen datasets and requiring fewer training parameters. QCNNs have also shown competitive performance on widely relevant problems \cite{jsupercomp2025, icassp2021, elhag2025, ieeeaccess2023} and are a promising application for quantum computing. In order for any ML-based algorithm to perform well on the quantum computer, the pre-trained circuit needs to perform well on the hardware and with low or reasonable latency. That is what this benchmark based on inference demonstrates.
\subsubsection{Algorithm Description} In the particular QCNN structure we used, the original images, which were of $28 \times 28$ shape, are resized into $3 \times 3$ and $4 \times 4$ images using standard Dataloader modules in pyTorch. These images are then vectorized and encoded into the circuit by Angle Encoding, where the pixel values are encoded in single-qubit rotation gates. It is then followed by layers of unitary gates, which act as the convolutional layers, and controlled rotation gates, which act as the pooling function [11]. Depending on the complexity of the task, we can add more layers of the trainable unitaries before the final pooling. At the end, we perform X, Y, and Z measurements on the first qubit, which are then fed into a final feed-forward neural network as shown in the diagram below. 

\begin{figure}[H]
    \centering
    \includegraphics[width=0.85\linewidth]{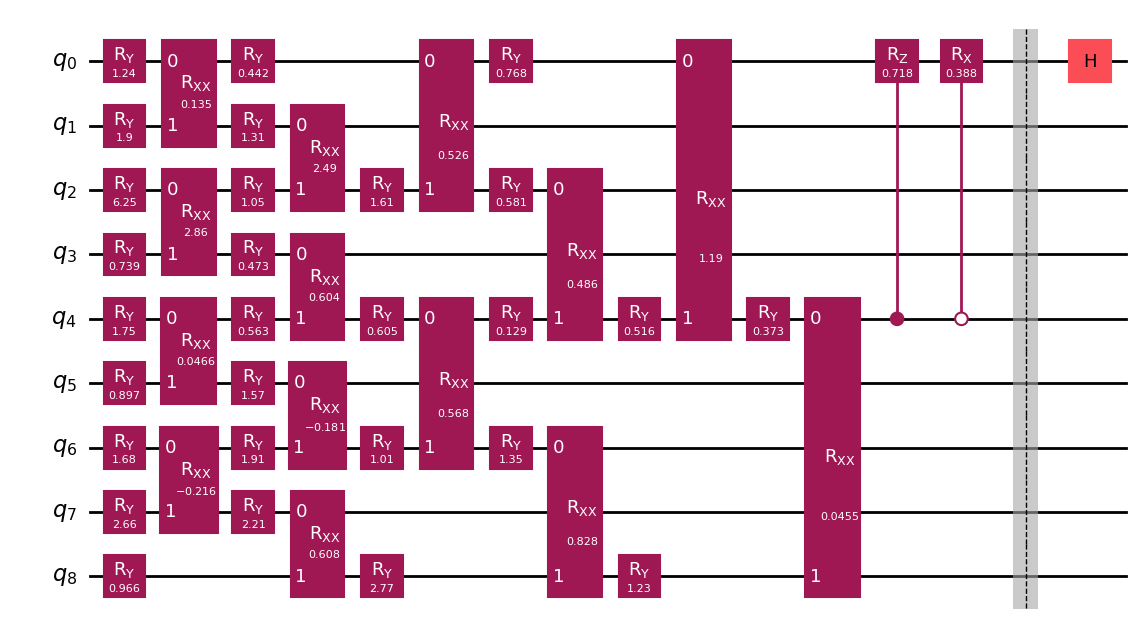}
    \caption{A typical QCNN circuit. Once the data is loaded onto the qubit register, it is transformed through a sequence of convolutional layers. Then measurements are performed on the top qubits in $X$, $Y$, and $Z$ bases. The final output is produced by a neural network layer which outputs the label.}
    \label{fig:qcnn_algorithm}
\end{figure}

The network is trained with 20-40 resized images and with cross-entropy loss. We used COBYLA as the optimizer. To bypass the problem of vanishing gradients, we restarted the optimizer after every 400 iterations. We ran the optimization through 10 such restarts, thus, for 4,000 iterations. We initialized all the parameters, both classical and quantum, randomly. 

\subsubsection{Problem Instances} We perform two sets of classification: one where we classify between $0$-$1$ classes and the other between $1$-$7$ classes in the MNIST dataset. In the ideal case, the 1-7 classification shows lower test accuracy than $0$-$1$ classification with the specific algorithm used. In each set, we have two circuit sizes. One with 9 qubits and $\sim 30-40$ two qubit gates. The other with 16 qubits and $\sim 70-80$ two qubit gates. We only use inference for benchmarking. There are 4 sets of problem instances, where each classification type has two circuit sizes. Each set consists of the final trained parameters of the corresponding QCNN model, the corresponding test images and their labels. Note that the 9 qubit circuit is trained on 3X3 images and 16 qubit circuits are trained on $4 \times 4$ images. The only pre-processing we use on the images before loading them on the quantum circuit is resizing. The quantum circuit has the QCNN structure with $\sim 40$ parameters in the 9 qubit instance and $\sim 80$ parameters in the 16 qubit instance. The output of the quantum circuit are single qubit X, Y and Z measurements, which are then post-processed by a small classical neural network (NN) with $\sim 14$ parameters to predict the class label. The models were trained on $20$-$40$ resized images and tested on $50$ resized images. The score of the benchmark is the test accuracy.

In this application, we benchmark the hardware results against ideal results. We do not benchmark performance scaling with increasing hardware size, because the ideal model itself did not show any particular trend between the two circuit sizes. The reason for this is that the two models are different, being trained on two different image sizes and, hence, image types.

\subsubsection{Results} 

\begin{figure}[H]
    \centering
\includegraphics[width=0.6\linewidth]{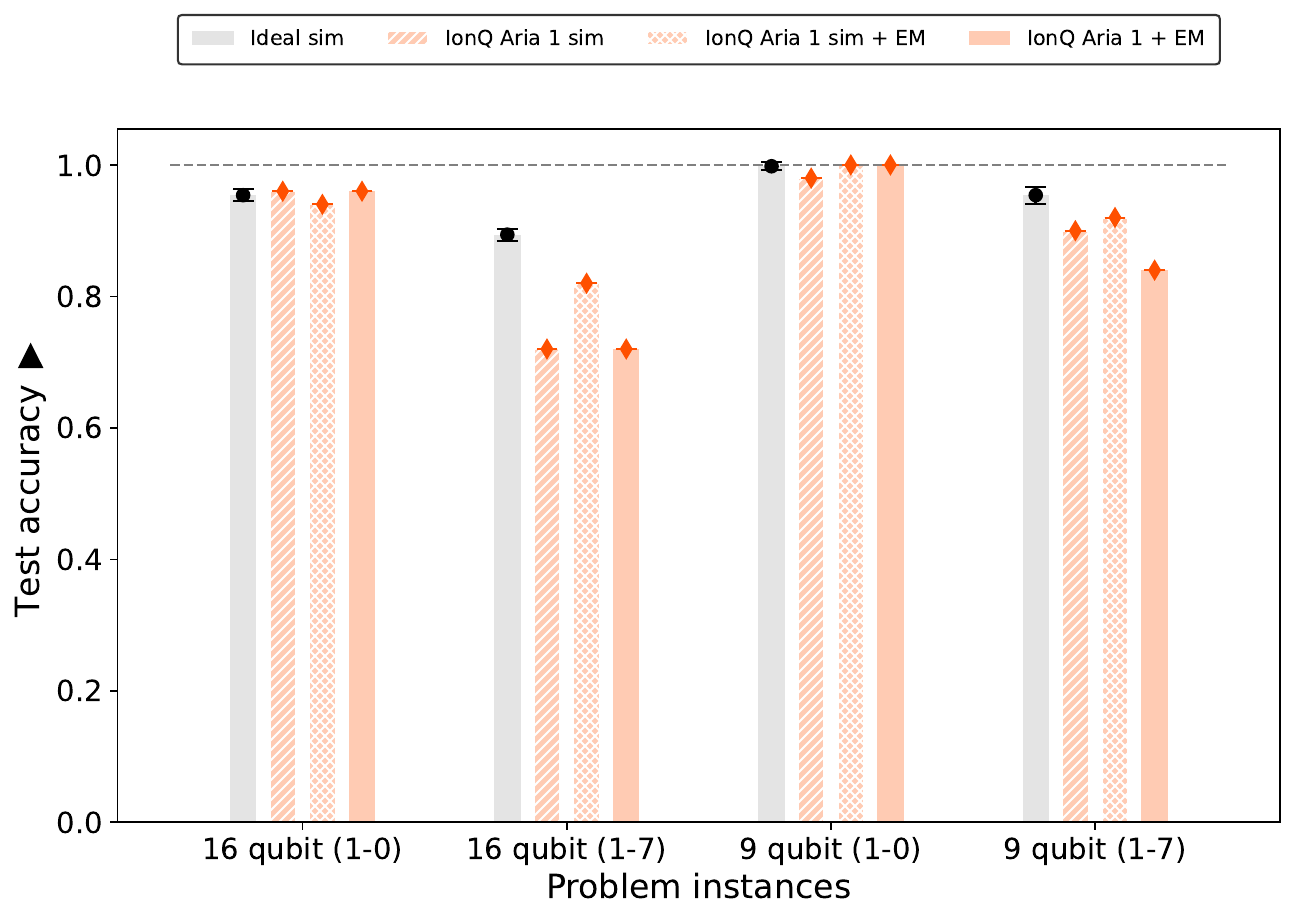}
    \caption{Test accuracy of the 4 QCNN problem instances with 50 test images.}
    \label{fig:qcnn_results}
\end{figure}

In \Cref{fig:qcnn_results}, we demonstrate the QPU performance along with noisy simulation and ideal results. The target of this benchmark is to achieve $100\%$ accuracy in each problem instance.

We see that Aria, with custom error mitigation, performs as well as the ideal simulation in the 1-0 classification task ($\sim 95 -100\%$ accuracy), for both the $9$q and $16$q models. In this category, we emphasize that the hardware has achieved the target, as shown by the two models. However, the 1-7 category of classification shows reduced performance on both ideal and noisy simulators, and expectedly, on the QPU as well.

\subsection{Portfolio Risk Analysis Using Quantum Copulas}

\subsubsection{Rationale} Copulas are mathematical tools used to model multivariate joint probability distributions \cite{Sklar1959FonctionsDR}. Copulas enable one to treat the marginal distributions of each of the random variables independently, and have been used in various fields such as quantitative finance, engineering, signal processing, and medicine \cite{repec:bla:jtsera:v:36:y:2015:i:4:p:599-600,cherubini2004copula,LEBRUN2009312,LAMBERT}.   

The three main classes of copulas are Archimedean, vine, and elliptical copulas. The most commonly used copulas in practice are empirical copulas, which are a mixture of copulas. Empirical copulas are usually modeled using parametric methods such as maximum likelihood estimation \cite{JoeHarry1997Mmad}. As a result, it becomes more computationally expensive to model empirical copulas in high dimensions.

It was shown that a Quantum Circuit Born Machine (QCBM) can be trained to model the joint probability distribution of a portfolio of up to 4 indices \cite{Daiwei2022}. It was also shown that the QCBM copula can outperform classical methods when used to estimate the value of at risk (VaR) of the portfolios. The VaR is a metric commonly used in the financial sector to estimate risk associated with a portfolio over a specific time frame at a given confidence interval. Given a portfolio with loss $L$, time horizon, and confidence interval $\alpha$, the VaR of the portfolio is the smallest number $l$ such that the probability that $L > l$ is at most $1-\alpha$. Mathematically, VaR is defined as
\begin{equation}
\mathrm{VaR}_{\alpha} = inf\{l\in \Re : F_L (l) > \alpha\},
\label{eqn:VaR}
\end{equation}
where $F_L$ is the cumulative distribution of the loss function. Note that the VaR is simply the quantile of the loss distribution.
In previous works, the QCBM copula was trained by minimizing the Kullback-Leibler (KL) divergence between the distribution of the training data and that of the data generated by the QCBM. This training method, however, does not scale because the KL divergence becomes intractable as the number of indices increases. In this work, we implement a novel training method that overcomes the scaling issues of the previous iterations of QCBM copulas, and show that quantum copulas can now be used to model the joint probability distributions of up to 10 indices.

\subsubsection{Algorithm Description} QCBM copulas are trained to estimate the out of sample 95\% VaR of equally weighted portfolios. Portfolios of 5-10 variables (stock indices) were built from the following individual returns: AAPL, ADBE , AMZN, FORD, INTC, JPM, KO, MCD, MSFT, and PFE. The trading period is from 01/04/2001 to 12/30/2020. The last 1,000 days of the trading period are used for VaR estimation, and the rest of the data is used to train the QCBM models. 

The quantum models were trained by minimizing the maximum mean discrepancy (MMD) between the training data and samples from the QCBM models in copula space. The MMD is a metric used to measure the distance between two distributions. Given distributions $X$ and $Y$, the MMD between the distributions is given by
\begin{eqnarray}
    \textit{MMD}(X,Y) = \mathbb{E}_{x,x' \sim X} [k(x,x')] \,+\, \mathbb{E}_{y,y' \sim Y} [k(y,y')] \,-\, 2 \mathbb{E}_{x\sim X,y \sim Y} [k(x,y)],
    \label{eqn:MMD_Score}
\end{eqnarray}
where $k(x,y)$ is a kernel function. In this work, we use the Gaussian kernel with width $\sigma = 1$ to train the QCBM models. The training data is converted into copula space by using marginal distribution $F_j$ for each index. In the financial industry, it is customary to use student's t-distributions to estimate the marginal distributions $F_j$.  Measurements from the QCBM circuit can be converted into copula space by using a specified number of qubits per variable $m$. As an example, for an $n$ variable portfolio with $m$ qubits per variable, the total number of qubits used in a QCBM circuit is $mn$. Sampling from the circuit will generate bit strings of length $mn$, and each $m$ bit strings are converted into fractions, yielding copula samples $u_1,...,u_n$ with $u_i \in [0,1)$. 

\begin{figure}[H]
    \centering
    \includegraphics[width=0.9\linewidth]{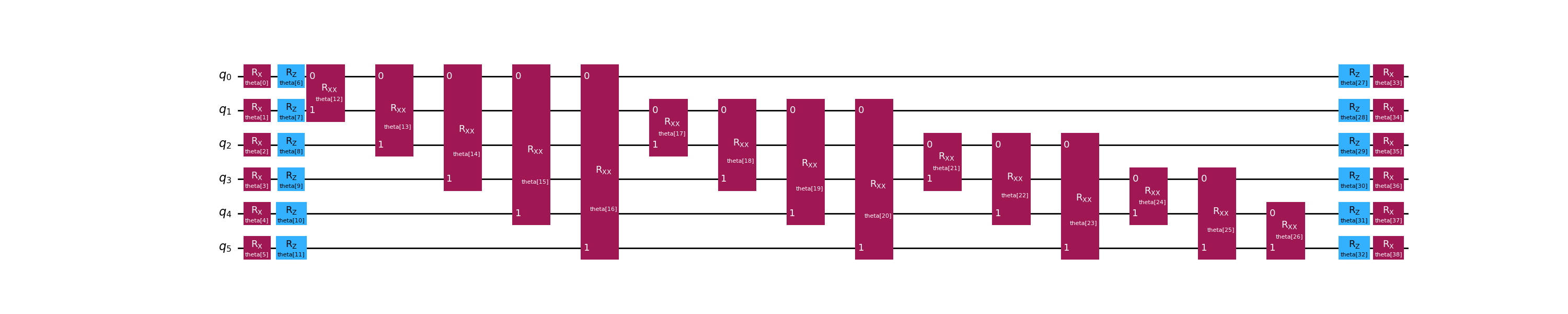}
    \caption{QCBM ansatz 1 used to model a 2-variable copula with 3 bits per variable.}
    \label{fig:QCBM0}
\end{figure}

\begin{figure}[H]
    \centering
    \includegraphics[width=0.9\linewidth]{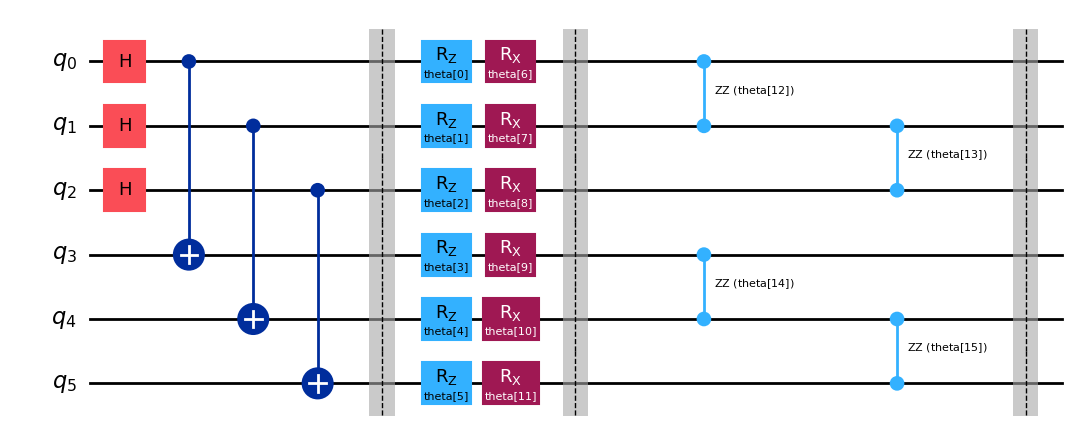}
    \caption{QCBM ansatz 2 used to model a 2-variable copula with 3 bits per variable.}
    \label{fig:QCBM1}
\end{figure}

We considered two types of ansatz for the QCBM models. The first type (ansatz 1) has layers of single qubit rotation gates $R_x(\theta)$, $R_z(\phi)$, and two qubit entangling gates $R_{xx}(\psi)$ with all-to-all connectivity. An example of ansatz 1 used to model a 2-variable copula using $m = 3$ qubits per variable is shown in \Cref{fig:QCBM0}. The second type of ansatz (ansatz 2) is constructed by forming a maximally entangled Greenberger–Horne–Zeilinger (GHZ) state with one qubit from each register (one register per variable) followed by layers of single qubit rotation gates $R_z(\phi)$, $R_x(\theta)$, and two qubit entangling gates $R_{zz}(\psi)$ with next-nearest neighbor connectivity. An example of ansatz 2 used to model a 2-variable copula using $m = 3$ qubits per variable is shown in \Cref{fig:QCBM1}.

Copula samples $u_j$ from the fully trained QCBM circuits are converted into real space $x_j$ by using the inverse transformation $x_j = F^{-1}_{j}(u_j)$. The VaR for the generated data is then calculated by using the generated loss distribution. This algorithm returns a score, which is the ratio between the VaR calculated from the data generated by QCBM to the actual VaR on the test data. The score is taken to be the minimum of VaR and 1/VaR so as to ensure it lies in the range [0,1].

\subsubsection{Problem Instances} 

\begin{table}[H]
\centering
\begin{tabular}{cccccc}
    \toprule
 \#vars & \#q & \#1q (ansatz 1) & \#2q (ansatz 1) & \#1q (ansatz 2) & \#2q (ansatz 2) \\
 \midrule
 5 & 15 & 60 & 105 & 33 & 22 \\
 6 & 18 & 72 & 153 & 39 & 27 \\ 
 7 & 21& 84 & 210 & 45 & 32 \\ 
 8 & 24 & 96 & 276 & 51 & 37 \\ 
 9 & 27 & 108 & 351 & 57 & 42 \\ 
 10 & 30 & 120 & 435 & 63 & 47 \\ 
    \bottomrule
\end{tabular}
\caption{Quantum copula: Overview of the number of variables (\#vars), number of qubits (\#q), as well as number of 1-qubit gates (\#1q), and 2-qubit gates (\#2q) for the problem instances and the 2 ans\"atze considered.}
\label{table:1}
\end{table}

This benchmark considers portfolios of 5-10 variables. QCBM models were trained for each model using the two ansatz discussed in the previous section, using $m = 3$ qubits per register (variable). As a result, the total number of problem instances in this benchmark is 12. Note that for portfolios of $n$ variables, the number of qubits required in the QCBM ansatz is $mn$. The number of qubits and gates used for each of the ansatz, for each problem instance, is shown in \Cref{table:1}.
\subsubsection{Results} 

\begin{figure}[H]
    \centering
    \includegraphics[width=0.9\linewidth]{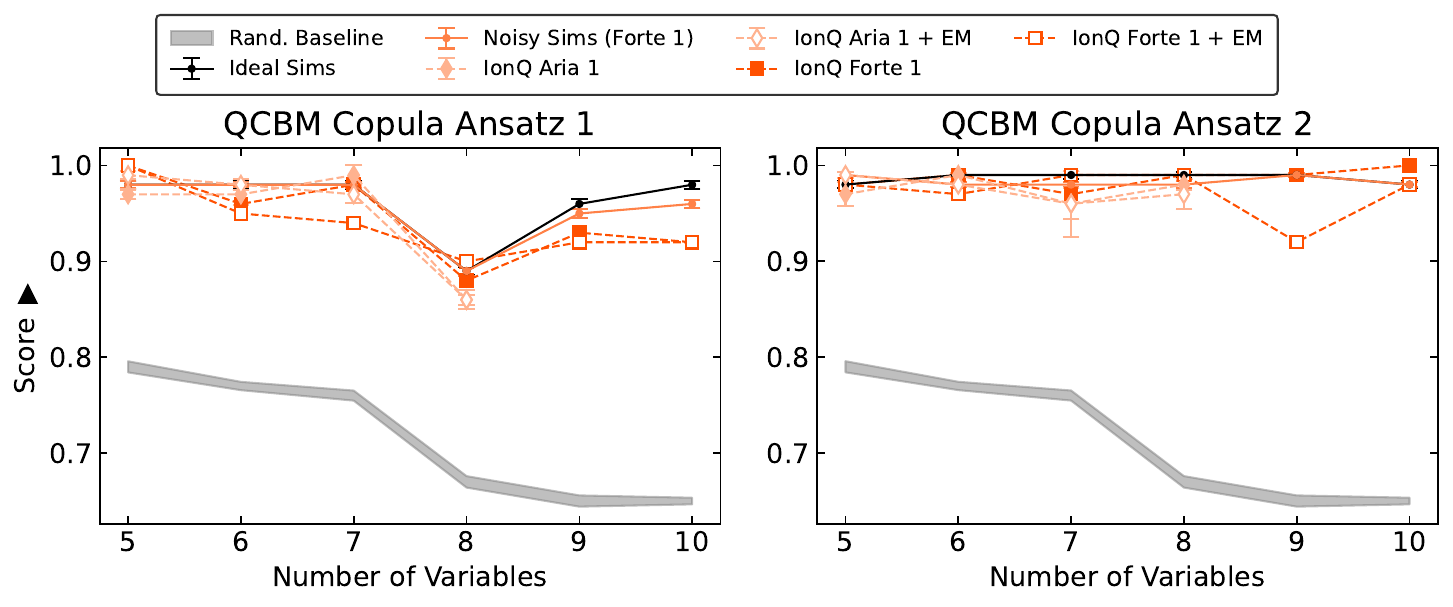}
    \caption{Benchmark scores versus number of portfolio variables for QCBM ansatz 1 (left) and 2 (right panel). Quantum copula results shown for ideal simulations, noisy simulations, and simulations on IonQ's quantum hardware. Results for the random sampling baseline are also included for comparison.}
    \label{fig:QCBM2}
\end{figure}

The performance of QCBM copulas trained using the two ansatz for each of the problem instances is shown in \Cref{fig:QCBM2}. The figure shows score versus number of variables for ideal simulations, noisy simulations, and simulations on IonQ's Aria-1 and Forte-1 quantum hardware (with and without error mitigation). Also shown on the plot is the performance of a classical baseline, which generates copula samples randomly from a uniform distribution. Due to the limited number of qubits, only portfolios of up to 8 variables were simulated on Aria-1 (see \Cref{table:1}). The error bars are obtained from 10 independent runs for ideal and noisy simulations, 3 runs for Aria-1 simulations of ansatz 1, 2 runs for Aria 1 simulations of ansatz 2, 5 runs for Aria-1 simulations of ansatz 1 with error mitigation (EM), 3 runs for Aria-1 simulations of ansatz 2 with error mitigation, and 10 runs for random sampling. The custom error mitigation parameters used for this benchmark are $power = 1.5$ and $threshold = 0.0$. 

\Cref{fig:QCBM2} shows that QCBM copulas outperform random sampling in estimating the out-of-sample VaRs of all the portfolios considered. This shows the effectiveness of trained QCBM copulas for portfolio risk assessment. For ansatz 1 (see left panel of \Cref{fig:QCBM2}), Forte 1 hardware results without error mitigation are statistically similar to ideal simulation results up to 8 variables. For 9 and 10 variables, the higher number of 2-qubit gates causes an increase in quantum errors, resulting in slightly lower Forte 1 scores compared to ideal simulations. For ansatz 2, all hardware scores are similar to ideal simulation scores for all portfolios. This result is expected because the significantly smaller number of 2-qubit gates makes ansatz 2 more resilient to quantum errors. \Cref{fig:QCBM2} also shows that simulations on quantum processing units (QPUs) with error mitigation do not seem to improve scores for either ansatz. 

\subsection{Image Loading with Tensor Network-Based Quantum Circuits}

\subsubsection{Rationale} Generating circuits to prepare an initial quantum wave function is a critical step in a wide range of quantum algorithms. Preparing an arbitrary wave function, $\ket{\Psi_0}$, is notoriously difficult in the most general case. Preparing an arbitrary initial state on $N$ qubits typically requires a circuit with O($2^N$) quantum gates.  This presents particular challenges in the field of quantum machine learning, where it is often needed to prepare an initial state that transforms classical data into quantum data by representing the data as a quantum state.  In the context of QML, this state preparation method is known as data loading. While there are many possible choices of how to encode this classical data, one particularly popular method, due to its qubit efficiency, is known as the dense amplitude encoding, where we prepare the wave function

\begin{align*}
\ket{\Psi_0} = \frac{1}{Z} \sum_{i=0}^M \sqrt{x_i} \ket{i},
\end{align*}

where $x_i$ are the $M$ real numbers which make up a classical data point $x$, and $Ζ$ is the normalization constant. 

For an arbitrary data point $x$, it is clear that $O(M)$ quantum gates are required to perform this data loading step. On NISQ devices, this data loading step quickly becomes impractical for even moderately large input data, and can be a computational bottleneck even on fully fault-tolerant devices. However, in many cases, the input data is not an arbitrary vector but instead has an internal structure that can be taken advantage of to greatly reduce the number of quantum operations needed to perform the data loading. This is the case in computer vision applications, where the input data is an image that may be fed into a supervised learning algorithm to perform, for example, a classification task. In this case, for the data point $x$, the index $i$ can represent the $(x,y)$ coordinates of the pixel location, and the value of $x_i$ represents the pixel intensity. In most images, the pixel intensities are not randomly distributed, but instead form a smooth function in the two-dimensional space represented by the $(x,y)$ pixel coordinates.  

Recent advances in quantum data loading have found that such smooth distributions can be approximately represented rather efficiently using mathematical objects known as tensor networks. A particular kind of tensor network, the matrix product state (MPS), has been shown to be particularly useful for these applications, both due to the efficiency of performing classical manipulations on these objects and also due to the development of algorithms to generate low-depth quantum circuits that can approximately prepare these MPS states. Using these  methods, a quantum circuit which approximately loads 2D images can be constructed with O(DN) quantum gates, where $D\sim O(1)$ is the depth of the circuit. The quality of approximation of the image representation improves as $D$ increases, however device noise also increases with $D$.

In this benchmark, we test the ability of the quantum device to load grayscale 2D images of various complexities. We measure the fidelity of the final reconstructed image, measured from the output of the MPS image loading circuits, with the original image and test the ability of the quantum hardware to improve the image fidelity as the circuit depth increases.

\subsubsection{Algorithm Description} The benchmark for the image loading algorithm involves generating a series of quantum circuits that approximately prepare a matrix product state wavefunction whose amplitudes encode a 2D grayscale image. 

A matrix product state wave function has the form
\begin{align*}
   \ket{\Psi_{mps}} = \frac{1}{Z} \sum_{\sigma_i}\prod_{i=0}^{N}{M^{\sigma_i}_{\alpha_i\beta_{i+1}}} \ket{\sigma},
\end{align*}
where $M^{\sigma_i}_{\alpha_i\beta_{i+1}}$ is a 3-component tensor for each qubit $\sigma_i$ . For a particular spin configuration of the $i^{th}$ qubit, this becomes a matrix with indices $\alpha_i$, $\beta_{i+1}$, which can take values $1, \ldots, \chi$ . The variable $\chi$, is the linear size of the matrix and is known as the bond dimension.  It has been shown that a fixed $\chi$ is sufficient to perform the image loading to a fixed fidelity, independent of the size of the image. However, the value of $\chi$ does depend on the complexity of the original image.

As described in Refs.~\cite{rudolph2023decomposition,iaconis2023tensor}, the matrices, M, can be viewed as isometries, which can then be expanded and converted directly into unitary matrices, and then directly form the set of quantum gates that prepare the MPS state on a quantum computer.  These unitary matrices act on $log_2(\chi)+1$, qubits. While this construction is already efficient in theory, in practice, we can reduce the circuit depth even further by applying an iterative algorithm for approximating a bond-dimension $\chi$ MPS state. 

\begin{figure}[H]
    \centering
    \includegraphics[width=0.95\linewidth]{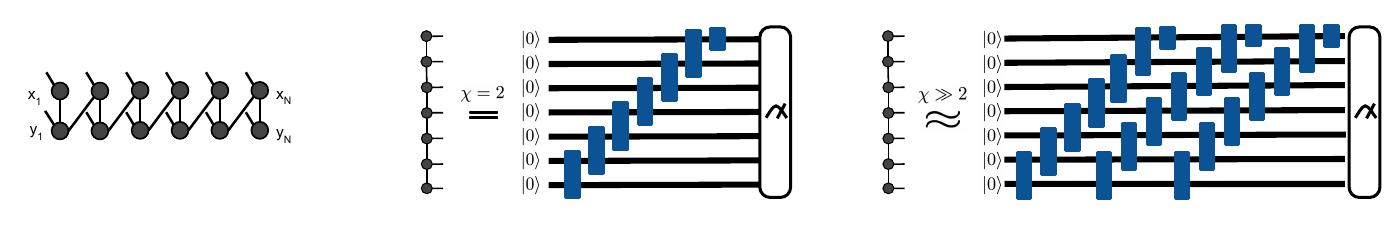}
   \caption{ The tensor network diagram used to represent a 2D image {\it (left) }. A single layer quantum circuit {\it (middle) } can exactly construct an MPS state with bond dimension $\chi =2$. A depth $D$ circuit {\it (right)} approximately constructs a $\chi >2$ MPS state. }
    \label{fig:imageloading_description}
\end{figure}

For each image, the method proceeds by first approximating the image as a matrix product state with bond dimension $\chi$. This MPS is then used to construct a quantum circuit with $D$ layers, each layer consisting of $N-1$ general $O(4)$ matrices. The $O(4)$ matrices are optimized using the methods of Ref.~\cite{iaconis2023tensor} to approximate the tensor network. As the number of layers $D$ increases, the quality of the MPS approximation and the fidelity of the quantum state measured in the computational basis with the input image also increases. However, on noisy quantum hardware, the increased circuit depth of higher $D$ circuits leads to increased levels of noise and a reduction in the fidelity with the input image. Therefore, we expect to see an optimal depth $D$, where both sources of error are comparable to each other. The optimal value of $D$ depends both on the size and complexity of the input image, which affects the quality of the MPS approximation, and on the hardware noise, which affects the ability to prepare the target quantum state. The final score is the mean-squared-error (MSE) between the reconstructed image and the original image. That is
\begin{eqnarray}
    MSE = \sum_{i=0}^{2^N} \left(x_i^{(Q)} - x_i^{(I)}\right)^2
\end{eqnarray}
where $x_i^{(Q)}$ is the measured amplitude from the quantum state of the bit string for pixel $i$, and $x_i^{(I)}$ is the corresponding pixel intensity of the real image $I$. In both cases, we normalize the pixel values so that $\sum_i (x_i^I)^2 = 1$. 

\subsubsection{Problem Instances}

We benchmark the MPS loading method on two grayscale images, which are respectively of low and medium image complexity. The low complexity image is a $32\times 32$ grayscale image of the number $5$ taken from the MNIST handwritten digit database. The medium complexity image is a $64\times64$ grayscale image of the outline of a shark taken from the ImageNet-Sketch database.  For both images, we generate circuits for increasing number of layers, $D$, and measure the output distribution in the computational basis, giving a score for each value of $D$. The quantum circuit for the $M \times M$ circuit with $D$ layers uses $2\log_2(M)$ qubits and approximately $2.5 D\times(\log_2(M)-1)$ CNOT gates. The score is the mean-squared-error between the reconstructed image and the original image. 

The overall score is the lowest MSE achieved for the given image, taken at the depth D that gives the best score. The hardware passes the benchmark for a given image if it is able to achieve a score below $MSE \sim 0.1$

\subsubsection{Results}

\paragraph{QPU Comparison:}

\begin{figure}[H]
    \centering
    \includegraphics[width=0.95\linewidth]{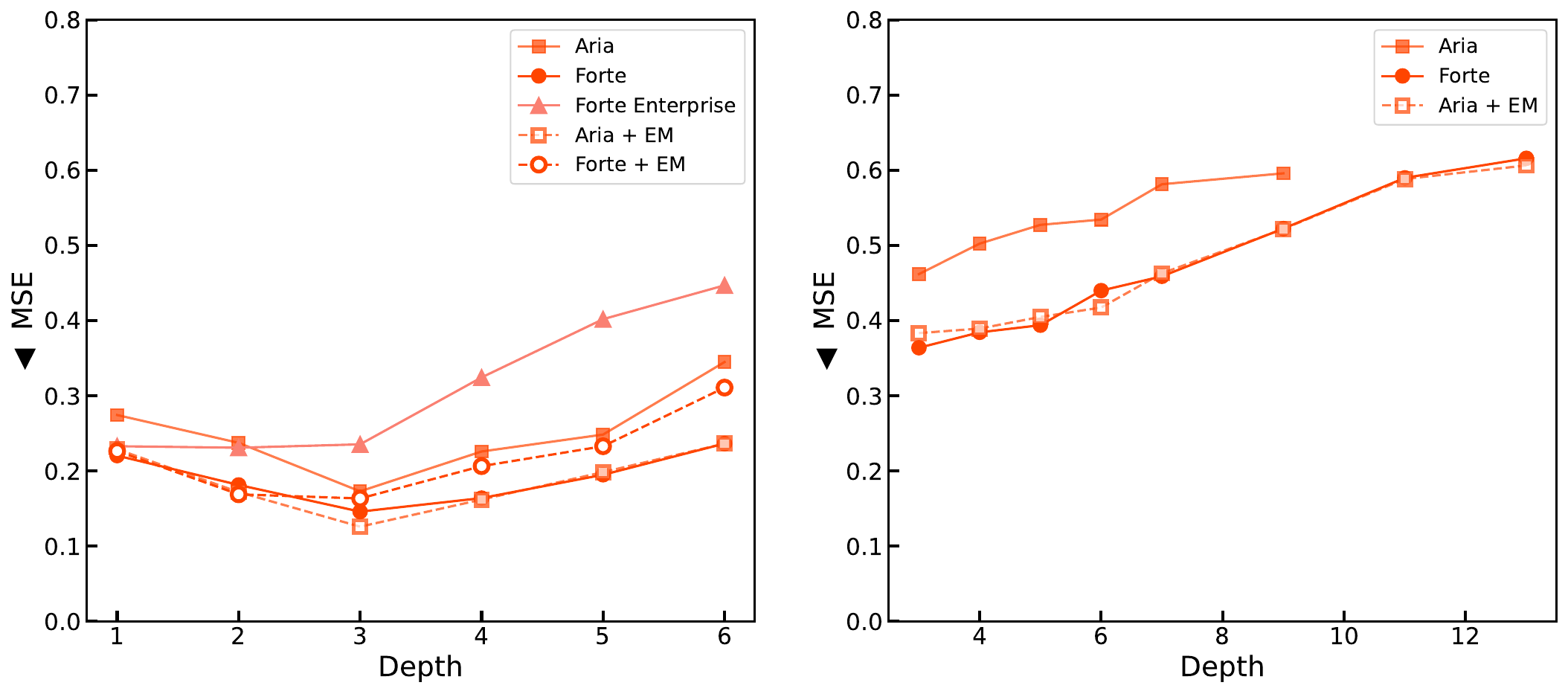}
    \caption{The Mean-Squared error score of the MNIST images (left) and the ImageNet-Sketch Images (right), prepared on the IonQ QPUs, as a function of depth. Note that lower is better.}
    \label{fig:imgload_score}
\end{figure}

A comparison of the hardware results between the IonQ machines is shown in \Cref{fig:imgload_score}. The left shows the performance on the easy difficulty MNIST handwritten digit test case, and the right shows the performance on the medium difficulty ImageNet-Sketch image.  The IonQ Forte QPU showed the best performance up to depth $D=3$ for the MNIST image, achieving an MSE $\sim 0.15$. The IonQ Aria QPU performed similarly to IonQ Forte when error-mitigation methods were applied to the result. In both cases, the best MSE score on the ImageNet-Sketch image was achieved for the lowest depth $D=3$. 

\begin{figure}[H]
    \centering
    \includegraphics[width=0.8\linewidth]{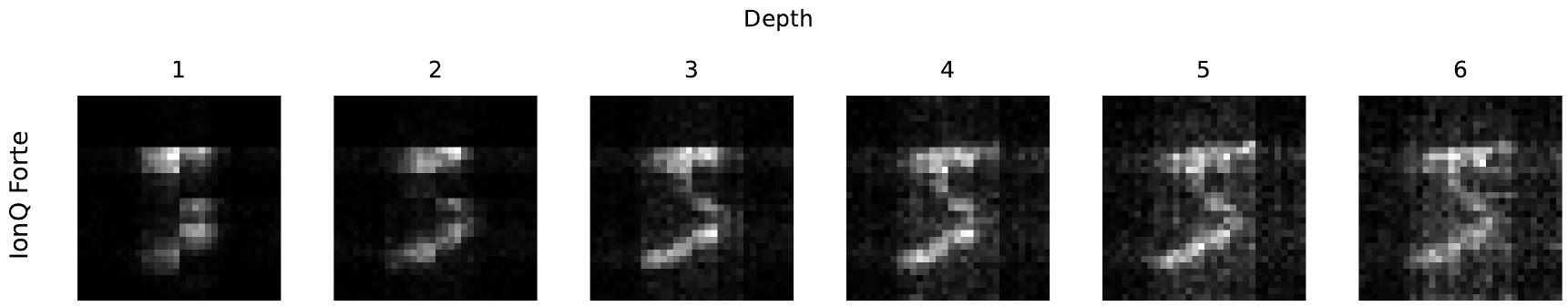}
    \caption{The images generated on IonQ Forte for the MNIST images at different depths. }
    \label{fig:mnist_img}
\end{figure}

\begin{figure}[H]
        \centering
        \includegraphics[width=0.95\linewidth]{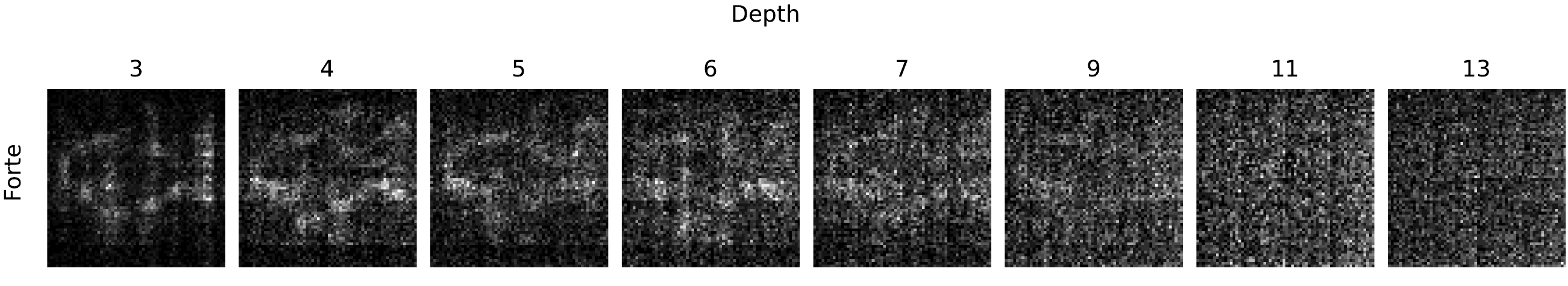}
        \caption{The generated ImageNet-Sketch images sampled from the output of the best-performing IonQ QPU backend.}
        \label{fig:sketch_img}
\end{figure}

In \Cref{fig:mnist_img,fig:sketch_img}, we show the output images at all layer depths $D$ for the best-performing QPUs on the IonQ  hardware. In the MNIST case, one can see the initial improvement in image quality as $D$ increases due to the improved MPS approximation, followed by a degradation of image quality at higher values of $D$ when hardware noise begins to dominate. For the ImageNet-Sketch image, the best performing model occurs at depth $D=3$.


\subsection{Quantum Fourier Transform (QFT)}

\subsubsection{Rationale} 

QFT implementations are of fundamental interest in the world of quantum computing. The QFT represents one of the few known subroutines for which fault-tolerant quantum computers are guaranteed to deliver computational advantage (from a complexity theory perspective) over the best known classical algorithms; concretely, quantum computers can compute Fourier transform operations on discrete vectors of length $N$ in $O(\log N)$ quantum operations, whereas the best classical algorithms run in $O(N \log N)$.

QFT subroutines feature prominently in well-known algorithms poised to deliver quantum advantage, like Shor's and HHL. Thus, benchmarking the ability of today's hardware to execute QFT subroutines is fundamental for gauging progress towards the practical implementation of algorithms set to deliver quantum advantage.

As suggested above, the goal of this benchmark is to measure the fidelity of a quantum computer's execution of a QFT subroutine. This is challenging for at least two reasons, and care must be taken in designing suitable benchmark challenges. On one hand, when the input state is supported on a few computational basis states, modern compilers employing advanced techniques can effectively compress the benchmark circuit, which results in a poor challenge that fails to reflect the performance a user might encounter when executing an algorithm that relies on QFT subroutines. In particular, compilers are particularly effective at dealing with subcircuits, placing QFT gates and their inverse in sequence, even when there are additional gates in between. On the other, it is non-trivial to design a complex input state that results in a compact output that can readily be compared against an expected, baseline result. Together, this means QFT subroutine benchmarks must balance the complexity of the input state with that of the output state, so that the input is sufficiently ``interesting,'' so the benchmark circuit is not ``compiled away,'' and yet the expected output state is sufficiently ``simple'' so that it can be readily interpreted into a consistent benchmark score.

We find that existing benchmarks are codified in quantum circuits whose gate count can be significantly reduced by advanced compilation techniques, and therefore fail to serve as adequate proxies for the performance of QFT subroutines employed in more complex workflows in practice. In this paper, we introduce three families of QFT challenges intended to address this concern, while producing output states whose correctness can be readily verified.

In each of the following benchmarks, each problem instance consists of a single quantum circuit that features at least one QFT gate.

\subsubsection{Algorithm Description}

The QFT benchmark actually consists of 2 separate problems, Cosine QFT, and Hidden Phase QFT, both of which have in common that they use the QFT circuit as the key subroutine. The difference between the tasks is the preparation of the initial state to which the QFT is applied and the definition of what success looks like for the respective tasks. 

\paragraph{Cosine QFT Challenge}

This challenge uses a quantum circuit to load a cosine plane wave onto the amplitudes of the qubit register and then applies a QFT gate to recover the frequency of the loaded cosine. The input to the final QFT gate is a plane wave, so the expected output state is supported on exactly two computational basis states; the benchmark score is computed by measuring the fidelity of the sampled output against the known reference.

The novelty here is how the challenge circuits load cosine waves onto the qubit registers. Mathematically, the implementation relies on properties of the Discrete Fourier Transform (DFT) that relate the real part of an $N$-vector $x_n$ in the time domain to its representation $\widehat{x}_k$ in the frequency domain, and conversely, the real part of $\widehat{x}_k$ to $x_n$:
\begin{equation}\label{eq:qft_real_pt_freq}
    \mathrm{Re}(x_n) = \frac{1}{2}\big(\widehat{x}_k + \widehat{x}_{N - k}^*\big),
    \quad\text{and}\quad
    \mathrm{Re}(\widehat{x}_k) = \frac{1}{2}\big(x_n + x_{N-n}\big).
\end{equation}

Concretely, the $n$-qubit challenge circuit loads the cosine plane wave with a given frequency $s$ as follows. First, the circuit prepares a uniform superposition of the states $\ket{0}$ and $\ket{N-1}$ using a single Hadamard gate and $n-1$ CNOT operations. Then the circuit uses up to $n-1$ additional CNOT gates to transform $\ket{0}$ into $\ket{N - 2s - 1}$ by applying $X$ gates targeting the active bits in the binary representation of $s$, controlled by the Most Significant Bit (MSB). We assume the target frequency $s$ satisfies $N/4 < s < N/2$, so that $N - 2s -1$ represents a positive integer in the two's complement representation. The latter guarantees that the MSB in the binary representation of $N - 2s -1$ is not active. In symbols, and keeping the two's complement representation for negative integers in mind, we can express this initial state preparation procedure as follows:
\begin{align*}
    \ket{0}^{\otimes n} 
    \to \frac{1}{\sqrt{2}}\big(\ket{0} + \ket{-1}\big)
    \to \frac{1}{\sqrt{2}}\big(\ket{N - 2s - 1} + \ket{-1}\big).
\end{align*}
The challenge circuit then performs Draper-style addition to load the desired cosine wave. In particular, the circuit features a QFT gate followed by a sequence of controlled single-qubit rotations to implement the following transformation:
\begin{align*}
    \to \frac{1}{\sqrt{2}}\big(\widehat{\ket{N - 2s - 1)}} + \widehat{\ket{-1}}\big) 
    \to \frac{1}{\sqrt{2}}\big(\widehat{\ket{-s}} + \widehat{\ket{s}}\big)
    = \sum_k \cos\bigg(\frac{2\pi k}{N} s\bigg) \ket{k}.
\end{align*}
The second transformation shifts each Fourier basis vector by $s + 1$, and the equality follows by the first relation in \Cref{eq:qft_real_pt_freq} and the definition of the QFT.

Finally, the challenge circuit applies a QFT gate to the loaded plane wave to recover the encoded frequency $s$:
\begin{align*}
    \sum_k \cos\bigg(\frac{2\pi k}{N} s\bigg) \ket{k}
    \to \sum_k \cos\bigg(\frac{2\pi k}{N} s\bigg) \widehat{\ket{k}}
    \to \frac{1}{\sqrt{2}} \big(\ket{s} + \ket{-s}\big).
\end{align*}
The last calculation follows from the second relation in \Cref{eq:qft_real_pt_freq} and the definition of the QFT.

\begin{figure}
    \centering
    \includegraphics[width=0.65\linewidth]{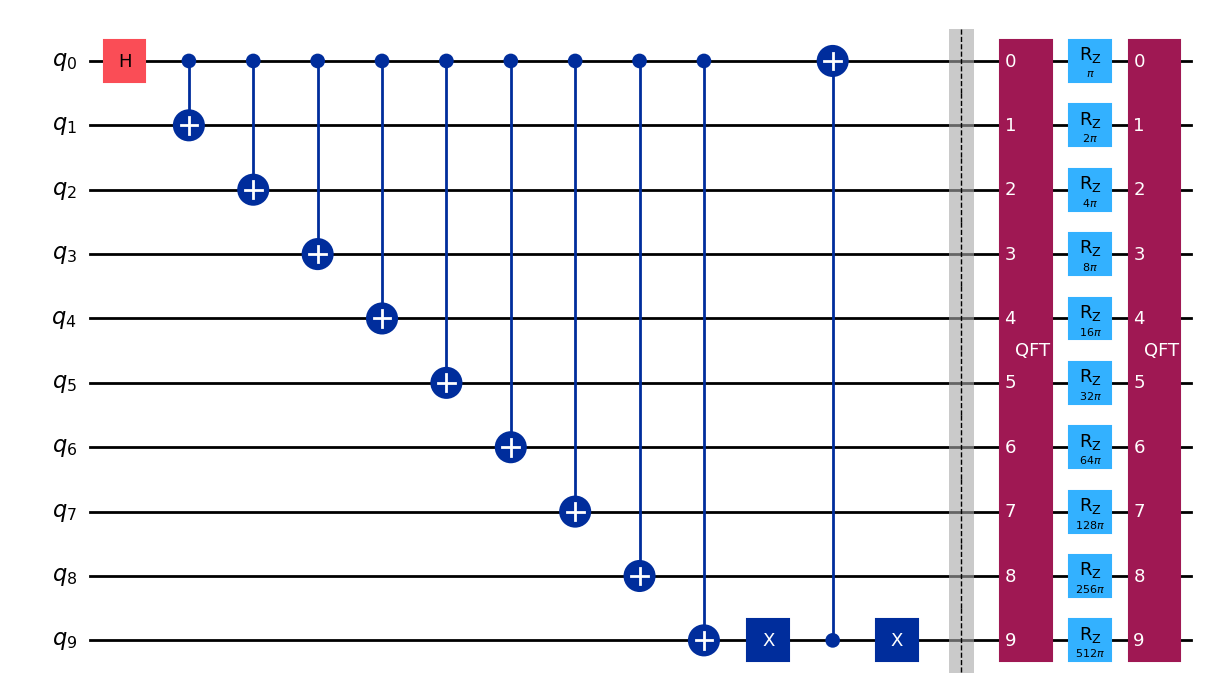}
    \caption{A typical Cosine QFT Challenge circuit. The circuit first loads a cosine plane wave of a given frequency onto the amplitudes of the qubit register, and then it applies a QFT gate to recover the encoded frequency of the cosine wave. The circuit loads the cosine wave by applying a number of entangling gates to a superposition of two computational basis states, and then performing Draper-style addition.}
    \label{fig:cosine_qft_challenge}
\end{figure}

\Cref{fig:cosine_qft_challenge} depicts a typical challenge circuit diagram. The identification $\frac{1}{\sqrt{2}}\big(\widehat{\ket{-s}} + \widehat{\ket{s}}\big)
    = \sum_k \cos\big(\frac{2\pi k}{N} s\big) \ket{k}$ makes it easy to produce a compactly supported     
output reference state in principle, and yet makes it difficult for compilers to render challenge circuits trivial.

\paragraph{Hidden Phase QFT Challenge} 

Each Hidden Phase QFT challenge circuit consists of a Hadamard and a Fourier transform, followed by a sequence of controlled single-qubit rotations, and an additional set of Fourier and Hadamard transforms. The overall operation is trivial for the work qubit register, with the core computation occurring on an ancilla qubit. This ancilla qubit's rotation frequency, which is set by the controlled single-qubit gates, must be determined by sampling the output state. The expected output state is thus compactly supported on the computational basis, and the benchmark is scored by estimating the fidelity of the output state against the reference state.

In symbols, we implement the following qubit state transformations. Given a work register with $n$ qubits and a hidden frequency $0 \leq k^\star < N$, with $N = 2^n$, we begin by loading a uniform superposition over all Fourier basis vectors $\widehat{\ket{x}}$. We define these vectors as the image of the computational basis vectors $\ket{x}$ under the QFT gate:
$$
    \ket{0}^{\otimes (n+1)}
    \to \frac{1}{\sqrt{2N}}\sum_{x=0}^N \big(\ket{x}\ket{0} + \ket{x}\ket{1}\big)
    \to \frac{1}{\sqrt{2N}}\sum_{x=0}^N \big(\widehat{\ket{x}}\ket{0} + \widehat{\ket{x}}\ket{1}\big).
$$
Next, we set $\lambda_k = \frac{2\pi}{N}k$ and apply cyclic shift operators (implemented as controlled single-qubit $R_Z$ rotations) to obtain
$$
    \to \frac{1}{\sqrt{2N}}\sum_x \big(\widehat{\ket{x + k^\star}}\ket{0} + \widehat{\ket{x - k^\star}}\ket{1}\big)
    = \frac{1}{\sqrt{2N}} \sum_x \big(\omega_N^{k^\star}\widehat{\ket{x}}\ket{0} + \omega_N^{-k^\star}\widehat{\ket{x}}\ket{1}\big),
$$
with $\omega_N = e^{2 \pi i/ N}$ denoting a primitive $N$-th root of unity. The last equality relies on well-known properties of the discrete Fourier transform. Finally, we apply additional Fourier and Hadamard transforms to recover the
initial state of the work qubit register:
\begin{align}\label{eq:hidden-phase-qft-final-state}
    &\to \frac{1}{\sqrt{2N}} \sum_x \big(\omega_N^{k^\star}\ket{x}\ket{0} + \omega_N^{-k^\star}\ket{x}\ket{1}\big)
    \to \frac{1}{\sqrt{2}} \ket{0} \otimes \big(\omega_N^{k^\star} H\ket{0} + \omega_N^{-k^\star} H \ket{1}\big) \\ \nonumber
    &\to \ket{0} \otimes \big(\cos(\lambda_{k^\star}) \ket{0} + \sin(\lambda_{k^\star}) \ket{1} \big).
\end{align}

When all is said and done, the work qubit register is mapped back to the initial $\ket{0}$ state, and the ancilla qubit has picked up the hidden phase; it's in the superposition
\begin{equation*}
    H R_Z(\lambda_{k^\star}) H \ket{0}.
\end{equation*}
In principle, we can recover the correct hidden phase by sampling the output state.

In practice, we compute the benchmark score as the Hellinger fidelity between the observed output distribution and the reference state, which is given by \Cref{eq:hidden-phase-qft-final-state}.

\begin{figure}[H]
    \centering
    \includegraphics[width=0.85\linewidth]{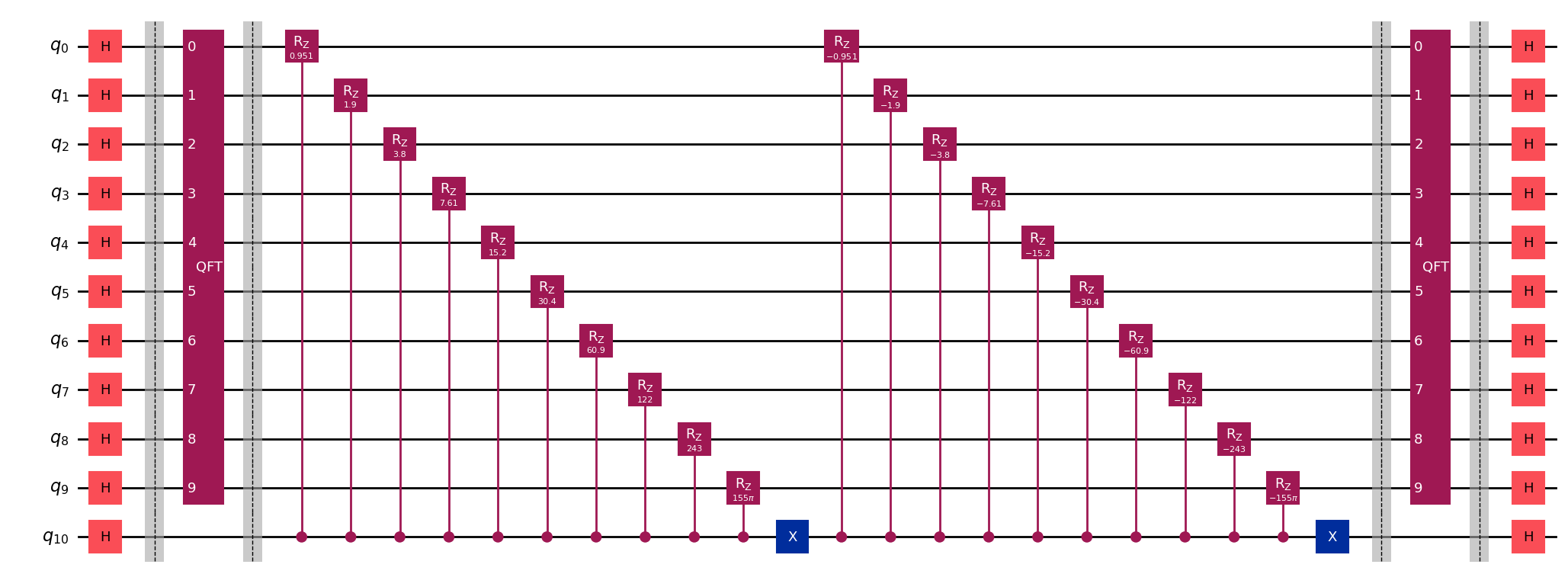}
    \caption{A typical Hidden Phase QFT challenge circuit. The circuit first loads a superposition over all qubits, then applies a QFT gate on the work register, here illustrated as the top 9 wires. Then the circuit applies two Draper-adder-type operations: it shifts the vectors with ancilla state $\ket{0}$ \textit{upwards} by the hidden phase $k^*$, and shifts those with ancilla state $\ket{1}$ \textit{downwards} by $k^*$. Then the circut applies another QFT gate on the work registered, followed by a Hadamard transformation on all the qubits. The work register returns to the initial $\ket{0}$ state, and the ancilla qubit has picked up a hidden phase as described above.}
    \label{fig:hidden-phase-qft-qc}
\end{figure}

\subsubsection{Problem Instances} 

For Cosine QFT, in principle each problem instance is characterized by two integers: a positive integer $n$ denoting the number of qubits in the challenge circuit, and an integer $2^{n/4} < s < 2^{n/2}$ denoting the cosine frequency. In practice, we've set the cosine frequency as $s = 2^{n/2} - 1$ as a safeguard to prevent artificially high scores on machines whose qubits tend to decohere to the $\ket{0}$ state.

For Hidden Phase QFT, each problem instance is characterized by two integers: a positive integer $n$ denoting the number of work qubits in the challenge circuit, and an integer $2^{n/2} \leq k^* < 2^n$ denoting the \textit{hidden frequency}. For each challenge, the hidden frequency $k^*$ was generated as a uniform random sample in $[2^{n/2}, 2^n)$. We chose to avoid frequencies in $[0, 2^{n/2})$ since lower frequency values could result in ancilla states with no significant amplitude in the $\ket{1}$ state.

\subsubsection{Results} 

\begin{figure}[H]
    \centering
    \includegraphics[width=0.75\linewidth]{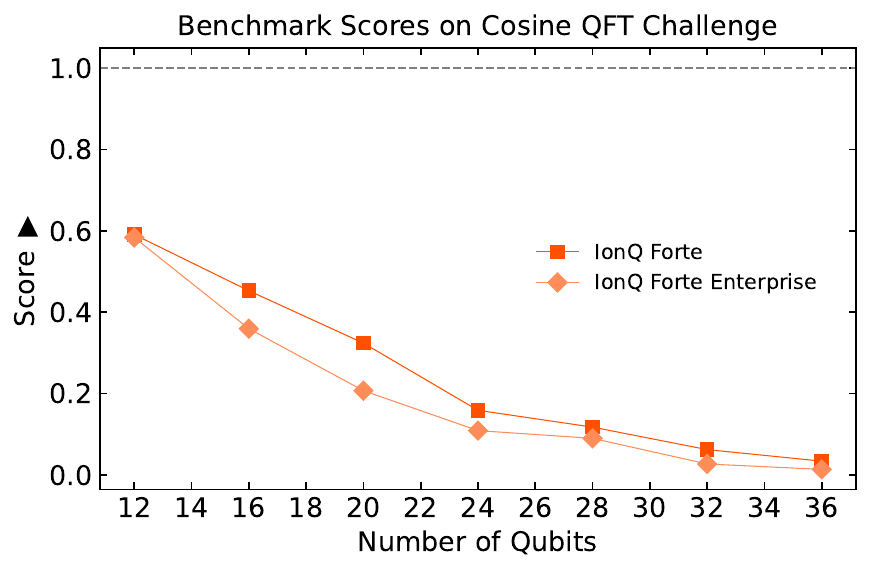}
    \caption{Benchmark score for the Cosine QFT challenge obtained using IonQ Forte and IonQ Forte Enterprise.}
    \label{fig:cosine-qft-score}
\end{figure}

For Cosine QFT, 
\Cref{fig:cosine-qft-score} plots the observed benchmark scores for the Cosine QFT challenge. Each circuit was measured $1,000$ times. Recall that a noiseless quantum device would always achieve the maximum score of $1.0$. IonQ Forte shows significant performance degradation with increasing qubit count, with fidelity scores nearing zero as we push the system to its maximum circuit width. The observed scores seem to follow an exponential trend.

\begin{figure}[H]
    \centering
    \includegraphics[width=0.75\linewidth]{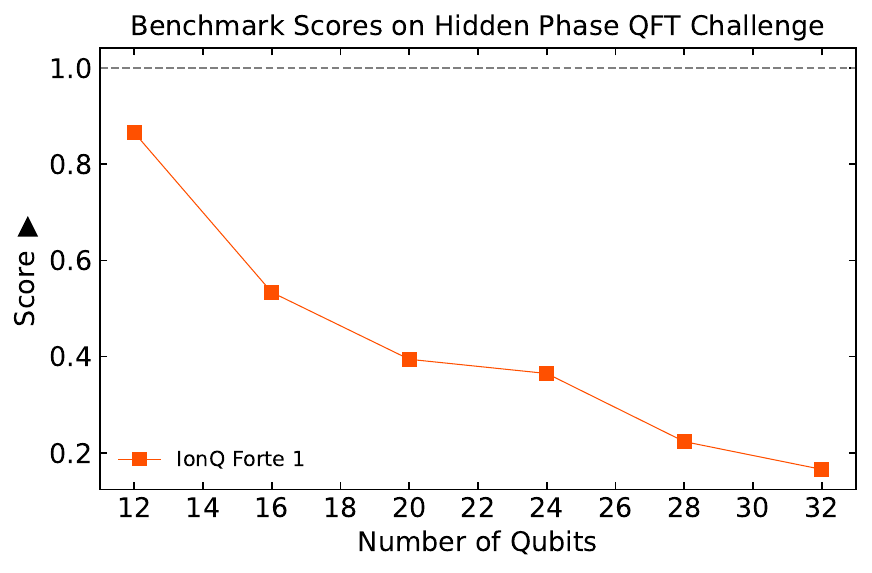}
    \caption{Benchmark score for the Hidden Phase QFT challenge.}
    \label{fig:hidden-phase-qft-score}
\end{figure}

For Hidden Phase QFT, 
\Cref{fig:hidden-phase-qft-score} plots the observed benchmark scores for the Hidden Phase QFT Challenge. Each circuit was measured $1,000$ times. Recall that a noiseless quantum device would always achieve the maximum score of $1.0$. IonQ Forte shows significant performance degradation with increasing qubit count, with fidelity scores nearing zero as we push the system to its maximum circuit width. The observed scores seem to follow an exponential trend, with some deviations. The latter are likely due to the slight differences in challenge difficulty that arise from the particular choice of hidden frequency.

\subsection{Hidden Shift Benchmark Problem (HSBP)}

\subsubsection{Rationale} The hidden shift problem is a natural source of challenges where a quantum computer may demonstrate advantage over classical computational resources. In particular, the hidden shift problem for Boolean functions introduced in~\cite{Roetteler2010hiddenshift} defines a blueprint for benchmark challenges whose difficulty can be tweaked at will, by increasing the number of CNOT and Toffoli gates used in permutation operations. The algorithm introduced in~\cite{Roetteler2010hiddenshift} shows an oracle separation between $P$ and $BQP$ and has since been widely used for benchmarking classical simulation algorithms.

The benchmark challenge we introduce below is based on Theorem $6$ in~\cite{Roetteler2010hiddenshift}, and it serves as a sort of universal benchmark for quantum computers in the sense that it tests the computer's ability to execute circuits with varying numbers of entangling gates in different configurations, albeit it is not directly connected to any practical application as of yet. While Theorem 6 in~\cite{Roetteler2010hiddenshift} applies to the class of bent functions, we focus on the subclass of Majorana-McFarland bent functions of the form 
\begin{equation*}
 h_\pi(x, y) = x \cdot \pi(y)^t = x_1 y_{\pi_1} \oplus \cdots \oplus x_n y_{\pi_n},
\end{equation*}
with $\pi\colon \mathbb{Z}^m \to \mathbb{Z}^m$ denoting an arbitrary permutation on $m$ points. The dual of $h_\pi$ is conveniently given by $\widetilde{h_\pi}(x, y) = \pi^{-1}(x) \cdot y^t$. In what follows below, we let $n = 2m$ and consider the function $f_\pi\colon \mathbb{Z}_2^n \to \mathbb{Z}_2$ given by $f(x) = h_\pi(x_o, x_e)$, with $x_o = (x_1, x_3, \ldots, x_{2m-1})$ denoting the odd-indexed components, and $x_e = (x_2, x_4, \ldots, x_{2m})$ denoting the even-indexed components of $x$. Explicitly,
$$
    f(x) = x_1 x_{\pi_2} \oplus \cdots \oplus x_{2m - 1} x_{\pi_{2m}}.
$$

\subsubsection{Algorithm Description} Each HSBP challenge consists of a number of quantum circuits. For each circuit width, the challenge circuits share a common structure, and they correspond to different choices of hidden shift and permutation. We use multiple circuits in order to improve the precision of performance metrics by reducing idiosyncratic noise in the benchmark scores associated with the particular choice of hidden shifts or permutation gates.

Each challenge circuit is constructed as follows. First, we rotate the qubits into a uniform superposition over all computational basis states. Next we compute a shifted version of $f_\pi$ into the amplitudes of the qubit register, apply a Hadamard transform, and then compute the dual function $\widetilde{f_\pi}$ into the amplitudes of the resulting state. Finally, we apply another Hadamard transform to recover the hidden shift. 

The output state is just that: a pure state supported on exactly one bit-string, which describes the hidden shift used to calculate the shifted version of $f_\pi$. The benchmark score is computed as the probability of sampling the desired hidden shift bit-string in the output state.

The permutation $\pi$ is a key component of our benchmarking algorithm: it provides a mechanism for tuning the computational complexity, and thereby the difficulty, of each benchmark challenge. In practice, we implement every given permutation using a collection of CNOT and Toffoli gates; each problem instance is in part defined by the choice of $\pi$, or equivalently, the choice of CNOT and Toffoli gates used to implement it in the challenge circuit. 

\begin{figure}[H]
    \centering
    \includegraphics[width=0.5\linewidth]{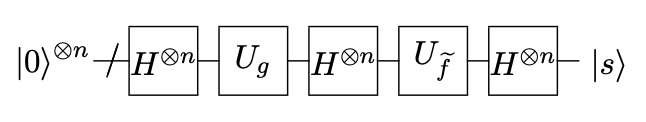}
    \caption{Structure of a hidden shift circuit. The objective is to recover the hidden shift $s$ through a sequence of operations involving Hadamard transforms and the computation of bent functions into the amplitudes of a qubit register.}
    \label{fig:hidden-shift-qc}
\end{figure}

\Cref{fig:hidden-shift-qc} illustrates the anatomy of a typical challenge circuit. Given an even number of qubits $n = 2m$, a hidden shift $s \in \{0, 1\}^m$, and a permutation $\pi$ on $m$ points, the circuit first applies a Hadamard transform to prepare a uniform superposition over all computational basis states, and then it computes the shifted bent function $g(x) = f_\pi(x \oplus s)$ into the amplitudes of the qubit register, resulting in the transformation
$$
    \ket{0}^{\otimes n} \to \frac{1}{\sqrt{2^n}} \sum_x \ket{x}
    \to \frac{1}{\sqrt{2^n}} \sum_x (-1)^{f_\pi(x \oplus s)} \ket{x}.
$$
The $U_g$ block represents the computation of $g$. The circuit then applies a Hadamard transformation to obtain
\begin{align*}
    \frac{1}{\sqrt{2^n}} \sum_x (-1)^{f_\pi(x \oplus s)} \ket{x}
    &\to \frac{1}{2^n} \sum_{x, y} (-1)^{x y^t + f_\pi(x + \oplus s)} \ket{y} \\
    &= \sum_y (-1)^{y s^t} \bigg(\frac{1}{2^n} \sum_z (-1)^{y z^t + f(z)} \bigg) \ket{y} \\
    &= \sum_y (-1)^{y s^t} \widehat{f_\pi(y)} \ket{y} \\
    &= \frac{1}{\sqrt{2^n}}\sum_y (-1)^{y s^t} (-1)^{\widetilde{f_\pi}(y)} \ket{y},
\end{align*}
where $\widehat{f_\pi(y)} = \frac{1}{2^n} \sum_z (-1)^{y z^t + f(z)}$ denotes a Fourier coefficient. Note that the last equality follows by definition of the dual bent function~\cite{Roetteler2010hiddenshift}. The circuit then computes the dual function $\widetilde{f_\pi}$ onto the amplitudes of the qubit register and finally it applies a third Hadamard transformation to recover the hidden shift $s$:
\begin{align*}
    \frac{1}{\sqrt{2^n}}\sum_y (-1)^{y s^t} (-1)^{\widetilde{f_\pi}(y)} \ket{y}
    &\to \frac{1}{\sqrt{2^n}}\sum_y (-1)^{y s^t} \ket{y} \\
    &\to \frac{1}{2^n} \sum_{x, y} (-1)^{(x + s) y^t} \ket{x} \\
    &= \sum_x \widehat{\delta}(x + s) \ket{x} \\ 
    &= \ket{s}.
\end{align*}
In the first equality $\delta$ denotes the discrete delta function whose output is trivial unless the input is null.

In practice, each challenge circuit computes $f_\pi$ and related functions onto the amplitude of its qubit register using a collection of CNOT and Toffoli gates to implement the permutation $\pi$, and a stack of CZ gates to compute the required signs: note that the CZ gate transforms $\ket{x_1 x_2}$ into $(-1)^{x_1 x_2} \ket{x_1 x_2}$.

\subsubsection{Problem Instances} Each problem instance consists of a number of challenge circuits, as described above, and each circuit is defined by an even integer $n = 2m$ together with a hidden shift $s \in \{0, 1\}^m$ and a permutation $\pi$ on $m$ points. For every challenge, we generate hidden shifts using a sampling procedure where we generate each bit independently, drawing Bernouilli variables with probability of success set to $0.75$.

While arbitrary permutations can be used to define HSBP challenges, in this paper, we focus on four families of permutations for ease of characterization. The first family consists of those permutations whose action is implemented by a single multi-controlled X gate. The associated challenges are labeled by MCX and their permutation blocks consist of a single multi-controlled X gate targeting the bottom wire and controlled by the rest of the even-indexed qubits, which is then transpiled into a sequence of CNOT gates and single-qubit rotations. 

The second family consists of those permutations whose action is implemented by a ``ladder'' of CNOT gates; for these challenges, 
\begin{equation*}
    \pi = \prod_{j = 1}^{m-1} CNOT(x_{2j}, x_{2j+2}).
\end{equation*}

The third family is similar, but instead of a CNOT ladder, we use a ladder of Toffoli, or CCX gates. Concretely, in these challenges,
\begin{equation*}
    \pi = \prod_{j = 1}^{m-2} CCX(x_{2j}, x_{2j+2}, x_{2j + 4}).
\end{equation*}

Each problem instance in each of these challenge families consists of $10$ quantum circuits of the same width, each circuit corresponding to a different hidden shift. 

The fourth challenge family leverages randomly generated permutations. Each of these permutations consists of a fixed number of CNOT gates. Each such permutation was generated by sampling pairs of integers representing the target and control qubits from the possible universe uniformly at random. Each problem instance in this challenge family consists of $9 = 3 \times 3$ quantum circuits: we use three hidden shifts and three random permutations, and each circuit corresponds to a possible hidden shift and permutation pair. This structure allows for averaging benchmark scores across a minimal number of hidden shift and random permutation choices. For the purposes of this paper we tested random permutation challenges with $10, 20, \ldots, 100$ CNOT gates.

\subsubsection{Results} 
\begin{figure}[H]
    \centering
    \includegraphics[width=0.7\linewidth]{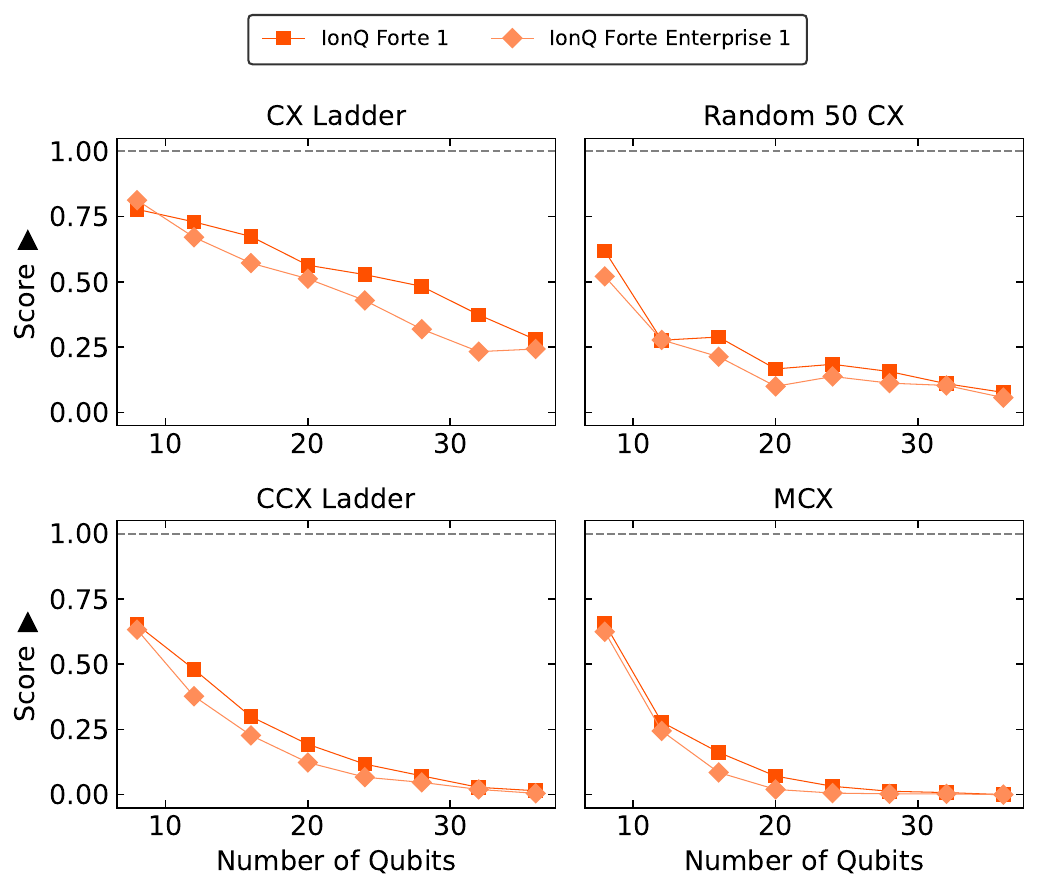}
    \caption{Benchmark scores computed on IonQ Forte class systems for the HSBP challenge families considered: CX ladder, random permutation with 50 CNOT gates, CCX ladder, and MCX permutations.}
    \label{fig:hsbp-results}
\end{figure}

\Cref{fig:hsbp-results} plots the benchmark scores obtained on IonQ Forte for each of the challenge families considered. The MCX challenge is the most difficult, as expected, since the corresponding circuits require the highest number of entangling gates.

\begin{figure}[H]
    \centering
    \includegraphics[width=0.85\linewidth]{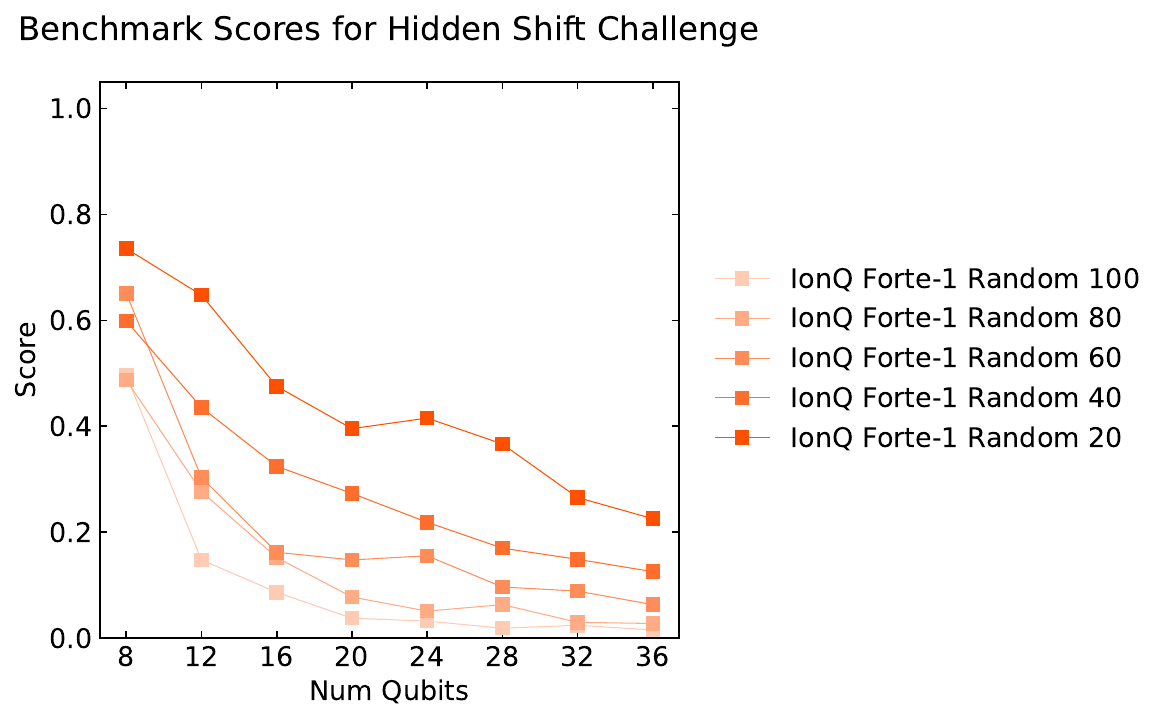}
    \caption{Benchmark scores obtained on IonQ Forte when executing HSBP random permutation challenges with $20, 40, 60, 80$, and $100$ CNOT gates. This plot illustrates the increasing difficulty of benchmark challenges as a function of the number of entangling in the permutations.}
\end{figure}

\section{Results: Open Benchmarks}
\label{sec:open}

\subsection{Variational Quantum Imaginary-Time Evolution (varQITE)}

\subsubsection{Rationale}
The variational quantum imaginary-time evolution approach is an interesting alternative to QAOA for solving combinatorial optimization problems~\cite{morris2024varqite}. Like QAOA, varQITE is a general-purpose variational algorithm that seeks to minimize the energy of a Hamiltonian by tuning circuit parameters in order to obtain the optimal solution to a difficult combinatorial problem. 

Two main advantages of varQITE over QAOA include
\begin{itemize}
    \item it can leverage shallow problem- and hardware-specific ans\"atze; and
    \item it obviates the need for a classical optimizer to tune circuit parameters by providing an update rule based on time-marching a linear differential equation.
\end{itemize}

Aside from being sensitive to hardware characteristics like qubit connectivity, gate fidelities, and coherence times, the performance for varQITE on a given problem heavily depends on the choice of ansatz. For this benchmark, we again focus on the MaxCut problem, using the same quantum Hamiltonian formulation that we leverage in our QAOA and related benchmarks. In fact, we use the same problem instances as well, for the sake of comparing the performance of different algorithms on the same problems. 

We leverage the ansatz layout that was introduced in~\cite{aboumrad2025ansys} in the context of the graph partitioning problem (GPP).

\subsubsection{Algorithm Description}

Much like QAOA, varQITE is a variational approach that finds approximate solutions to optimization problems by encoding them into a Hamiltonian $H$ and minimizing the cost function $C(\theta)$ defined by the expectation value
\begin{align*}
    C(\theta) = \bra{\psi(\theta)}{H_c}\ket{\psi(\theta)},
\end{align*}
with $\ket{\theta}$ denoting the chosen problem-specific ansatz. 

The varQITE algorithm defines an explicit parameter update rule that does not rely on classical optimization routines; instead, the update rule enforces quantum imaginary-time evolution under the action of the problem Hamiltonian $H$ for a chosen set of observables. For details, see~\cite{morris2024varqite}.

For our benchmark challenges, we use the so-called Heavy Neighbors Ansatz introduced in~\cite{aboumrad2025ansys}. This ansatz is well-suited for the MaxCut problem since it selectively reinforces entanglement between qubits by leveraging structural properties of the underlying problem graph. Concretely, the ansatz features parametrized entangling gates connecting qubits that represent nodes connected by edges with high weight, which allows for effective cut size tuning by tweaking gate parameters.

In any case, each benchmark challenge consists of a single quantum circuit. The benchmark challenge measures the performance of quantum hardware upon executing an optimized ansatz to retrieve high-quality approximations to the MaxCut problem.

We prepare each challenge circuit by running the variational portion of the varQITE algorithm on an ideal simulator until convergence. This ``pre-training'' yields the parameter values that are fixed for each challenge circuit. 

Upon executing each challenge circuit, we measure the resulting AR as defined by \Cref{eq:apx-ratio} and report the value as the challenge score.

\subsubsection{Problem Instances} For the sake of comparing the performance of different solution algorithms on the same problems, we consider the universe of MaxCut instances described in \Cref{sec:qaoa}. In particular, we focus on the subset of $3$-regular graphs and their corresponding MaxCut instances. 

To execute varQITE, we use the two-layer Heavy Neighbors Ansatz defined by the problem graph with $3n$ entangling gates per layer, with $n$ denoting the number of nodes in the problem graph. The ansatz parameters are obtained by executing varQITE in an ideal simulator until the absolute change in Hamiltonian energy falls below $10^{-4}$.

\subsubsection{Results} 
\begin{figure}[H]
    \centering
    \includegraphics[width=0.6\linewidth]{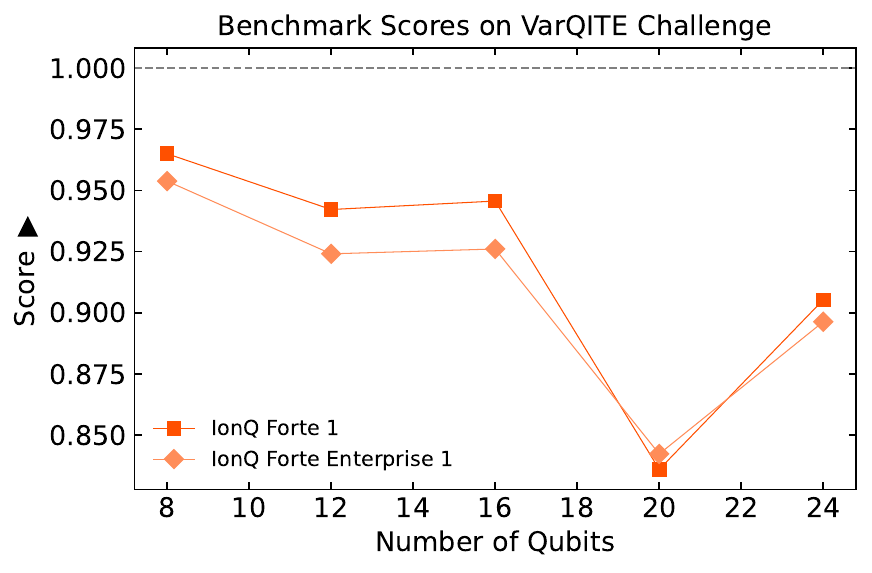}
    \caption{Benchmark score for varQITE challenges testing the solution quality of MaxCut problems on $3$-regular graphs of varying sizes.}
    \label{fig:varqite-results}
\end{figure}

\Cref{fig:varqite-results} plots benchmark scores obtained on IonQ Forte. Each challenge circuit was executed $1{,}000$ times. The plot illustrates a decreasing trend in performance as the circuit width, and hence the circuit depth, increases. 

\subsection{Quantum-Classical Auxiliary Field Quantum Monte Carlo (QC-AFQMC)}

\subsubsection{Rationale}
Auxiliary-field quantum Monte Carlo (AFQMC) is a projector-based quantum Monte Carlo approach particularly effective in accurately modeling ground-state electronic structures, especially when electron correlation effects are significant \cite{Lee2022-ep, Motta2018-ft, Zhang2020-wu, Shee2023-gs}. Its strength stems from systematically improvable accuracy achievable through enhanced trial wavefunctions, controlled imaginary-time propagation, and careful walker population management. Quantum-classical AFQMC (QC-AFQMC) is an extension of classical AFQMC that utilizes efficient quantum circuits for the trial state that would otherwise be inefficient to simulate classically \cite{Huggins2022-lh, Wan2023-df, Kiser2024-be, Huang2024-ya, Zhao2025-mw}. Because the quantum trial states can be made systematically more expressive at polynomial circuit depth, they can capture multi-reference and strong-correlation structure that is prohibitively expensive for classical ans{\"a}tze, while still entering AFQMC only through quantities that are efficient to estimate on quantum hardware (notably overlaps and local energies). In this way, the quantum device supplies a compact, high-fidelity trial state that improves the phaseless constraint and reduces bias, and the classical Monte Carlo projector supplies the long-imaginary-time refinement needed to recover the ground state with controllable cost and error.

To align with this benchmarking framework's emphasis on practical performance metrics, QC-AFQMC is evaluated explicitly by measuring the end-to-end wall time (TTS) required to attain a predefined energy accuracy ($\Delta E$) relative to a classical benchmark, along with the corresponding energy utilized (EPS). Rather than relying on proxy circuit metrics, this approach directly assesses the ability of the quantum-classical workflow to achieve stated accuracy goals and quantifies the computational resources (time and energy) needed to do so.

The hybrid QC-AFQMC structure clearly separates tasks between quantum and classical processors. The quantum device efficiently prepares highly correlated trial wavefunctions ($\ket{\Psi_T}$) and generates measurement data through matchgate shadows, which are a specialized form of randomized measurement suitable for efficient classical post-processing \cite{Wan2023-df,Huang2024-ya,Zhao2025-mw}. Classical computational infrastructure then performs the AFQMC projection steps, implements importance sampling, and accumulates statistical estimators. This task division ensures practical applicability by explicitly exposing the entire quantum-classical algorithm stack to rigorous TTS and EPS benchmarking.

\subsubsection{Algorithm Description}
QC-AFQMC is fundamentally built upon the phaseless AFQMC methodology, employing imaginary-time propagation to project onto the ground state. The Hubbard–Stratonovich transformation decomposes the propagation into stochastic one-body operations, sampled using an ensemble of walkers represented as Slater determinants. Importance sampling strategies involving force biases are used to reduce statistical variance. To manage phase instabilities inherent in quantum Monte Carlo methods, a phaseless constraint is introduced that projects walker phases toward the trial wavefunction $\ket{\Psi_T}$, creating a controllable, trial-dependent bias that diminishes as trial wavefunction quality improves. Energies are computed primarily using mixed estimators, with optional back-propagation methods available to enhance accuracy. Operational parameters include imaginary-time step size ($\Delta \tau$), walker populations, reconfiguration frequency, and orbital reorthogonalization intervals.

The quantum enhancement in QC-AFQMC is realized through matchgate shadows,\cite{Wan2023-df} which facilitate efficient quantum-classical integration. The quantum processor generates high-quality trial wavefunctions, typically the pair-coupled cluster doubles variant of the unitary coupled-cluster (UpCCD) ans{\"a}tze. Random matchgate basis rotation operations are appended to the trial circuit in order to operationalize the entire circuit for (partial) shadow tomography and observable estimation. Matchgate shadows yield randomized quantum measurements from which classical processing reconstructs essential quantities, such as overlaps $\braket{\Psi_T|\phi}$, force biases, and local energies. Classical reconstruction exploits Pfaffian algebra, ensuring computational feasibility even as system sizes scale. This procedure scales polynomially, at approximately $\mathcal{O}(n^5)$ per shadow, driven mainly by Pfaffian and tensor contractions. Empirically, maintaining variance at fixed accuracy requires the number of shadows $N_{\mathrm{sh}}$ to scale modestly as $n^{0.5}\log n$, resulting in a total classical processing complexity of approximately $\mathcal{O}(n^{5.5}\log n)$ \cite{Zhao2025-mw}.

\subsubsection{Problem Instances}
We benchmark QC-AFQMC on linear hydrogen chains in the STO-3G basis at \SI{2.0}{\angstrom}. The instances reported here are \ce{H4} and \ce{H6}, with exact FCI references \(-1.898\)~Ha and \(-2.847\)~Ha, respectively. The \ce{H4} instance is treated in a CAS(4,4) active space mapped to 8 qubits, and \ce{H6} in CAS(6,6) mapped to 12 qubits. Trial states are prepared with the UpCCD circuit family and accessed through matchgate shadows for overlap, force-bias, and local-energy estimation. 

Unless otherwise noted, all QC-AFQMC runs use an imaginary-time step \(\Delta\tau=0.01\) with 256 walkers and a projection length of 80 blocks \(\times\) 20 steps (\(\tau_{\max}=16\)~a.u.). The first 10 blocks are discarded as equilibration. Statistical uncertainties are computed by reblocking over the final 70 blocks and reported as 95\% confidence intervals.

Two-body operators are represented by a Cholesky decomposition with tolerance \(10^{-7}\); this yields 7 auxiliary fields for \ce{H4} and 11 for \ce{H6}. Per platform, we acquire \(N_{\mathrm{sh}}=21{,}080\) matchgate shadow samples for \ce{H4} and \(N_{\mathrm{sh}}=28{,}289\) samples for \ce{H6}.

\subsubsection{Results}

\Cref{tab:qcafqmc_energies_2A} summarizes QC-AFQMC total energies for linear \ce{H4} and \ce{H6} chains in STO-3G at \SI{2.0}{\angstrom}. The corresponding FCI references are \(-1.898\)~Ha (\ce{H4}) and \(-2.847\)~Ha (\ce{H6}). The \ce{H4} calculations employ a CAS(4,4) active space mapped to 8 qubits with 21{,}080 matchgate shadow samples, while \ce{H6} uses CAS(6,6) on 12 qubits with 28{,}289 samples. 

On the ideal simulator, QC-AFQMC yields \(-1.868\)~Ha for \ce{H4} (\(18.79\)~kcal/mol above FCI) and \(-2.768\)~Ha for \ce{H6} (\(49.98\)~kcal/mol above FCI). These residuals are expected: the phaseless constraint introduces a trial-state-dependent bias that vanishes only as \(\ket{\Psi_T}\to\ket{\Psi_0}\), and the UpCCD trial ansatz is restricted to paired excitations. The larger residual for \ce{H6} is consistent with stronger multireference character in the CAS(6,6) manifold. These simulator results therefore provide noise-free baselines for assessing QPU-prepared trial states.

\begin{figure}[H]
  \centering
  \includegraphics[width=1.0\linewidth]{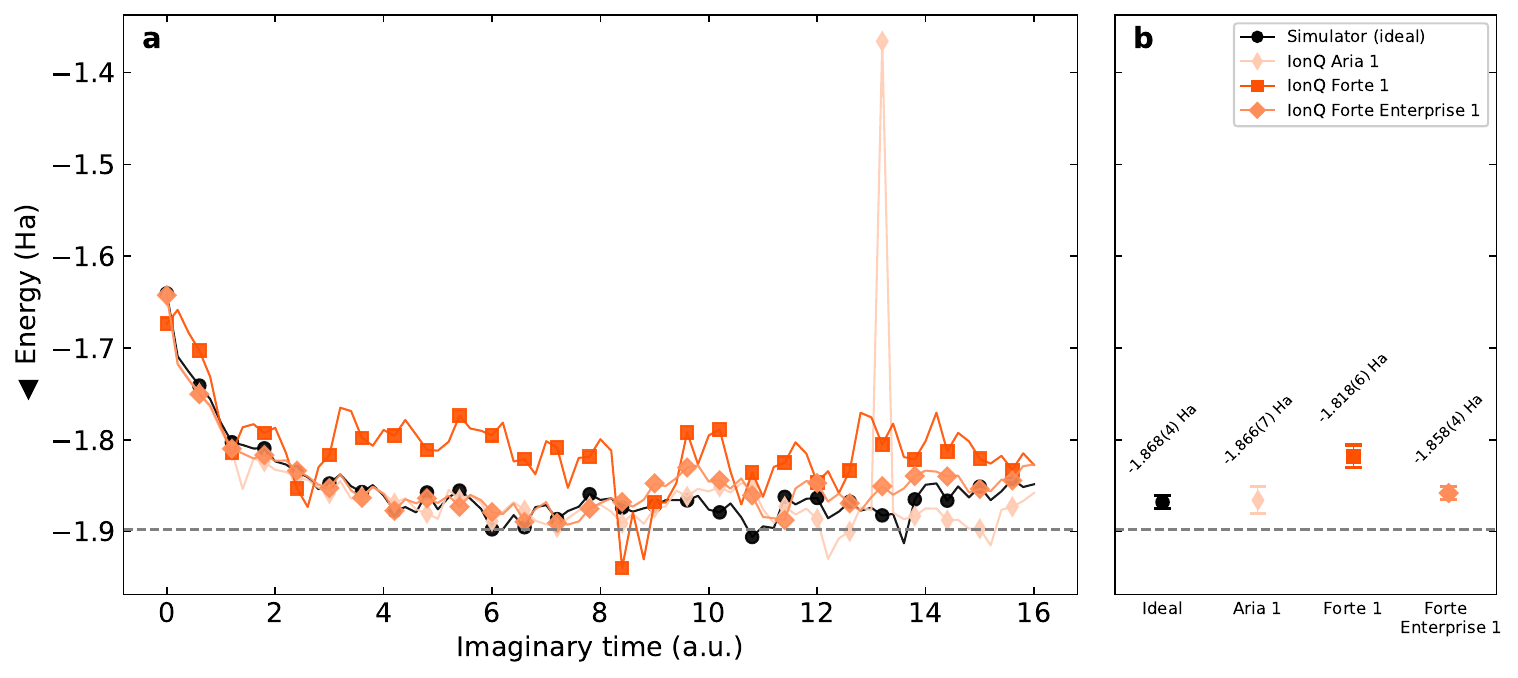}
    \caption{QC-AFQMC results for linear \ce{H4} in STO-3G at \SI{2.0}{\angstrom} using CAS(4,4) (8 qubits). (a)~Imaginary-time evolution of the total energy; the dashed line marks the FCI reference (\(-1.898\)~Ha). (b)~Final reblocked energies for the ideal simulator and the \texttt{Aria-1}, \texttt{Forte-1}, and \texttt{Forte-Enterprise-1} QPUs; error bars denote 95\% confidence intervals (numbers in parentheses indicate the corresponding uncertainty in the last digits).} 
  \label{fig:h4_qcafqmc}
\end{figure}

For \ce{H4}, Aria-1 agrees with the simulator within uncertainty, giving \(-1.866\)~Ha (95\% CI \(14.5\)~mHa). Forte-Enterprise-1 reaches \(-1.858\)~Ha (\(25.05\)~kcal/mol above FCI), while Forte-1 yields \(-1.818\)~Ha (\(50.09\)~kcal/mol above FCI). The imaginary-time traces in \Cref{fig:h4_qcafqmc}(a) show larger projection-time fluctuations for Forte-1. Aria-1 exhibits a transient spike near \(\tau=13\)--\(14\)~a.u., attributable to numerical instability in the walker renormalization matrix \(R\) during finite-difference force-bias evaluation \cite{Zhao2025-mw}.

\begin{figure}[H]
  \centering
  \includegraphics[width=1.0\linewidth]{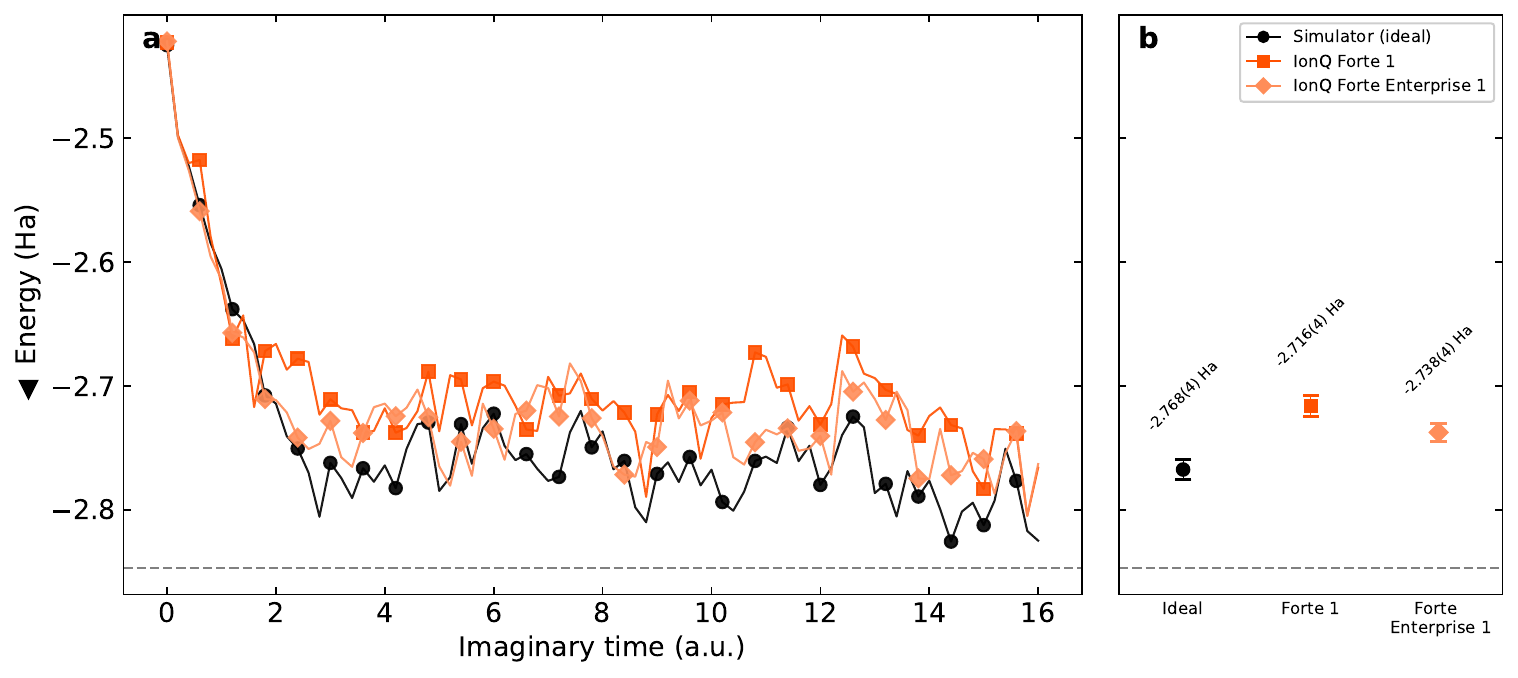}
    \caption{QC-AFQMC results for linear \ce{H6} in STO-3G at \SI{2.0}{\angstrom} using CAS(6,6) (12 qubits). (a)~Imaginary-time evolution of the total energy; the dashed line marks the FCI reference (\(-2.847\)~Ha). (b)~Final reblocked energies for the ideal simulator and the \texttt{Forte-1} and \texttt{Forte-Enterprise-1} QPUs; error bars denote 95\% confidence intervals (numbers in parentheses indicate the corresponding uncertainty in the last digits).}
  \label{fig:h6_qcafqmc}
\end{figure}

For \ce{H6}, Forte-Enterprise-1 reaches \(-2.738\)~Ha (\(68.80\)~kcal/mol above FCI; 95\% CI \(7.4\)~mHa) and Forte-1 yields \(-2.716\)~Ha (\(82.07\)~kcal/mol above FCI; 95\% CI \(8.6\)~mHa). Relative to the simulator baseline, the QPU-derived trial states lie \(18.82\)~kcal/mol (Forte-Enterprise-1) and \(32.09\)~kcal/mol (Forte-1) higher, a larger degradation than for \ce{H4}, consistent with the deeper circuits required at 12 qubits. Backend ordering is preserved across both systems, with Forte-Enterprise-1 outperforming Forte-1. The \ce{H6} convergence traces in \Cref{fig:h6_qcafqmc}(a) show an equilibration transient over the first \(\sim 2\)~a.u.\ before reaching steady-state fluctuations.

\Cref{fig:h4_qcafqmc,fig:h6_qcafqmc}(b) report the final reblocked estimates and 95\% confidence intervals for each backend.

\begin{table}[H]
  \centering
  \begin{threeparttable}
  \caption{QC-AFQMC total energies for linear \ce{H4} and \ce{H6} chains in STO-3G at \SI{2.0}{\angstrom} for simulated and QPU-generated trial states.}
  \label{tab:qcafqmc_energies_2A}

  \begin{tabular}{
    l
    l
    S[table-format=-1.6]
    S[table-format=2.1]
    S[table-format=2.2]
  }
    \toprule
    \multicolumn{1}{c}{System} &
    \multicolumn{1}{c}{Backend} &
    \multicolumn{1}{c}{\(E\) (Ha)} &
    \multicolumn{1}{c}{95\% CI (mHa)} &
    \multicolumn{1}{c}{\(\Delta E_{\mathrm{FCI}}\) (kcal/mol)} \\
    \midrule

    \multirow{4}{*}{\ce{H4}} &
      Simulator (ideal)              & -1.868049 &  7.2 & 18.79 \\
    & Aria-1 QPU                     & -1.865905 & 14.5 & 20.14 \\
    & Forte-1 QPU                    & -1.818176 & 12.4 & 50.09 \\
    & Forte-Enterprise-1 QPU         & -1.858087 &  7.4 & 25.05 \\

    \addlinespace[3pt]
    \cmidrule(lr){1-5}
    \addlinespace[3pt]

    \multirow{3}{*}{\ce{H6}} &
      Simulator (ideal)              & -2.767543 &  8.4 & 49.98 \\
    & Forte-1 QPU                    & -2.716399 &  8.6 & 82.07 \\
    & Forte-Enterprise-1 QPU         & -2.737551 &  7.4 & 68.80 \\

    \bottomrule
  \end{tabular}

  \begin{tablenotes}[flushleft]\footnotesize
    \item Uncertainties are 95\% confidence intervals from reblocking over the final 70 blocks (first 10 discarded as equilibration).
    \item \(\Delta E_{\mathrm{FCI}}\) is relative to exact FCI references: \(-1.898\) Ha (\ce{H4}) and \(-2.847\) Ha (\ce{H6}).
  \end{tablenotes}
  \end{threeparttable}
\end{table}

\subsection{Quantum Lattice Boltzmann Method for Differential Equations}
\subsubsection{Rationale}
This benchmark is based on a Quantum Lattice Boltzmann (QLBM) algorithm for solving the advection diffusion equation in 2D~\cite{ionq-ansys-qlbm-2025}. The QLBM is an iterative algorithm where a macroscopic density function is described as a sum of individual microscopic particle densities. The algorithm pipeline for each time step includes: initial state encoding, collision and streaming operations, recovering the macroscopic density from the individual densities, and implementing the appropriate boundary conditions. The final state obtained after each such evolution at time $t$, serves as the initial state for the next iteration. To benchmark the quantum hardware we use quantum state fidelity during the evolution as the metric.

\begin{figure*}[t]
    \centering
    \includegraphics[width=0.85\linewidth]{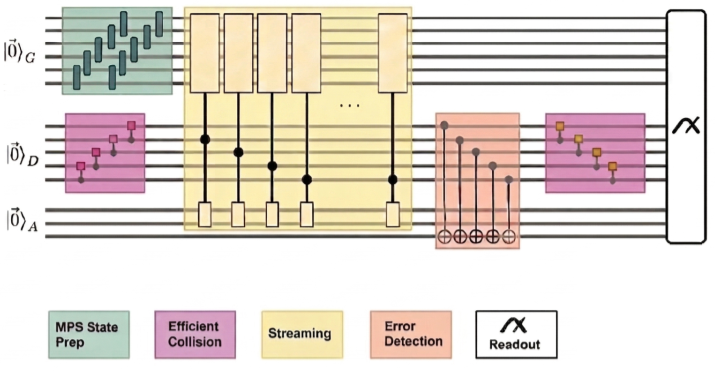}
    \caption{Overview of the QLBM pipeline, showing the D2Q5 lattice Boltzmann formulation to simulate macroscopic fluid dynamics. The quantum algorithm takes as input a smooth density distribution (a 2D Gaussian in this case) and applies the QLBM quantum circuit for T iterations. The input distribution is loaded using a MPS state preparation circuit. The collision operator is prepared using a unary state preparation circuit on the direction qubits. The streaming step involves a series of operations on the grid qubits which are controlled on the direction qubits. Ancilla qubits are used to reduce the total gate count of the multi-control unitary gates. We perform error detection via additional ancilla qubits during the circuit execution.}
    \label{fig: qlbm_circuit}
\end{figure*}

\subsubsection{Algorithm Description}
We follow the algorithm developed and implemented in~\cite{ionq-ansys-qlbm-2025}, represented in Fig.~\ref{fig: qlbm_circuit}, which is based on the quantum lattice Boltzmann method (qLBM) for simulating a density function $\Phi(\textbf{x},t)$ evolving under a linear advection-diffusion equation~\cite{budinski2021, budinski2022, wawrzyniak2025a}. The equation is given by,
\begin{equation}\label{eq:advecdiff}
    \partial_t \Phi = D\nabla^2\Phi - \nabla \left(\textbf{u}\Phi\right),
\end{equation}
where, $\textbf{u}(\textbf{x})$ is the velocity field and $D$ is a constant diffusion coefficient.

In this approach, the macroscopic density $\Phi\left(\textbf{x}, t\right)$ evolution is modeled as that of individual species $f_i\left(\textbf{x}, t\right)$ of microscopic particle densities, $\Phi(x,t) = \sum_i f_i\left(\textbf{x}, t\right)$. At time $t$, this approximation is given by ~\cite{mohamad2019}, 
\begin{eqnarray}
\label{eq:lbm}
    f_i(\textbf{x}+c_i \Delta t, t+ \Delta t) = (1-\omega)f_i(\textbf{x},t) + \omega f_i^{eq}(\textbf{x},t),
\end{eqnarray}, 
where $\Delta t$ is the time step, $f_i^{eq}$ is the equilibrium density distribution and $\omega = \Delta t / \tau$ with relaxation time $\tau$. $f_i^{eq}$ is given by the Bhatnagar-Gross-Krook (BGK) formulation \cite{BGK} as,
\begin{equation}
\label{eq:eq_func}
    f_i^{eq} = \omega_i \Phi\left[1+\frac{c_i\cdot u_i}{c_s^2}\right].
\end{equation}

We set the relaxation constant $\omega=1$, which simplifies \Cref{eq:lbm,eq:eq_func} to,

\begin{eqnarray}
    \Phi(x,t) &=&\sum_i k_i\Phi\left[\textbf{x}-c_i\Delta t, t-\Delta t\right], \\
    \text{where } k_i&=&\omega_i\left[1+\frac{c_i\cdot u_i}{c_s^2}\right].\label{eq:k_define}
\end{eqnarray}

Following are the modules in the final algorithm:

\begin{enumerate}
    \item \textbf{\textit{Encoding:}} $\Phi(\textbf{x},t)$ is encoded as a quantum state $|\Phi_t\rangle_G$,
    \begin{equation}
        |\Phi_t\rangle_G := \frac{1}{\|\Phi_t\|_2}\sum_\textbf{x}\Phi(\textbf{x},t)|x\rangle_G,
    \end{equation}
    where $\|\Phi_t\|_2:=\sqrt{\sum_\textbf{x}\Phi(\textbf{x},t)^2}$
    where each gridpoint $\textbf{x}$ is represented by a computational basis state $|x\rangle$. This state is prepared on $n_G$ `grid qubits' to represent $N=L^d$ lattice sites in $d$ dimensions.
    \item \textbf{\textit{Collision: }} The velocity is considered to be uniform and the collision operator is applied via state preparation of a quantum state $|k\rangle$ with the help of an auxiliary qubit, where
    \begin{equation}
        |k\rangle=\frac{1}{\sqrt{2^{n_D}}}\sum_i k_i |i\rangle_D|0\rangle_{a}+|\chi_0\rangle_D|1\rangle_a,
    \end{equation}
    for some orthogonal state $|\chi_0\rangle$.
    The state $|k\rangle$ is prepared on $n_D$ `direction qubits' for $M$ directions.  
    
    \item \textbf{\textit{Streaming: }} The streaming operator $U_S$ is created via controlled unitaries,
    \begin{equation}\label{eq:stream_defined}
        U_S=\prod_i \left(\mathbf{1}_G\otimes (\mathbf{1}-|i\rangle\langle i|)_D+S_i\otimes(|i\rangle\langle i|)_D\right).
    \end{equation}
    where the controls are on direction qubits and 
    unitaries $S_i$ are given by,
    $S_i|x\rangle_G=|x+c_i\Delta t\rangle_G$. 

    Applying $U_S$ on the prepared input state $|\Phi_t\rangle_G|k\rangle_D$ yields,
    \begin{align}
        \frac{1}{\|\Phi_t\|_2\sqrt{2^{n_D}}}&\sum_{x,i} k_i|i\rangle_D\Phi(x-c_i\Delta t, t)|x\rangle_G|0\rangle_a\nonumber\\
        &+|\chi_1\rangle_{D,G}|1\rangle_a,
    \end{align}
    where $|\chi_1\rangle_{D,G}$ is some orthogonal state.

    \item \textbf{\textit{Macroscopic quantities: }} The final step involves applying Hadamard gates on all the direction qubits which essentially adds all the amplitudes over the direction qubits. The final state is given by:
    \begin{align}
        &\frac{1}{2^{n_D}}\sum_x\left(\sum_i k_i\Phi(x-c_i\Delta t, t) \right)|x\rangle_G|0\rangle_{D,a}\nonumber+ |\chi_2\rangle\\
        =&\frac{1}{2^{n_D}}\sum_x \Phi(x,t+\Delta t)|x\rangle_G|0\rangle_{D,a}\nonumber+|\chi_2\rangle\\
        \label{eq:final_state_old}
        =&\frac{\|\Phi_{t+1}\|_2}{2^{n_D}\|\Phi_t\|_2}|\Phi_{t+1}\rangle_G|0\rangle_{D,a}+|\chi_2\rangle
    \end{align}
    $|\chi_2\rangle$ is an orthogonal state such that, $(\mathbbm{1}_G\otimes\langle0|_{D,a})|\chi_2\rangle=0$. The final evolved state $|\Phi_{t+1}\rangle$ is obtained by post-selecting on the $|0\rangle_D$ state.
    
\end{enumerate}

\subsubsection{Problem Instances}
\begin{figure*}[t]
    \centering
     \includegraphics[width=0.55\linewidth]{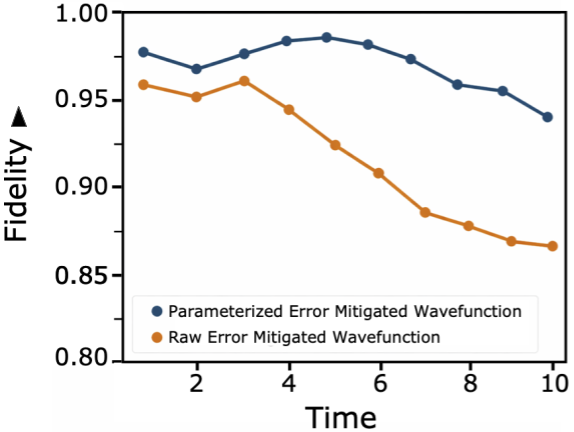}
    \caption{The overlap, $|\langle \psi | \phi\rangle|$, between the QPU density and the exact classical LBM solution (orange curve). Further improvement in fidelity is achieved by measuring observables (instead of doing full state tomography from raw shots) and reconstructing the 2D Gaussian distribution, where fidelity remains $\sim 94\%$ even after $10$ time steps (blue curve). }
    \label{fig:qlbm_qpu_results}
\end{figure*}

We implemented the QLBM algorithm on hardware to solve the advection-diffusion equation for a 2D initial Gaussian density distribution on a $ 16 \times 16$ grid. The quantum circuit used $19 $ qubits and $\sim 260$ two-qubit gates. The results of this execution are shown in \Cref{fig:qlbm_qpu_results}, and demonstrate agreement with the classical LBM simulation as shown by the fidelity values. We used $10$k shots and custom error mitigation strategies to achieve this performance.

\subsubsection{Results}
The benchmark metric in this algorithm is fidelity, which is measured with respect to the exact classical solution at every time step. Note that the fidelity values differ from the target value of $1$ not only because of the hardware noise but also due to shot noise and approximate state loading techniques like tensor network methods. The orange curve in \Cref{fig:qlbm_qpu_results} demonstrates fidelity when the quantum state is constructed using shots from the quantum circuit readout. The fidelity can be further improved by reading out observables like mean, variance and covariance from the shots and reconstructing the state from these observables, as demonstrated by the blue curve. This algorithm achieves fidelity upto $0.95$ at the end of $10$ time steps under the advection-diffusion equation on IonQ Forte.

\subsection{High Energy Physics}

\subsubsection{Rationale}
Future quantum computers could enable \textit{ab initio} investigations in nuclear and high-energy physics, revolutionizing these fields, particularly in simulating nuclear reactions and dynamics on yocto-second timescales, similar to femtosecond imaging in chemistry.
Meanwhile, current NISQ computers, despite limitations, are valuable for simulating out-of-equilibrium dynamics and quantum many-body phenomena.

The work presented in \cite{chernyshev2026pathfinding} establishes a benchmark for exploring a potential exotic nuclear decay relevant to new physics searches. The stability of matter provides strong constraints on physics beyond the Standard Model, with observations of proton decay, neutron-antineutron oscillations, and nuclear beta decay setting limits on fundamental symmetry breaking. Specifically, neutrinoless double beta decay ($0u\beta\beta$-decay) of certain nuclei indicates lepton number violation. Calculating these decay rates is challenging due to the Majorana-neutrino induced process, which involves two charged-current weak interactions connected by a near-massless neutrino within the nucleus. Simulating this doubly-weak decay processes is a significant challenge for classical computing due to the necessity of coherently tracking the dynamics in a strongly-interacting nucleus. While progress has been made with Euclidean-space lattice QCD simulations, quantum computation may be better suited for tracking low-energy nuclear excitations and strong correlations.

In this benchmark we test the ability of  quantum computers to simulate rare fundamental physics processes by simulating the $0\nu\beta\beta$-decay of a simple nucleus in 1+1D lattice QCD and observing the lepton-number violating dynamics.
The simulation involves two baryons confined to two spatial sites, incorporating strong and weak interactions, and explicitly violating lepton-number conservation with a Majorana neutrino mass term.

\subsubsection{Algorithm Description} 
We perform the simulation of the neutrinoless double-beta ($0\nu\beta\beta$) decay of a nucleus using a 1+1D lattice Quantum Chromodynamics (QCD) model. This model incorporates dynamical quarks (up and down) and leptons (electrons and neutrinos). The simulation utilizes a lattice with periodic boundary conditions (PBCs) and two spatial lattice sites, which are mapped onto 32 qubits. A minimum of two spatial sites is essential to accommodate the degrees of freedom generated during $0\nu\beta\beta$-decay.

Weak interactions are represented by an effective four-Fermi interaction that locally couples quarks and leptons. The hadronic states in 1+1D QCD form isospin multiplets, with the lowest-lying baryon multiplet having an isospin of $I=3/2$. This multiplet is labeled $\Delta^{++}, \Delta^+, \Delta^0, \Delta^-$ due to its resemblance to the $\Delta$ resonance observed in 3+1D QCD.

Parameters within the Hamiltonian, including a Majorana mass term, are tuned to facilitate the $0\nu\beta\beta$-decay of an initial $|\Delta^- \Delta^-\rangle$ state. To simulate this decay, a quantum circuit is employed. This circuit first initializes the lepton vacuum and the $|\Delta^- \Delta^-\rangle$ state. Subsequently, it undergoes time evolution driven by a Hamiltonian that encompasses strong and weak interactions, as well as free fermion terms. The inclusion of a lepton-number breaking neutrino Majorana mass in the free fermion Hamiltonian enables the $0\nu\beta\beta$ decay channel. Following the time evolution, the lepton qubits are measured to detect the occurrence of weak decays.

We include techniques like symmetrization, twirling, amplification, extrapolation, regression, and post-selection to approximate ideal circuit outputs by leveraging application and device symmetries to reduce noise-induced biases with lower overhead. Combining specific circuit variants with observable-specific post-selection rules allows for precise error detection and higher shot efficiency.

The overall approach integrates debiasing through symmetrization, post-selection via symmetry checks, and a new parameterized nonlinear filtering method on the lepton qubit register, utilizing spare qubits for flag-based mid-circuit symmetry checks and leakage error detection.

Specific for the hardware, the circuits were optimized by merging two-qubit gate blocks thanks to the all-to-all connectivity and native RZZ($\theta$) gates, reducing entangling gates by 15\% and efficiently reducing bias through varying qubit-to-ion assignments, and employing phase-flip twirling.
The circuit also utilizes bit-flip symmetrization of readout.

In post-processing, measurement statistics from different twirled variants are combined, filtering out outlier bit strings based on a filter threshold. This threshold depends on device noise, the number of twirled variants, and shots per variant; a higher threshold better mitigates hardware noise but requires more resources, while a too-high threshold for fixed resources can lead to information loss.

\subsubsection{Problem Instances}
We simulate $0\nu\beta\beta$-decay by first preparing the initial state $| \psi_{\text{init}}\rangle=|\psi^{(\text{lep})}_{\text{vac}} \rangle , |\Delta^-\Delta^-\rangle$ using the Scalable Circuit ADAPT-VQE (SC-ADAPT-VQE) algorithm. This prepares the lepton vacuum and the quark wavefunction, where $|0\rangle^{\otimes 6}$ represents the fully occupied $d$ quark register. The SC-ADAPT-VQE uses a single parameterized circuit $e^{i \theta \hat{O}}$ on a two-site lattice, built from the commutator of mass and kinetic terms for $u$ quarks, ensuring QCD Hamiltonian symmetries.

Next, we time-evolve the initial state with $e^{-i \hat{H} t}$ using first-order Trotterization, simplifying the first step as the initial state is an eigenstate of $(\hat{H}_{\text{free}} + \hat{H}_{\text{Maj}}+\hat{H}_{\text{glue}})$. We developed a new, highly parallelizable method for constructing time-evolution circuits, overcoming Jordan-Wigner transformation challenges by designing circuits without Pauli-$\hat{Z}$ strings, then incorporating them via $CZ$ conjugation.

The quantum circuit for $0\nu\beta\beta$-decay simulation involves preparing the initial state with SC-ADAPT-VQE, followed by two Trotterized time evolution steps, and finally flagging leakage events with ancilla-coupled lepton qubits before z-basis measurement.

In hardware experiments on 36 qubit quantum computers, we used 32 qubits to prepare the initial state and perform two steps of first-order Trotterized time evolution. We reduced two-qubit gates by truncating chromoelectric interaction and only keeping weak interaction terms acting on valence fermions, and removed small-angle two-qubit rotations. The compiled circuits require 470 $R_{ZZ}$ gates.

We optimized circuit performance using noise tailoring, error mitigation (Pauli twirling, XY4 dynamic decoupling, measurement twirling via bit flipping), and error detection. Each circuit was compiled into 96 twirled variants with unique qubit assignments and bit flips for readout error symmetrization, each with 150 shots (160 for $t=2$), totaling 24,000 shots. The remaining four qubits flagged qubit leakage, and measured bit strings were post-selected based on ancilla states to indicate no leakage and conserve color and total electric charges.

\subsubsection{Results} 

\begin{figure}[H]
    \centering
    \includegraphics[width=\linewidth]{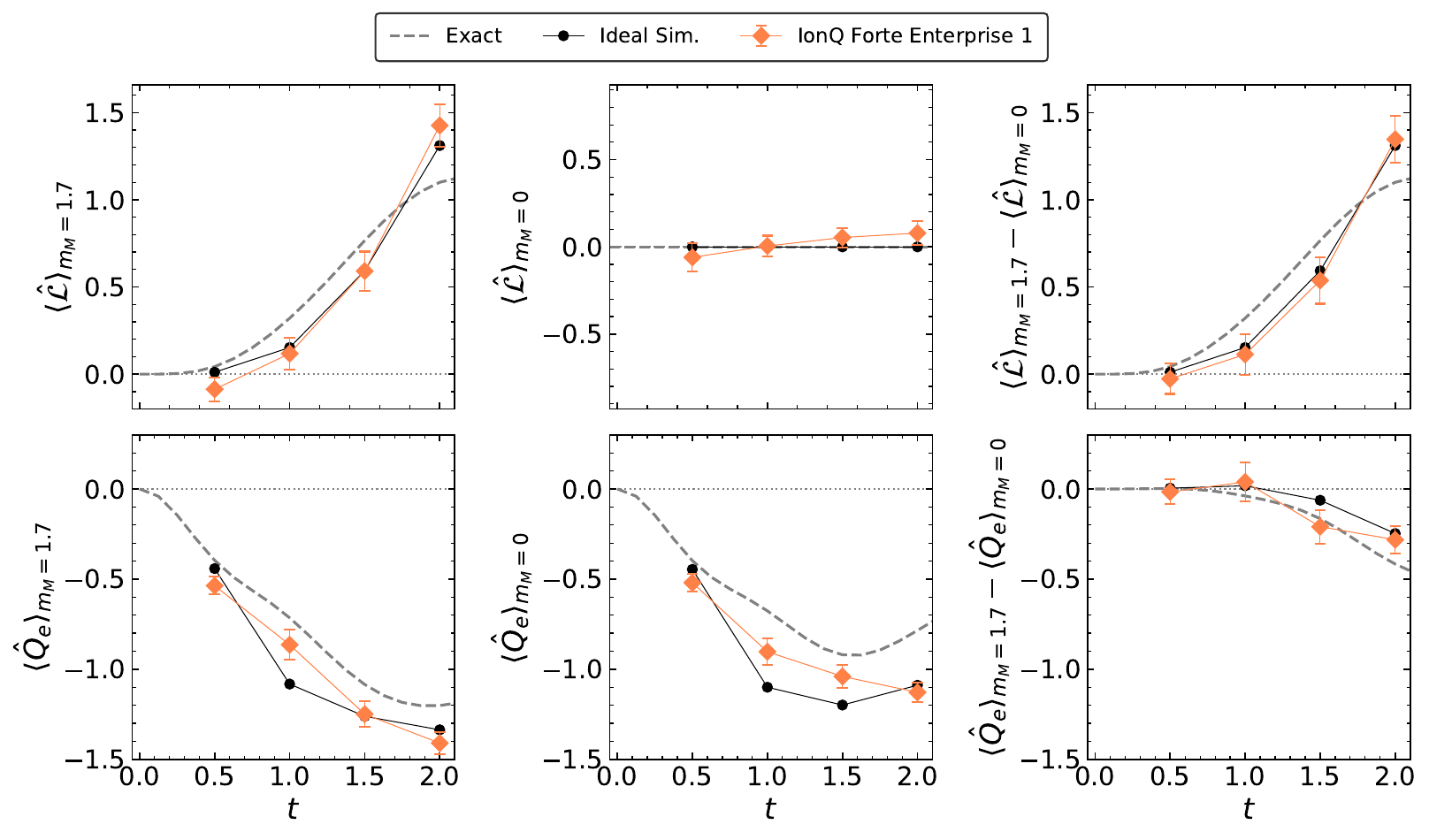}
    \caption{Lepton number time evolution during $|\Delta^-\Delta^-\rangle$ two-baryon state decay in 1+1D QCD. The circuit used two first-order Trotter steps with approximate valence-fermion weak interactions (470 two-qubit gates). Panels show results for $m_M=1.7$ (left), $m_M=0$ (center), and the difference between them (right). Data was obtained from IonQ Forte Enterprise. Black points indicate ideal simulations; the gray dotted line is for reference.}
    \label{fig:hep_results}
\end{figure}

Weak decays can be detected and classified by measuring the total electric charge of the electrons $\hat{Q}_e$ and the lepton number $\hat{{\cal L}}$. Both are zero in the initial state; non-zero $\hat{Q}_e$ signals a decay, while non-zero $\hat{{\cal L}}$ indicates a neutrino Majorana mass effect. 

\Cref{fig:hep_results} displays IonQ Forte Enterprise results for $t=\{0.5,1.0,1.5,2.0\}$ and $m_M=\{0.0,1.7\}$. We utilized 14,400 shots for $t \leq 1.5$ and 24,000 shots for $t=2.0$. Roughly $10\%$ of shots survived filtering based on leakage and charge conservation. Error bars represent bootstrap resampling \cite{chernyshev2026pathfinding}. Ideal noiseless classical simulations, also plotted, show excellent agreement with the experimental data. Notably, for $m_M=1.7$, we find a statistically significant signal for lepton number violation—a $10\sigma$ difference at $t=2.0$ compared to the $m_M=0$ case.

For $m_M=0$, the lepton number remains zero, consistent with single $\beta$-decay and $2\nu\beta\beta$-decay, although electric charge is produced. Due to the confined simulation volume, baryons and leptons interact continuously, preventing separation and sustaining a non-zero lepton electric charge density. Analyzing the time dependence of $\hat{{\cal L}}$ and $\hat{Q}_e$ could allow us to probe the underlying reaction mechanism. Predicting these values with accuracy requires the coherent summation of interfering reaction pathways—a task for which quantum computers are uniquely suited.  Wavefunction evolution at the yocto-scale can provide key information to identify dominant decay pathways.

\section{Results: Time-to-Solution Benchmarks}

\subsection{Time-To-Solution for Hidden Shift and Cosine QFT}

The time it takes to hit high quality solutions for the first time when iteratively trying to sample solutions can be used to assess TTS performance on various benchmarks. For instance, in the Hidden Shift with permutations (HSBP) challenge at 36 qubits, we use permutations consisting of $80$ randomly assigned CX gates. Sampling bit-strings close to the target shift becomes increasingly difficult for leading superconducting chips, and the confidence time-to-first-sample increases super exponentially. We compare against binonmial random samplers where each measured qubit is a perfect $p=0.5$ bit-flip.

The Cosine QFT challenge at 36 qubits similarly shows that sampling bit-strings close to the known target solutions becomes increasingly difficult for leading superconducting chips, and the confidence time-to-first-sample increases super exponentially. Upon 1 million circuit executions, the random sampler failed to sample a single bit-string with less than 3 bit-flip errors, while IonQ Forte manages to sample the correct solution in minutes.

These examples illustrate that while raw shot-rate remains a valid indicator of hardware-level performance, shot ``quality'' as measured by the confidence time-to-first solution proxy, is the critical metric of system utility for real applications. The per-shot time for the Cosine QFT was set to 472us per shot and for Hidden Shift to 337us per shot. This corresponds to a 2q-gate time of around 200ns. 

\begin{figure}[H]
    \centering
    \includegraphics[width=0.6\textwidth]{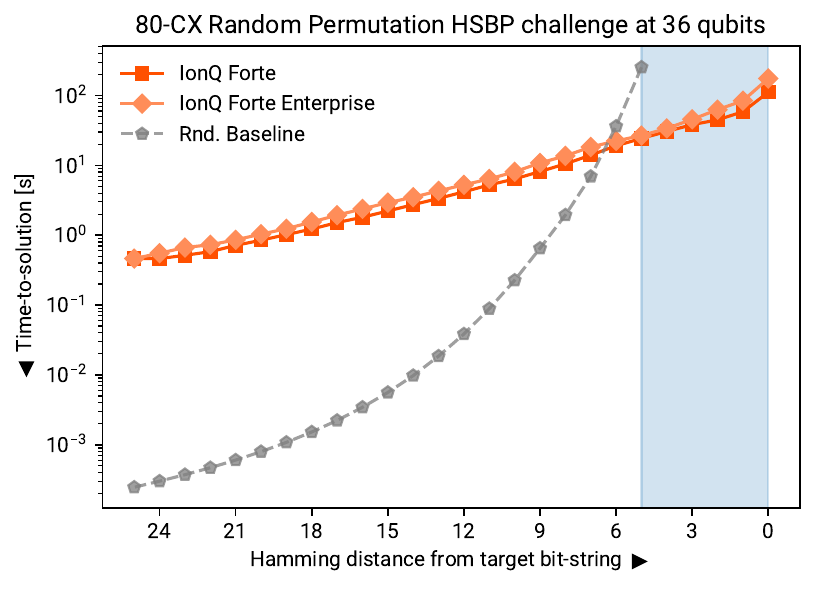}
    \caption{\label{fig:hiddenshiftTTS} Time-to-solution for the HSBP challenge at 36 qubits with permutations consisting of $80$ randomly assigned CX gates. The plot illustrates that sampling bit-strings close to the target shift becomes increasingly difficult for leading superconducting chips and the confidence time-to-first-sample increases super exponentially. The random baseline was generated by sampling 1 million bit-string uniformly at random and computing the Hamming distance from the target shift; the distribution is binomial. On the contrary, IonQ Forte manages to sample the correct solution in minutes.}
\end{figure}

\begin{figure}[H]
    \centering
    \includegraphics[width=0.6\textwidth]{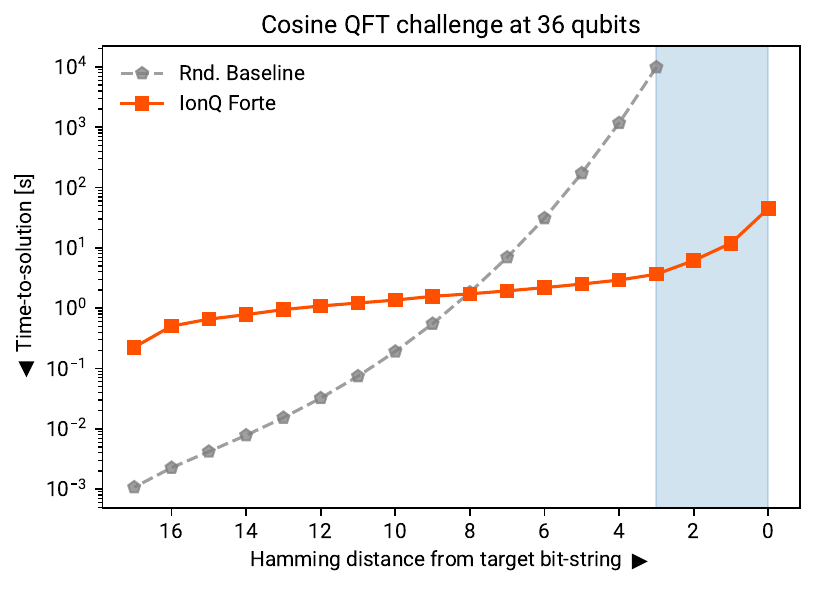}
    \caption{Time-to-solution for the Cosine QFT challenge at 36 qubits. This plot illustrates that sampling bit-strings close to the known target solutions to the Cosine QFT challenge becomes increasingly difficult for leading superconducting chips and the confidence time-to-first-sample increases super exponentially. On the contrary, IonQ Forte manages to sample the correct solution in minutes.}
\end{figure}

\begin{figure}[H]
    \centering
    \includegraphics[width=0.6\textwidth]{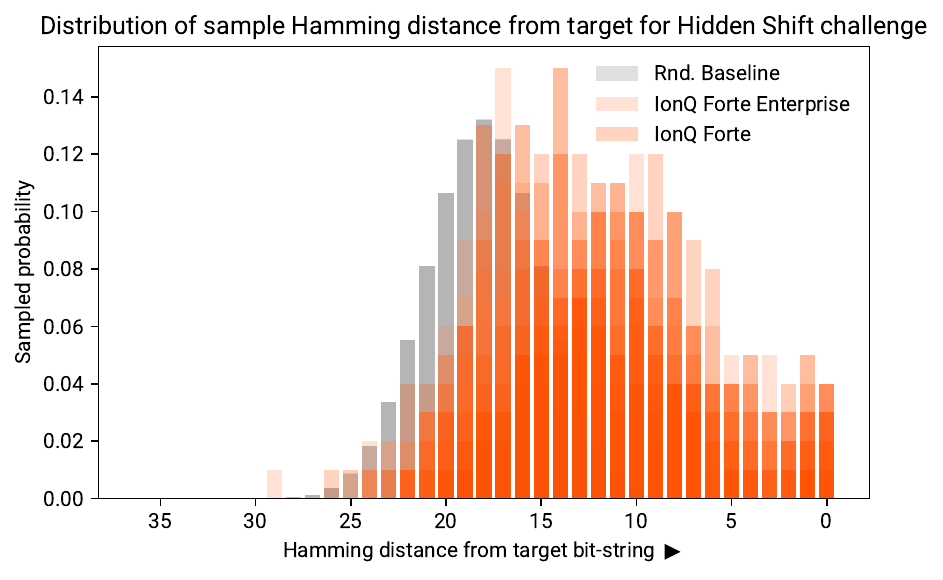}
    \caption{Shown are the quasi-probability distributions of the Hamming weight of bitvectors for the hidden shift problem. The shown distributions are taken from IonQ Forte (5k shots, orange) and the random sampler (1M shots, blue). As can be seen from the distributions, the IonQ device is able to sample from a heavy tail near approximation ratio (AR) of 1.0, whereas the random sampler has no mass there at all, even after taking 1M samples.  
}
    \label{fig:hiddenShiftSamples}
\end{figure}

The underlying reason for the advantage illustrated in Fig.~\ref{fig:hiddenshiftTTS} is the difference in the ability to sample from long tails of the hidden shift bitvector distribution. As shown in Fig.~\ref{fig:hiddenShiftSamples}, the IonQ device is able to find `needles-in-a-hackstack' bitvectors that have high approximation ratio, whereas the random sampler fails, i.e., the long tails are sampled with exponentially low probability. 

\subsection{Time-To-Solution for LR-QAOA}

\begin{figure}[H]
    \centering
    \includegraphics[width=0.6\textwidth]{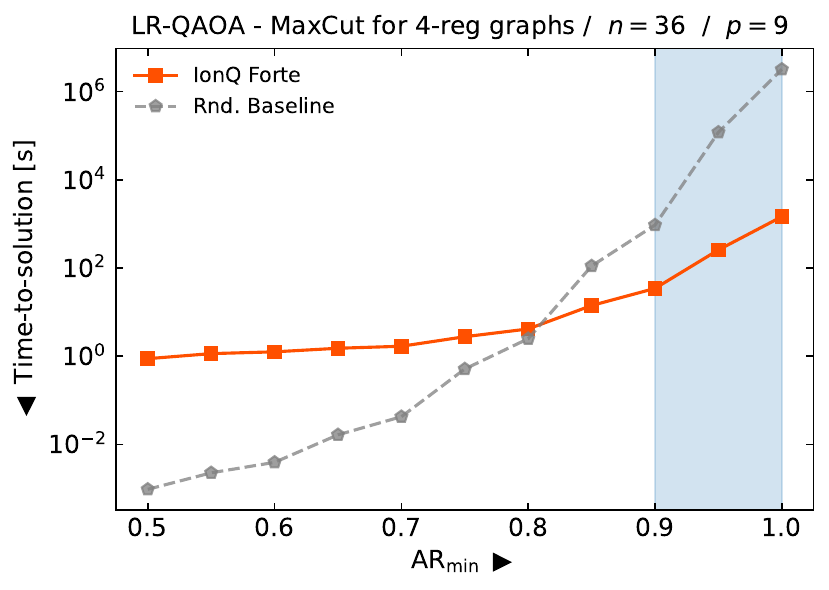}
    \caption{Time-to-solution (TTS) at 99\% confidence as a function of the minimum approximation ratio $\mathrm{AR}_{\min}$ for $p=9$ LR-QAOA on a 36-qubit, 4-regular MaxCut instance. The figure compares IonQ Forte data at 5,000 shots against the exact random-sampling baseline. IonQ Forte achieves finite TTS up to $\mathrm{AR}_{\min}=1.0$, while the random baseline grows rapidly at high target approximation ratios.}
    \label{fig:lrqaoa_tts}
\end{figure}

To quantify the practical performance of quantum hardware relative to a non-algorithmic baseline, we compute the Time-to-Solution (TTS) metric at 99\% confidence (see \Cref{fig:lrqaoa_tts}). For a given success threshold $\mathrm{AR}_{\min}$, let $p_s$ denote the probability that a single circuit execution yields an approximation ratio $\mathrm{AR} \geq \mathrm{AR}_{\min}$. The number of independent shots required to observe at least one such solution with a confidence level of $0.01$ (i.e., 99\% confidence) is
\begin{align}
    N_{\text{shots}} = \frac{\ln 0.01}{\ln(1 - p_s)},
\end{align}
and the Time-to-Solution is then $\mathrm{TTS} = N_{\text{shots}} \cdot t_{\text{shot}}$, where $t_{\text{shot}}$ is the wall-clock time per circuit execution. The TTS thus answers the question: how long must one run the device before being 99\% certain of having sampled at least one bit string with approximation ratio at or above a given target?

We benchmark a 4-regular MaxCut instance on $n = 36$ qubits using a $p = 9$ layer LR-QAOA circuit. The depth $p = 9$ was selected because it yielded the highest raw approximation ratio in the IonQ Forte dataset among all tested depths, without any error mitigation applied. The IonQ Forte data comprises
$5,000$ shots with $t_{\text{shot}} = 0.89$~s.

As a point of comparison, we also compute the TTS for a uniformly random sampler. Rather than drawing finite random samples, we evaluate the approximation ratio of every bit string in the full $2^{36}$-dimensional Hilbert space and construct the exact AR distribution, so that $p_s = (\text{number of states with } \mathrm{AR} \geq \mathrm{AR}_{\min}) / 2^{36}$. The per-shot time for this baseline is set to $t_{\text{shot}} = 445~\mu$s to give the random sampler a deliberately generous advantage. This corresponds to a 2q-gate time of around 200ns. 

Relative to this exact random baseline, IonQ Forte remains strongly separated at high AR thresholds. With only $5,000$ shots, the device samples 14 bit strings that achieve the optimal cut ($\mathrm{AR}=1.0$), yielding a finite TTS even at the most stringent threshold. At $\mathrm{AR}_{\min} = 0.90$, Forte's TTS is approximately $34$~s, compared to $942$~s for the random baseline. The gap widens further as the target threshold approaches the optimum, showing that the observed output distribution retains meaningful algorithmic structure rather than collapsing toward uniform random sampling. This TTS analysis therefore complements the raw AR and $\mathrm{AR}_{\mathrm{eff}}$ metrics by translating the separation from random sampling into an operational measure of solution cost.

To illustrate the projected scaling of the superconducting qubit system TTS beyond its observed range, we fit a model of the form $\mathrm{TTS}(\mathrm{AR}_{\min}) = e^{a\,x^2 + b\,x + c}$ to its data in the regime $\mathrm{AR}_{\min} \in [0.50, 0.90]$, where $x \equiv \mathrm{AR}_{\min}$. Extrapolating this fit to $\mathrm{AR}_{\min} = 1.0$ yields a projected TTS on the order of $10^5$\~s, underscoring the impracticality of the noisy superconducting backend for this circuit depth and problem size. The shaded region in \Cref{fig:lrqaoa_tts} highlights the extrapolation domain where the superconducting qubit system has no observed data, reinforcing the conclusion that the trapped-ion device maintains a substantial advantage in solution quality for this benchmark.

\begin{figure}[H]
    \centering
    \includegraphics[width=0.6\textwidth]{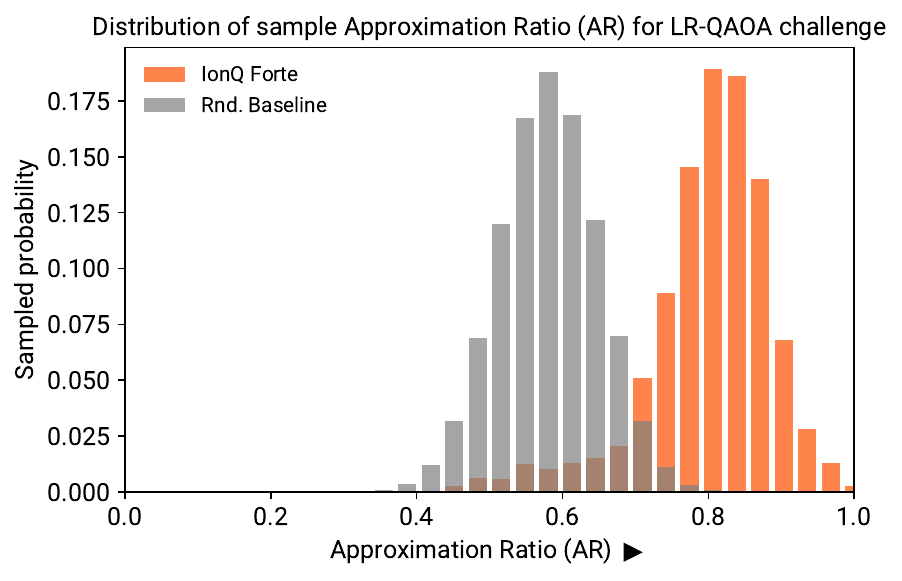}
    \caption{Shown are the quasi-probability distributions of the approximation ratio of bitvectors for the MaxCut problem. The shown distributions are taken from IonQ Forte (5k shots, orange) and the random sampler (1M shots, blue). As can be seen from the distributions, the IonQ device is able to sample from a heavy tail near approximation ratio (AR) of 1.0, whereas the random sampler has no mass there at all, even after taking 1M samples. 
}
    \label{fig:lrqaoa_tts_samples}
\end{figure}

The underlying reason for the advantage illustrated in Fig.~\ref{fig:lrqaoa_tts} is the difference in the ability to sample from long tails. As shown in Fig.~\ref{fig:lrqaoa_tts_samples}, the IonQ device is able to find `needles-in-a-hackstack' bitvectors that have high approximation ratio, whereas the random sampler fails, i.e., the long tails are sampled with exponentially low probability. 

Finally, in Fig.~\ref{fig:comparison}, we show that the IonQ device's ability to sample more effectively from tails of distributions translates into advantages over architectures such as superconducting qubits, which due to the mapping overhead to their 2D architecture behave closely to random samplers, i.e., they are not able to sample from valuable solutions from the long tails.  

\section{Conclusions}

We introduced a scalable and principled framework for application-level quantum benchmarking that addresses the diverse needs of internal development, customer and community engagements, and comparative evaluation across quantum computing platforms. By focusing on application-driven metrics such as solution quality, Time-to-Solution, and cost proxies, the proposed framework provides a more meaningful characterization of system performance than traditional hardware-centric benchmarks.

A key contribution of this work is the identification of guiding principles—measurability, simplicity, scalability, and extensibility -- that ensure benchmarks remain actionable, interpretable, and reproducible. These principles enable the construction of benchmark suites that can evolve alongside hardware improvements while maintaining consistency and credibility over time. Code that allow to independently run and extend the provided benchmarks was provided along with a detailed discussion of the benchmarks themselves. 

The framework also highlights the importance of standardization and openness in benchmarking practices. By enabling full disclosure of implementation details and supporting reproducibility, it fosters trust among customers and facilitates collaboration across the broader quantum ecosystem. Moreover, the use of benchmark families allows for systematic scaling across problem sizes, ensuring relevance as quantum systems grow in capability.

Looking forward, this framework provides a foundation for the development of industry-wide benchmarking standards and the integration of application-level metrics into decision-making processes. It also opens the door to future extensions, including the incorporation of new application domains, improved cost models, and tighter integration with emerging quantum software and hardware stacks.

Ultimately, measuring what matters requires aligning benchmarking efforts with real-world use cases. The approach presented here represents a step toward that goal, enabling a more accurate and transparent assessment of quantum computing performance and its impact on practical applications.

\clearpage
\bibliographystyle{unsrt}
\bibliography{bibliography}
\clearpage
\begin{appendices}

\section{Format and Requirements for Closed and Open Categories}
\label{app:format}

The following tables collect results from a benchmarking campaign carried out on  IonQ's Aria, Forte, and Forte Enterprise systems. 
Each line corresponds to a particular instance of the benchmark, characterized by a unique identifier as provided in the accompanying code repository at \texttt{\href{https://github.com/ionq-publications/apps-benchmark}{github.com/ionq-publications/apps-benchmark}}. The categories of data reported are the Domain to which the benchmark broadly belongs, the specific type of Problem that is addressed, and which Algorithm is used. Regarding the specific implementation, the total number of qubits (\#q), the total number of different circuits (\#qc), the total number of single-qubit gates (\#1q) and two-qubit gates (\#2q), as well as the total number of shots (\#s) is reported. Backend specifies against which specific quantum machine the benchmark was run. The field EM indicates whether an error mitigation method was used. The score is a single numerical value that is defined for each benchmark and can be computed using the code provided. Exec.\,Time and Energy are the total QPU time and total energy spent to complete the benchmark as defined in Section~\ref{sec:exec}. 
Execution time is reported in total number of seconds (s), Energy is reported in total kiloWatt-hours (kWh). Note that only Forte and Forte-Enterprise are equipped with instrumentation to measure the energy consumption, i.e., the corresponding values for Aria are left blank (``-''). 

The framework is designed to be extensible and we provide a guide (see \texttt{README.md} and \texttt{ DIY\_BACKEND.md}) on how to make your own backend connector for other QPUs. The following are requirements for reporting results for the Closed and Open divisions: 

\begin{itemize}
    \item The Closed Division is designed for fair, direct comparisons of hardware and software frameworks. This means the following: 
{\bf Pre-processing/Post-processing:} Must be equivalent to the reference.
{\bf Algorithm change:} Prohibited. {\bf Compliance Runs:} Must pass mandatory compliance tests to ensure score, which are verified by the benchmarking harness. 
{\bf Dataset:} Must be identical.
{\bf Reproducibility:} Code used for the benchmark must be open-sourced. 
\item 
The Open Division encourages innovation in the approaches to tackle the problems. This means the following: {\bf Quantum algorithm flexibility}: Allows changing the quantum circuits, error mitigation, model architecture, etc. 
{\bf Pre-processing/Post-processing:} Allows arbitrary pre-processing or post-processing as long as it does not rely the solution of the problem. 
{\bf Target Quality:} The model must still meet the defined target quality (accuracy) metric.
{\bf Software/Framework:} Can use different methods. 
\item
General requirements for both Divisions:
{\bf System Availability:} The hardware and software stack must be available (or soon available) for purchase or access purchase, although preview hardware might be considered as long as the required metrics mentioned below are reported.
{\bf Reproducibility:} Methods that allow reproduction of the results (ideally, based on source code) must be shared.
{\bf Auditability:} Submissions are subject to a random or committee-selected audit process to verify compliance, especially for the Closed Division.
{\bf Metrics:} Must report all categories provided in the headers, in particular Score and Execution Time. Energy is optional, but reporting it is preferred.  
\end{itemize}
For contributions to the framework, participants may open tickets (github Issues) against the underlying open source repository. 

\section{Tables with Numerical Benchmark Results}
\label{app:tables}

{
\tiny
\sffamily
\begin{longtable}{lllrrrrrllrrr}
\toprule
Domain & Problem & Algorithm & \#q & \#qc & \#1q & \#2q & Shots & Backend & EM? & Score & Exec. Time (s) & Energy (kWh) \\
\midrule
\endfirsthead
\toprule
Domain & Problem & Algorithm & \#q & \#qc & \#1q & \#2q & Shots & Backend & EM? & Score & Exec. Time (s) & Energy (kWh) \\
\midrule
\endhead
\midrule
\multicolumn{13}{r}{Continued on next page} \\
\midrule
\endfoot
\bottomrule
\endlastfoot
hidden shift & ccx ladder & hidden shift & 8 & 10 & 1,510 & 80 & 1,000 & IonQ Forte & N & 0.651 & 73.385 & 0.10 \\
hidden shift & ccx ladder & hidden shift & 8 & 10 & 1,510 & 80 & 1,000 & IonQ Forte-E & N & 0.632 & 81.848 & 0.11 \\
hidden shift & ccx ladder & hidden shift & 12 & 10 & 2,200 & 120 & 1,000 & IonQ Forte & N & 0.48 & 127.524 & 0.17 \\
hidden shift & ccx ladder & hidden shift & 12 & 10 & 2,200 & 120 & 1,000 & IonQ Forte-E & N & 0.377 & 141.159 & 0.19 \\
hidden shift & ccx ladder & hidden shift & 16 & 10 & 2,940 & 160 & 1,000 & IonQ Forte & N & 0.299 & 180.439 & 0.24 \\
hidden shift & ccx ladder & hidden shift & 16 & 10 & 2,940 & 160 & 1,000 & IonQ Forte-E & N & 0.227 & 200.323 & 0.27 \\
hidden shift & ccx ladder & hidden shift & 20 & 10 & 3,700 & 200 & 1,000 & IonQ Forte & N & 0.193 & 234.91 & 0.32 \\
hidden shift & ccx ladder & hidden shift & 20 & 10 & 3,700 & 200 & 1,000 & IonQ Forte-E & N & 0.123 & 258.84 & 0.35 \\
\end{longtable}

\newpage
\begin{longtable}{lllrrrrrllrrr}
\toprule
Domain & Problem & Algorithm & \#q & \#qc & \#1q & \#2q & Shots & Backend & EM? & Score & Exec. Time (s) & Energy (kWh) \\
\midrule
\endfirsthead
\toprule
Domain & Problem & Algorithm & \#q & \#qc & \#1q & \#2q & Shots & Backend & EM? & Score & Exec. Time (s) & Energy (kWh) \\
\midrule
\endhead
\midrule
\multicolumn{13}{r}{Continued on next page} \\
\midrule
\endfoot
\bottomrule
\endlastfoot
hidden shift & random 020 cx & hidden shift & 8 & 9 & 1,422 & 792 & 900 & IonQ Forte & N & 0.7356 & 67.425 & 0.09 \\
hidden shift & random 020 cx & hidden shift & 8 & 9 & 1,422 & 792 & 900 & IonQ Forte-E & N & 0.7144 & 45.529 & 0.06 \\
hidden shift & random 020 cx & hidden shift & 12 & 9 & 2,058 & 828 & 900 & IonQ Forte & N & 0.6478 & 84.328 & 0.11 \\
hidden shift & random 020 cx & hidden shift & 12 & 9 & 2,058 & 828 & 900 & IonQ Forte-E & N & 0.5622 & 63.751 & 0.09 \\
hidden shift & random 020 cx & hidden shift & 16 & 9 & 2,784 & 864 & 900 & IonQ Forte & N & 0.4756 & 116.68 & 0.16 \\
hidden shift & random 020 cx & hidden shift & 16 & 9 & 2,784 & 864 & 900 & IonQ Forte-E & N & 0.4222 & 90.778 & 0.12 \\
hidden shift & random 020 cx & hidden shift & 20 & 9 & 3,450 & 900 & 900 & IonQ Forte & N & 0.3956 & 132.349 & 0.18 \\
hidden shift & random 020 cx & hidden shift & 20 & 9 & 3,450 & 900 & 900 & IonQ Forte-E & N & 0.4078 & 92.105 & 0.13 \\
\end{longtable}

\newpage
\begin{longtable}{lllrrrrrllrrr}
\toprule
Domain & Problem & Algorithm & \#q & \#qc & \#1q & \#2q & Shots & Backend & EM? & Score & Exec. Time (s) & Energy (kWh) \\
\midrule
\endfirsthead
\toprule
Domain & Problem & Algorithm & \#q & \#qc & \#1q & \#2q & Shots & Backend & EM? & Score & Exec. Time (s) & Energy (kWh) \\
\midrule
\endhead
\midrule
\multicolumn{13}{r}{Continued on next page} \\
\midrule
\endfoot
\bottomrule
\endlastfoot
hidden shift & random 070 cx & hidden shift & 8 & 9 & 1,392 & 2,592 & 900 & IonQ Forte & N & 0.5133 & 125.738 & 0.17 \\
hidden shift & random 070 cx & hidden shift & 8 & 9 & 1,392 & 2,592 & 900 & IonQ Forte-E & N & 0.4267 & 99.999 & 0.14 \\
hidden shift & random 070 cx & hidden shift & 12 & 9 & 2,178 & 2,628 & 900 & IonQ Forte & N & 0.2167 & 190.175 & 0.26 \\
hidden shift & random 070 cx & hidden shift & 12 & 9 & 2,178 & 2,628 & 900 & IonQ Forte-E & N & 0.2367 & 161.242 & 0.22 \\
hidden shift & random 070 cx & hidden shift & 16 & 9 & 2,844 & 2,664 & 900 & IonQ Forte & N & 0.1456 & 254.056 & 0.34 \\
hidden shift & random 070 cx & hidden shift & 16 & 9 & 2,844 & 2,664 & 900 & IonQ Forte-E & N & 0.1067 & 206.415 & 0.28 \\
hidden shift & random 070 cx & hidden shift & 20 & 9 & 3,510 & 2,700 & 900 & IonQ Forte & N & 0.09222 & 300.686 & 0.41 \\
hidden shift & random 070 cx & hidden shift & 20 & 9 & 3,510 & 2,700 & 900 & IonQ Forte-E & N & 0.06222 & 224.705 & 0.31 \\
\end{longtable}

\newpage
\begin{longtable}{lllrrrrrllrrr}
\toprule
Domain & Problem & Algorithm & \#q & \#qc & \#1q & \#2q & Shots & Backend & EM? & Score & Exec. Time (s) & Energy (kWh) \\
\midrule
\endfirsthead
\toprule
Domain & Problem & Algorithm & \#q & \#qc & \#1q & \#2q & Shots & Backend & EM? & Score & Exec. Time (s) & Energy (kWh) \\
\midrule
\endhead
\midrule
\multicolumn{13}{r}{Continued on next page} \\
\midrule
\endfoot
\bottomrule
\endlastfoot
machine learning & img loading & mps load & 10 & 1 & 2,084 & 113 & 10,000 & IonQ Aria & N & 0.497 & 3,774.4 & NaN \\
machine learning & img loading & mps load & 10 & 1 & 2,503 & 136 & 10,000 & IonQ Aria & N & 0.69 & 4,505.64 & NaN \\
machine learning & img loading & mps load & 10 & 25 & 1,718 & 550 & 10,000 & IonQ Aria & Y & 0.458 & 376.091 & NaN \\
machine learning & img loading & mps load & 10 & 25 & 2,275 & 550 & 10,000 & IonQ Forte & Y & 0.453 & 454.812 & 0.62 \\
machine learning & img loading & mps load & 10 & 25 & 2,821 & 1,125 & 10,000 & IonQ Aria & Y & 0.344 & 592.506 & NaN \\
machine learning & img loading & mps load & 10 & 25 & 3,935 & 1,625 & 10,000 & IonQ Aria & Y & 0.252 & 786.974 & NaN \\
machine learning & img loading & mps load & 10 & 25 & 4,225 & 1,125 & 10,000 & IonQ Forte & Y & 0.338 & 761.946 & 1.03 \\
machine learning & img loading & mps load & 10 & 25 & 5,157 & 2,275 & 10,000 & IonQ Aria & Y & 0.324 & 1,021.82 & NaN \\
\end{longtable}

\newpage
\begin{longtable}{lllrrrrrllrrr}
\toprule
Domain & Problem & Algorithm & \#q & \#qc & \#1q & \#2q & Shots & Backend & EM? & Score & Exec. Time (s) & Energy (kWh) \\
\midrule
\endfirsthead
\toprule
Domain & Problem & Algorithm & \#q & \#qc & \#1q & \#2q & Shots & Backend & EM? & Score & Exec. Time (s) & Energy (kWh) \\
\midrule
\endhead
\midrule
\multicolumn{13}{r}{Continued on next page} \\
\midrule
\endfoot
\bottomrule
\endlastfoot
optimization & maxcut & qaoa & 12 & 1 & 24 & 66 & 5,000 & IonQ Aria & N & 0.563 & 399.837 & NaN \\
optimization & maxcut & qaoa & 12 & 1 & 24 & 66 & 5,000 & IonQ Forte & N & 0.5599 & 616.513 & 0.83 \\
optimization & maxcut & qaoa & 12 & 1 & 24 & 66 & 5,000 & IonQ Forte-E & N & 0.6213 & 519.795 & 0.71 \\
optimization & maxcut & qaoa & 12 & 1 & 48 & 198 & 5,000 & IonQ Aria & N & 0.7748 & 991.815 & NaN \\
optimization & maxcut & qaoa & 12 & 1 & 48 & 198 & 5,000 & IonQ Forte & N & 0.7598 & 1,291.07 & 1.75 \\
optimization & maxcut & qaoa & 12 & 1 & 48 & 198 & 5,000 & IonQ Forte-E & N & 0.7855 & 1,426.39 & 1.94 \\
optimization & maxcut & qaoa & 12 & 1 & 72 & 330 & 5,000 & IonQ Aria & N & 0.821 & 1,570.59 & NaN \\
optimization & maxcut & qaoa & 12 & 1 & 72 & 330 & 5,000 & IonQ Forte & N & 0.8081 & 2,097.38 & 2.84 \\
\end{longtable}

\newpage
\begin{longtable}{lllrrrrrllrrr}
\toprule
Domain & Problem & Algorithm & \#q & \#qc & \#1q & \#2q & Shots & Backend & EM? & Score & Exec. Time (s) & Energy (kWh) \\
\midrule
\endfirsthead
\toprule
Domain & Problem & Algorithm & \#q & \#qc & \#1q & \#2q & Shots & Backend & EM? & Score & Exec. Time (s) & Energy (kWh) \\
\midrule
\endhead
\midrule
\multicolumn{13}{r}{Continued on next page} \\
\midrule
\endfoot
\bottomrule
\endlastfoot
optimization & maxcut & qaoa & 16 & 25 & 3,000 & 8,550 & 5,000 & IonQ Forte-E & Y & 0.7384 & 2,323.72 & 3.16 \\
optimization & maxcut & qaoa & 16 & 25 & 4,600 & 14,050 & 5,000 & IonQ Forte & Y & 0.7861 & 3,780.31 & 5.11 \\
optimization & maxcut & qaoa & 16 & 25 & 4,600 & 14,050 & 5,000 & IonQ Forte-E & Y & 0.7679 & 3,759.05 & 5.12 \\
optimization & maxcut & qaoa & 16 & 25 & 6,250 & 19,525 & 5,000 & IonQ Forte & Y & 0.8575 & 5,228.57 & 7.07 \\
optimization & maxcut & qaoa & 16 & 25 & 6,250 & 19,525 & 5,000 & IonQ Forte-E & Y & 0.8079 & 5,206.85 & 7.09 \\
optimization & maxcut & qaoa & 16 & 25 & 7,800 & 25,050 & 5,000 & IonQ Forte & Y & 0.8778 & 6,713.93 & 9.08 \\
optimization & maxcut & qaoa & 16 & 25 & 7,800 & 25,050 & 5,000 & IonQ Forte-E & Y & 0.8092 & 6,668.29 & 9.08 \\
optimization & maxcut & qaoa & 16 & 25 & 9,450 & 30,550 & 5,000 & IonQ Forte & Y & 0.8572 & 7,964.77 & 10.77 \\
\end{longtable}

\newpage
\begin{longtable}{lllrrrrrllrrr}
\toprule
Domain & Problem & Algorithm & \#q & \#qc & \#1q & \#2q & Shots & Backend & EM? & Score & Exec. Time (s) & Energy (kWh) \\
\midrule
\endfirsthead
\toprule
Domain & Problem & Algorithm & \#q & \#qc & \#1q & \#2q & Shots & Backend & EM? & Score & Exec. Time (s) & Energy (kWh) \\
\midrule
\endhead
\midrule
\multicolumn{13}{r}{Continued on next page} \\
\midrule
\endfoot
\bottomrule
\endlastfoot
optimization & maxcut & qaoa & 24 & 25 & 6,348 & 5,400 & 5,000 & IonQ Aria & Y & 0.9087 & 1,155.15 & NaN \\
optimization & maxcut & qaoa & 24 & 25 & 6,900 & 4,500 & 5,000 & IonQ Forte & Y & 0.8321 & 1,365.86 & 1.85 \\
optimization & maxcut & qaoa & 24 & 25 & 6,900 & 4,500 & 5,000 & IonQ Forte & Y & 0.9935 & 1,349.54 & 1.83 \\
optimization & maxcut & qaoa & 24 & 25 & 6,900 & 6,000 & 5,000 & IonQ Forte & Y & 0.8467 & 1,769.74 & 2.39 \\
optimization & maxcut & qaoa & 24 & 25 & 7,381 & 6,300 & 5,000 & IonQ Aria & Y & 0.9075 & 1,335.37 & NaN \\
optimization & maxcut & qaoa & 24 & 25 & 8,100 & 5,400 & 5,000 & IonQ Forte & Y & 0.9947 & 1,600.98 & 2.17 \\
optimization & maxcut & qaoa & 24 & 25 & 8,235 & 7,200 & 5,000 & IonQ Aria & Y & 0.8902 & 1,505.79 & NaN \\
optimization & maxcut & qaoa & 24 & 25 & 9,245 & 8,100 & 5,000 & IonQ Aria & Y & 0.8867 & 1,680.09 & NaN \\
\end{longtable}

\newpage
\begin{longtable}{lllrrrrrllrrr}
\toprule
Domain & Problem & Algorithm & \#q & \#qc & \#1q & \#2q & Shots & Backend & EM? & Score & Exec. Time (s) & Energy (kWh) \\
\midrule
\endfirsthead
\toprule
Domain & Problem & Algorithm & \#q & \#qc & \#1q & \#2q & Shots & Backend & EM? & Score & Exec. Time (s) & Energy (kWh) \\
\midrule
\endhead
\midrule
\multicolumn{13}{r}{Continued on next page} \\
\midrule
\endfoot
\bottomrule
\endlastfoot
optimization & maxcut & qaoa & 36 & 1 & 1,080 & 1,566 & 5,000 & IonQ Forte & N & 0.6889 & 11,479.8 & 15.53 \\
optimization & maxcut & qaoa & 36 & 1 & 1,080 & 2,088 & 5,000 & IonQ Forte & N & 0.6886 & 14,450.6 & 19.55 \\
optimization & maxcut & qaoa & 36 & 1 & 1,224 & 1,782 & 5,000 & IonQ Forte & N & 0.666 & 13,019.7 & 17.61 \\
optimization & maxcut & qaoa & 36 & 1 & 1,224 & 2,376 & 5,000 & IonQ Forte & N & 0.6627 & 16,519.7 & 22.35 \\
optimization & maxcut & qaoa & 36 & 1 & 1,368 & 1,998 & 5,000 & IonQ Forte & N & 0.6501 & 14,502.3 & 19.62 \\
optimization & maxcut & qaoa & 36 & 1 & 1,368 & 2,664 & 5,000 & IonQ Forte & N & 0.6455 & 18,737.4 & 25.35 \\
optimization & maxcut & qaoa & 36 & 1 & 1,512 & 2,214 & 5,000 & IonQ Forte & N & 0.6357 & 15,784.6 & 21.35 \\
optimization & maxcut & qaoa & 36 & 1 & 1,512 & 2,952 & 5,000 & IonQ Forte & N & 0.6294 & 20,825.3 & 28.17 \\
\end{longtable}

\newpage
\begin{longtable}{lllrrrrrllrrr}
\toprule
Domain & Problem & Algorithm & \#q & \#qc & \#1q & \#2q & Shots & Backend & EM? & Score & Exec. Time (s) & Energy (kWh) \\
\midrule
\endfirsthead
\toprule
Domain & Problem & Algorithm & \#q & \#qc & \#1q & \#2q & Shots & Backend & EM? & Score & Exec. Time (s) & Energy (kWh) \\
\midrule
\endhead
\midrule
\multicolumn{13}{r}{Continued on next page} \\
\midrule
\endfoot
\bottomrule
\endlastfoot
optimization & risk assessment & quantum copula & 18 & 1 & 39 & 27 & 5,000 & IonQ Forte & N & 0.9886 & 230.987 & 0.31 \\
optimization & risk assessment & quantum copula & 18 & 1 & 39 & 27 & 5,000 & IonQ Forte-E & N & 0.958 & 247.427 & 0.34 \\
optimization & risk assessment & quantum copula & 18 & 1 & 72 & 153 & 5,000 & IonQ Aria & N & 0.9591 & 481.078 & NaN \\
optimization & risk assessment & quantum copula & 18 & 1 & 72 & 153 & 5,000 & IonQ Aria & N & 0.9681 & 477.073 & NaN \\
optimization & risk assessment & quantum copula & 18 & 1 & 72 & 153 & 5,000 & IonQ Aria & N & 0.9971 & 474.985 & NaN \\
optimization & risk assessment & quantum copula & 18 & 1 & 72 & 153 & 5,000 & IonQ Forte & N & 0.9569 & 671.726 & 0.91 \\
optimization & risk assessment & quantum copula & 18 & 1 & 72 & 153 & 5,000 & IonQ Forte-E & N & 0.8695 & 757.515 & 1.03 \\
optimization & risk assessment & quantum copula & 18 & 25 & 1,575 & 375 & 5,000 & IonQ Forte & Y & 0.9725 & 245.716 & 0.33 \\
\end{longtable}

\newpage
\begin{longtable}{lllrrrrrllrrr}
\toprule
Domain & Problem & Algorithm & \#q & \#qc & \#1q & \#2q & Shots & Backend & EM? & Score & Exec. Time (s) & Energy (kWh) \\
\midrule
\endfirsthead
\toprule
Domain & Problem & Algorithm & \#q & \#qc & \#1q & \#2q & Shots & Backend & EM? & Score & Exec. Time (s) & Energy (kWh) \\
\midrule
\endhead
\midrule
\multicolumn{13}{r}{Continued on next page} \\
\midrule
\endfoot
\bottomrule
\endlastfoot
optimization & risk assessment & quantum copula & 30 & 25 & 4,250 & 3,525 & 5,000 & IonQ Forte-E & Y & 0.6466 & 1,183.8 & 1.61 \\
qft & qft & cosine qft & 12 & 1 & 99 & 156 & 1,000 & IonQ Forte & N & 0.591 & 263.161 & 0.36 \\
qft & qft & cosine qft & 12 & 1 & 99 & 156 & 1,000 & IonQ Forte-E & N & 0.5836 & 162.122 & 0.22 \\
qft & qft & cosine qft & 16 & 1 & 127 & 272 & 1,000 & IonQ Forte & N & 0.4529 & 401.26 & 0.54 \\
qft & qft & cosine qft & 16 & 1 & 127 & 272 & 1,000 & IonQ Forte-E & N & 0.3595 & 264.218 & 0.36 \\
qft & qft & cosine qft & 20 & 1 & 155 & 420 & 1,000 & IonQ Forte & N & 0.3241 & 522.35 & 0.71 \\
qft & qft & cosine qft & 20 & 1 & 155 & 420 & 1,000 & IonQ Forte-E & N & 0.2072 & 353.329 & 0.48 \\
qft & qft & cosine qft & 24 & 1 & 183 & 600 & 1,000 & IonQ Forte & N & 0.1589 & 636.525 & 0.86 \\
\end{longtable}

\newpage
\begin{longtable}{lllrrrrrllrrr}
\toprule
Domain & Problem & Algorithm & \#q & \#qc & \#1q & \#2q & Shots & Backend & EM? & Score & Exec. Time (s) & Energy (kWh) \\
\midrule
\endfirsthead
\toprule
Domain & Problem & Algorithm & \#q & \#qc & \#1q & \#2q & Shots & Backend & EM? & Score & Exec. Time (s) & Energy (kWh) \\
\midrule
\endhead
\midrule
\multicolumn{13}{r}{Continued on next page} \\
\midrule
\endfoot
\bottomrule
\endlastfoot
unstructured search & faa & faa & 4 & 1 & 530 & 216 & 1,000 & IonQ Forte & N & 0.209 & 319.8 & 0.43 \\
unstructured search & faa & faa & 4 & 1 & 530 & 216 & 1,000 & IonQ Forte & N & 0.227 & 318.093 & 0.43 \\
unstructured search & faa & faa & 4 & 1 & 530 & 216 & 1,000 & IonQ Forte & N & 0.231 & 322.206 & 0.44 \\
unstructured search & faa & faa & 4 & 1 & 530 & 216 & 1,000 & IonQ Forte & N & 0.242 & 322.098 & 0.44 \\
unstructured search & faa & faa & 4 & 1 & 530 & 216 & 1,000 & IonQ Forte-E & N & 0.144 & 325.697 & 0.44 \\
unstructured search & faa & faa & 4 & 1 & 530 & 216 & 1,000 & IonQ Forte-E & N & 0.146 & 326.171 & 0.44 \\
unstructured search & faa & faa & 4 & 1 & 530 & 216 & 1,000 & IonQ Forte-E & N & 0.165 & 329.334 & 0.45 \\
unstructured search & faa & faa & 4 & 1 & 530 & 216 & 1,000 & IonQ Forte-E & N & 0.181 & 323.262 & 0.44 \\
\end{longtable}

\newpage
\begin{longtable}{lllrrrrrllrrr}
\toprule
Domain & Problem & Algorithm & \#q & \#qc & \#1q & \#2q & Shots & Backend & EM? & Score & Exec. Time (s) & Energy (kWh) \\
\midrule
\endfirsthead
\toprule
Domain & Problem & Algorithm & \#q & \#qc & \#1q & \#2q & Shots & Backend & EM? & Score & Exec. Time (s) & Energy (kWh) \\
\midrule
\endhead
\midrule
\multicolumn{13}{r}{Continued on next page} \\
\midrule
\endfoot
\bottomrule
\endlastfoot
unstructured search & faa & faa & 5 & 1 & 753 & 324 & 1,000 & IonQ Forte-E & N & 0.057 & 392.828 & 0.53 \\
unstructured search & faa & faa & 5 & 1 & 753 & 324 & 1,000 & IonQ Forte-E & N & 0.058 & 398.24 & 0.54 \\
unstructured search & faa & faa & 5 & 1 & 753 & 324 & 1,000 & IonQ Forte-E & N & 0.059 & 397.937 & 0.54 \\
unstructured search & faa & faa & 5 & 1 & 753 & 324 & 1,000 & IonQ Forte-E & N & 0.063 & 396.782 & 0.54 \\
unstructured search & faa & faa & 5 & 1 & 753 & 324 & 1,000 & IonQ Forte-E & N & 0.064 & 432.084 & 0.59 \\
unstructured search & faa & faa & 5 & 1 & 763 & 324 & 1,000 & IonQ Aria & N & 0.109 & 265.77 & NaN \\
unstructured search & faa & faa & 5 & 1 & 763 & 324 & 1,000 & IonQ Aria & N & 0.134 & 272.212 & NaN \\
unstructured search & faa & faa & 5 & 1 & 763 & 324 & 1,000 & IonQ Aria & N & 0.153 & 266.879 & NaN \\
\end{longtable}

\newpage
\begin{longtable}{lllrrrrrllrrr}
\toprule
Domain & Problem & Algorithm & \#q & \#qc & \#1q & \#2q & Shots & Backend & EM? & Score & Exec. Time (s) & Energy (kWh) \\
\midrule
\endfirsthead
\toprule
Domain & Problem & Algorithm & \#q & \#qc & \#1q & \#2q & Shots & Backend & EM? & Score & Exec. Time (s) & Energy (kWh) \\
\midrule
\endhead
\midrule
\multicolumn{13}{r}{Continued on next page} \\
\midrule
\endfoot
\bottomrule
\endlastfoot
unstructured search & faa & faa & 5 & 1 & 1,613 & 684 & 1,000 & IonQ Forte-E & N & 0.033 & 777.817 & 1.06 \\
unstructured search & faa & faa & 5 & 1 & 1,613 & 684 & 1,000 & IonQ Forte-E & N & 0.034 & 756.06 & 1.03 \\
unstructured search & faa & faa & 5 & 1 & 1,613 & 684 & 1,000 & IonQ Forte-E & N & 0.034 & 779.631 & 1.06 \\
unstructured search & faa & faa & 5 & 1 & 1,613 & 684 & 1,000 & IonQ Forte-E & N & 0.034 & 818.611 & 1.11 \\
unstructured search & faa & faa & 5 & 1 & 1,613 & 684 & 1,000 & IonQ Forte-E & N & 0.035 & 772.356 & 1.05 \\
unstructured search & faa & faa & 5 & 1 & 1,613 & 684 & 1,000 & IonQ Forte-E & N & 0.035 & 794.754 & 1.08 \\
unstructured search & faa & faa & 5 & 1 & 1,613 & 684 & 1,000 & IonQ Forte-E & N & 0.04 & 801.446 & 1.09 \\
unstructured search & faa & faa & 5 & 1 & 1,613 & 684 & 1,000 & IonQ Forte-E & N & 0.043 & 813.788 & 1.11 \\
\end{longtable}

\newpage
\begin{longtable}{lllrrrrrllrrr}
\toprule
Domain & Problem & Algorithm & \#q & \#qc & \#1q & \#2q & Shots & Backend & EM? & Score & Exec. Time (s) & Energy (kWh) \\
\midrule
\endfirsthead
\toprule
Domain & Problem & Algorithm & \#q & \#qc & \#1q & \#2q & Shots & Backend & EM? & Score & Exec. Time (s) & Energy (kWh) \\
\midrule
\endhead
\midrule
\multicolumn{13}{r}{Continued on next page} \\
\midrule
\endfoot
\bottomrule
\endlastfoot
unstructured search & faa & faa & 6 & 1 & 1,006 & 432 & 1,000 & IonQ Aria & N & 0.106 & 316.995 & NaN \\
unstructured search & faa & faa & 6 & 1 & 1,006 & 432 & 1,000 & IonQ Aria & N & 0.107 & 316.918 & NaN \\
unstructured search & faa & faa & 6 & 1 & 1,006 & 432 & 1,000 & IonQ Aria & N & 0.118 & 318.627 & NaN \\
unstructured search & faa & faa & 6 & 1 & 1,016 & 432 & 1,000 & IonQ Aria & N & 0.07 & 319.763 & NaN \\
unstructured search & faa & faa & 6 & 1 & 1,016 & 432 & 1,000 & IonQ Aria & N & 0.071 & 319.819 & NaN \\
unstructured search & faa & faa & 6 & 1 & 1,016 & 432 & 1,000 & IonQ Aria & N & 0.075 & 313.121 & NaN \\
unstructured search & faa & faa & 6 & 1 & 1,016 & 432 & 1,000 & IonQ Aria & N & 0.084 & 309.759 & NaN \\
unstructured search & faa & faa & 6 & 1 & 1,016 & 432 & 1,000 & IonQ Aria & N & 0.085 & 319.232 & NaN \\
\end{longtable}

\newpage
\begin{longtable}{lllrrrrrllrrr}
\toprule
Domain & Problem & Algorithm & \#q & \#qc & \#1q & \#2q & Shots & Backend & EM? & Score & Exec. Time (s) & Energy (kWh) \\
\midrule
\endfirsthead
\toprule
Domain & Problem & Algorithm & \#q & \#qc & \#1q & \#2q & Shots & Backend & EM? & Score & Exec. Time (s) & Energy (kWh) \\
\midrule
\endhead
\midrule
\multicolumn{13}{r}{Continued on next page} \\
\midrule
\endfoot
\bottomrule
\endlastfoot
unstructured search & faa & faa & 6 & 1 & 2,146 & 912 & 1,000 & IonQ Aria & N & 0.058 & 599.946 & NaN \\
unstructured search & faa & faa & 6 & 1 & 2,146 & 912 & 1,000 & IonQ Aria & N & 0.059 & 600.803 & NaN \\
unstructured search & faa & faa & 6 & 1 & 2,166 & 912 & 1,000 & IonQ Aria & N & 0.038 & 600.81 & NaN \\
unstructured search & faa & faa & 6 & 1 & 2,166 & 912 & 1,000 & IonQ Aria & N & 0.042 & 594.795 & NaN \\
unstructured search & faa & faa & 6 & 1 & 2,166 & 912 & 1,000 & IonQ Aria & N & 0.045 & 599.712 & NaN \\
unstructured search & faa & faa & 6 & 1 & 2,166 & 912 & 1,000 & IonQ Aria & N & 0.047 & 607.013 & NaN \\
unstructured search & faa & faa & 6 & 1 & 2,166 & 912 & 1,000 & IonQ Aria & N & 0.052 & 597.946 & NaN \\
unstructured search & faa & faa & 6 & 1 & 2,166 & 912 & 1,000 & IonQ Aria & N & 0.068 & 598.842 & NaN \\
unstructured search & faa & faa & 6 & 1 & 2,186 & 912 & 1,000 & IonQ Aria & N & 0.043 & 602.782 & NaN \\
\end{longtable}

}

\end{appendices}

\end{document}